\documentclass[12pt,preprint]{aastex}
\usepackage{lineno}
\usepackage{url}
\shorttitle{\Fermi view of Mrk\,501}
\shortauthors{Abdo et al.}

\newcommand{\Fermic}{\textit{Fermi}}
\newcommand{\Fermi}{\Fermic\ }
\newcommand{\FermiLATc}{\Fermic-LAT}
\newcommand{\FermiLAT}{\FermiLATc\ }

\newcommand{\Swiftc}{\textit{Swift}}
\newcommand{\Swift}{\Swiftc\ }

\newcommand{\RXTEc}{\textit{RXTE}}
\newcommand{\RXTE}{\RXTEc\ }

\begin{document}

\title{Insights Into the High-energy $\gamma$-ray Emission of Markarian~501 from Extensive Multifrequency Observations in the \Fermi Era}
\author{
A.~A.~Abdo\altaffilmark{2,3}, 
M.~Ackermann\altaffilmark{4}, 
M.~Ajello\altaffilmark{4}, 
A.~Allafort\altaffilmark{4}, 
L.~Baldini\altaffilmark{5}, 
J.~Ballet\altaffilmark{6}, 
G.~Barbiellini\altaffilmark{7,8}, 
M.~G.~Baring\altaffilmark{9}, 
D.~Bastieri\altaffilmark{10,11}, 
K.~Bechtol\altaffilmark{4}
% ...
\\
(The \FermiLAT collaboration) \\
J.~Aleksi\'c\altaffilmark{64}, 
L.~A.~Antonelli\altaffilmark{65}, 
P.~Antoranz\altaffilmark{66}, 
M.~Backes\altaffilmark{67}, 
J.~A.~Barrio\altaffilmark{68}, 
J.~Becerra~Gonz\'alez\altaffilmark{69,70}, 
W.~Bednarek\altaffilmark{71}, 
A.~Berdyugin\altaffilmark{72}, 
K.~Berger\altaffilmark{70}, 
E.~Bernardini\altaffilmark{73}
% ...
\\
(The MAGIC collaboration) \\ 
V.~A.~Acciari\altaffilmark{88}, 
T.~Arlen\altaffilmark{89}, 
T.~Aune\altaffilmark{50}, 
W.~Benbow\altaffilmark{88}, 
D.~Boltuch\altaffilmark{90}, 
S.~M.~Bradbury\altaffilmark{91}, 
J.~H.~Buckley\altaffilmark{92}, 
V.~Bugaev\altaffilmark{92}, 
A.~Cannon\altaffilmark{21}, 
A.~Cesarini\altaffilmark{93}, 
L.~Ciupik\altaffilmark{94}, 
W.~Cui\altaffilmark{95}, 
R.~Dickherber\altaffilmark{92}, 
M.~Errando\altaffilmark{96}, 
A.~Falcone\altaffilmark{97}, 
J.~P.~Finley\altaffilmark{95}, 
G.~Finnegan\altaffilmark{98}, 
L.~Fortson\altaffilmark{94}, 
A.~Furniss\altaffilmark{50}, 
N.~Galante\altaffilmark{88}, 
D.~Gall\altaffilmark{95}, 
G.~H.~Gillanders\altaffilmark{93}, 
S.~Godambe\altaffilmark{98}, 
J.~Grube\altaffilmark{94}, 
R.~Guenette\altaffilmark{99}, 
G.~Gyuk\altaffilmark{94}, 
D.~Hanna\altaffilmark{99}, 
J.~Holder\altaffilmark{90}, 
D.~Huang\altaffilmark{100}, 
C.~M.~Hui\altaffilmark{98}, 
T.~B.~Humensky\altaffilmark{101}, 
P.~Kaaret\altaffilmark{102}, 
N.~Karlsson\altaffilmark{94}, 
M.~Kertzman\altaffilmark{103}, 
D.~Kieda\altaffilmark{98}, 
A.~Konopelko\altaffilmark{100}, 
H.~Krawczynski\altaffilmark{92}, 
F.~Krennrich\altaffilmark{104}, 
M.~J.~Lang\altaffilmark{93}, 
G.~Maier\altaffilmark{73,99}, 
S.~McArthur\altaffilmark{92}, 
A.~McCann\altaffilmark{99}, 
M.~McCutcheon\altaffilmark{99}, 
P.~Moriarty\altaffilmark{105}, 
R.~Mukherjee\altaffilmark{96}, 
R.~Ong\altaffilmark{89}, 
A.~N.~Otte\altaffilmark{50}, 
D.~Pandel\altaffilmark{102}, 
J.~S.~Perkins\altaffilmark{88}, 
A.~Pichel\altaffilmark{106}, 
M.~Pohl\altaffilmark{73,107}, 
J.~Quinn\altaffilmark{21}, 
K.~Ragan\altaffilmark{99}, 
L.~C.~Reyes\altaffilmark{108}, 
P.~T.~Reynolds\altaffilmark{109}, 
E.~Roache\altaffilmark{88}, 
H.~J.~Rose\altaffilmark{91}, 
A.~C.~Rovero\altaffilmark{106}, 
M.~Schroedter\altaffilmark{104}, 
G.~H.~Sembroski\altaffilmark{95}, 
G.~D.~Senturk\altaffilmark{110}, 
D.~Steele\altaffilmark{94,111}, 
S.~P.~Swordy\altaffilmark{112,101}, 
G.~Te\v{s}i\'c\altaffilmark{99}, 
M.~Theiling\altaffilmark{88}, 
S.~Thibadeau\altaffilmark{92}, 
A.~Varlotta\altaffilmark{95}, 
S.~Vincent\altaffilmark{98}, 
S.~P.~Wakely\altaffilmark{101}, 
J.~E.~Ward\altaffilmark{21}, 
T.~C.~Weekes\altaffilmark{88}, 
A.~Weinstein\altaffilmark{89}, 
T.~Weisgarber\altaffilmark{101}, 
D.~A.~Williams\altaffilmark{50}, 
M.~Wood\altaffilmark{89}, 
B.~Zitzer\altaffilmark{95}\\
(The VERITAS collaboration) \\
M.~Villata\altaffilmark{122}, 
C.~M.~Raiteri\altaffilmark{122}, 
H.~D.~Aller\altaffilmark{113}, 
M.~F.~Aller\altaffilmark{113}, 
A.~A.~Arkharov\altaffilmark{114}, 
D.~A.~Blinov\altaffilmark{114}, 
P.~Calcidese\altaffilmark{115}, 
W.~P.~Chen\altaffilmark{116}, 
N.~V.~Efimova\altaffilmark{114,117}, 
G.~Kimeridze\altaffilmark{118}, 
T.~S.~Konstantinova\altaffilmark{117}, 
E.~N.~Kopatskaya\altaffilmark{117}, 
E.~Koptelova\altaffilmark{116}, 
O.~M.~Kurtanidze\altaffilmark{118}, 
S.~O.~Kurtanidze\altaffilmark{118}, 
A.~L\"ahteenm\"aki\altaffilmark{119}, 
V.~M.~Larionov\altaffilmark{120,114,117}, 
E.~G.~Larionova\altaffilmark{117}, 
L.~V.~Larionova\altaffilmark{117}, 
R.~Ligustri\altaffilmark{121}, 
D.~A.~Morozova\altaffilmark{117}, 
M.~G.~Nikolashvili\altaffilmark{118}, 
L.~A.~Sigua\altaffilmark{118}, 
I.~S.~Troitsky\altaffilmark{117}, 
E.~Angelakis\altaffilmark{34}, 
M.~Capalbi\altaffilmark{23}, 
A.~Carrami\~nana\altaffilmark{123}, 
L.~Carrasco\altaffilmark{123}, 
P.~Cassaro\altaffilmark{124}, 
E.~de~la~Fuente\altaffilmark{137},
M.~A.~Gurwell\altaffilmark{125}, 
Y.~Y.~Kovalev\altaffilmark{126,34}, 
Yu.~A.~Kovalev\altaffilmark{126}, 
T.~P.~Krichbaum\altaffilmark{34}, 
H.~A.~Krimm\altaffilmark{24,40}, 
P.~Leto\altaffilmark{127}, 
M.~L.~Lister\altaffilmark{95}, 
G.~Maccaferri\altaffilmark{128}, 
J.~W.~Moody\altaffilmark{129}, 
Y.~Mori\altaffilmark{130}, 
I.~Nestoras\altaffilmark{34}, 
A.~Orlati\altaffilmark{128}, 
C.~Pagani\altaffilmark{131}, 
C.~Pace\altaffilmark{129}, 
R.~Pearson III\altaffilmark{129}, 
M.~Perri\altaffilmark{23}, 
B.~G.~Piner\altaffilmark{132}, 
A.~B.~Pushkarev\altaffilmark{133,34,114}, 
E.~Ros\altaffilmark{34,134}, 
A.~C.~Sadun\altaffilmark{135}, 
T.~Sakamoto\altaffilmark{20}, 
M.~Tornikoski\altaffilmark{119}, 
Y.~Yatsu\altaffilmark{130}, 
A.~Zook\altaffilmark{136}
}
\altaffiltext{1}{Corresponding authors: D.~Paneque,
dpaneque@mppmu.mpg.de, ~{\L}.~Stawarz, stawarz@astro.isas.jaxa.jp.}
\altaffiltext{2}{Space Science Division, Naval Research Laboratory, Washington, DC 20375, USA}
\altaffiltext{3}{National Research Council Research Associate, National Academy of Sciences, Washington, DC 20001, USA}
\altaffiltext{4}{W. W. Hansen Experimental Physics Laboratory, Kavli Institute for Particle Astrophysics and Cosmology, Department of Physics and SLAC National Accelerator Laboratory, Stanford University, Stanford, CA 94305, USA}
\altaffiltext{5}{Istituto Nazionale di Fisica Nucleare, Sezione di Pisa, I-56127 Pisa, Italy}
\altaffiltext{6}{Laboratoire AIM, CEA-IRFU/CNRS/Universit\'e Paris Diderot, Service d'Astrophysique, CEA Saclay, 91191 Gif sur Yvette, France}
\altaffiltext{7}{Istituto Nazionale di Fisica Nucleare, Sezione di Trieste, I-34127 Trieste, Italy}
\altaffiltext{8}{Dipartimento di Fisica, Universit\`a di Trieste, I-34127 Trieste, Italy}
\altaffiltext{9}{Rice University, Department of Physics and Astronomy, MS-108, P. O. Box 1892, Houston, TX 77251, USA}
\altaffiltext{10}{Istituto Nazionale di Fisica Nucleare, Sezione di Padova, I-35131 Padova, Italy}
\altaffiltext{11}{Dipartimento di Fisica ``G. Galilei", Universit\`a di Padova, I-35131 Padova, Italy}
\altaffiltext{12}{Istituto Nazionale di Fisica Nucleare, Sezione di Perugia, I-06123 Perugia, Italy}
\altaffiltext{13}{Dipartimento di Fisica, Universit\`a degli Studi di Perugia, I-06123 Perugia, Italy}
\altaffiltext{14}{Centre d'\'Etude Spatiale des Rayonnements, CNRS/UPS, BP 44346, F-30128 Toulouse Cedex 4, France}
\altaffiltext{15}{Department of Physics, Center for Cosmology and Astro-Particle Physics, The Ohio State University, Columbus, OH 43210, USA}
\altaffiltext{16}{Dipartimento di Fisica ``M. Merlin" dell'Universit\`a e del Politecnico di Bari, I-70126 Bari, Italy}
\altaffiltext{17}{Istituto Nazionale di Fisica Nucleare, Sezione di Bari, 70126 Bari, Italy}
\altaffiltext{18}{Laboratoire Leprince-Ringuet, \'Ecole polytechnique, CNRS/IN2P3, Palaiseau, France}
\altaffiltext{19}{Institut de Ciencies de l'Espai (IEEC-CSIC), Campus UAB, 08193 Barcelona, Spain}
\altaffiltext{20}{NASA Goddard Space Flight Center, Greenbelt, MD 20771, USA}
\altaffiltext{21}{University College Dublin, Belfield, Dublin 4, Ireland}
\altaffiltext{22}{INAF-Istituto di Astrofisica Spaziale e Fisica Cosmica, I-20133 Milano, Italy}
\altaffiltext{23}{Agenzia Spaziale Italiana (ASI) Science Data Center, I-00044 Frascati (Roma), Italy}
\altaffiltext{24}{Center for Research and Exploration in Space Science and Technology (CRESST) and NASA Goddard Space Flight Center, Greenbelt, MD 20771, USA}
\altaffiltext{25}{Department of Physics and Center for Space Sciences and Technology, University of Maryland Baltimore County, Baltimore, MD 21250, USA}
\altaffiltext{26}{George Mason University, Fairfax, VA 22030, USA}
\altaffiltext{27}{Laboratoire de Physique Th\'eorique et Astroparticules, Universit\'e Montpellier 2, CNRS/IN2P3, Montpellier, France}
\altaffiltext{28}{Department of Physics, Stockholm University, AlbaNova, SE-106 91 Stockholm, Sweden}
\altaffiltext{29}{The Oskar Klein Centre for Cosmoparticle Physics, AlbaNova, SE-106 91 Stockholm, Sweden}
\altaffiltext{30}{Royal Swedish Academy of Sciences Research Fellow, funded by a grant from the K. A. Wallenberg Foundation}
\altaffiltext{31}{Universit\'e Bordeaux 1, CNRS/IN2p3, Centre d'\'Etudes Nucl\'eaires de Bordeaux Gradignan, 33175 Gradignan, France}
\altaffiltext{32}{Dipartimento di Fisica, Universit\`a di Udine and Istituto Nazionale di Fisica Nucleare, Sezione di Trieste, Gruppo Collegato di Udine, I-33100 Udine, Italy}
\altaffiltext{33}{Osservatorio Astronomico di Trieste, Istituto Nazionale di Astrofisica, I-34143 Trieste, Italy}
\altaffiltext{34}{Max-Planck-Institut f\"ur Radioastronomie, Auf dem H\"ugel 69, 53121 Bonn, Germany}
\altaffiltext{35}{Department of Physical Sciences, Hiroshima University, Higashi-Hiroshima, Hiroshima 739-8526, Japan}
\altaffiltext{36}{INAF Istituto di Radioastronomia, 40129 Bologna, Italy}
\altaffiltext{37}{Center for Space Plasma and Aeronomic Research (CSPAR), University of Alabama in Huntsville, Huntsville, AL 35899, USA}
\altaffiltext{38}{Dr. Remeis-Sternwarte Bamberg, Sternwartstrasse 7, D-96049 Bamberg, Germany}
\altaffiltext{39}{Erlangen Centre for Astroparticle Physics, D-91058 Erlangen, Germany}
\altaffiltext{40}{Universities Space Research Association (USRA), Columbia, MD 21044, USA}
\altaffiltext{41}{Research Institute for Science and Engineering, Waseda University, 3-4-1, Okubo, Shinjuku, Tokyo, 169-8555 Japan}
\altaffiltext{42}{Cahill Center for Astronomy and Astrophysics, California Institute of Technology, Pasadena, CA 91125, USA}
\altaffiltext{43}{Department of Physics and Department of Astronomy, University of Maryland, College Park, MD 20742, USA}
\altaffiltext{44}{Istituto Nazionale di Fisica Nucleare, Sezione di Roma ``Tor Vergata", I-00133 Roma, Italy}
\altaffiltext{45}{Department of Physics and Astronomy, University of Denver, Denver, CO 80208, USA}
\altaffiltext{46}{Hiroshima Astrophysical Science Center, Hiroshima University, Higashi-Hiroshima, Hiroshima 739-8526, Japan}
\altaffiltext{47}{Institute of Space and Astronautical Science, JAXA, 3-1-1 Yoshinodai, Sagamihara, Kanagawa 229-8510, Japan}
\altaffiltext{48}{Max-Planck Institut f\"ur extraterrestrische Physik, 85748 Garching, Germany}
\altaffiltext{49}{Institut f\"ur Astro- und Teilchenphysik and Institut f\"ur Theoretische Physik, Leopold-Franzens-Universit\"at Innsbruck, A-6020 Innsbruck, Austria}
\altaffiltext{50}{Santa Cruz Institute for Particle Physics, Department of Physics and Department of Astronomy and Astrophysics, University of California at Santa Cruz, Santa Cruz, CA 95064, USA}
\altaffiltext{51}{Department of Physics, University of Washington, Seattle, WA 98195-1560, USA}
\altaffiltext{52}{Space Sciences Division, NASA Ames Research Center, Moffett Field, CA 94035-1000, USA}
\altaffiltext{53}{NYCB Real-Time Computing Inc., Lattingtown, NY 11560-1025, USA}
\altaffiltext{54}{Astronomical Observatory, Jagiellonian University, 30-244 Krak\'ow, Poland}
\altaffiltext{55}{Department of Chemistry and Physics, Purdue University Calumet, Hammond, IN 46323-2094, USA}
\altaffiltext{56}{Partially supported by the International Doctorate on Astroparticle Physics (IDAPP) program}
\altaffiltext{57}{Instituci\'o Catalana de Recerca i Estudis Avan\c{c}ats (ICREA), Barcelona, Spain}
\altaffiltext{58}{Consorzio Interuniversitario per la Fisica Spaziale (CIFS), I-10133 Torino, Italy}
\altaffiltext{59}{INTEGRAL Science Data Centre, CH-1290 Versoix, Switzerland}
\altaffiltext{60}{Dipartimento di Fisica, Universit\`a di Roma ``Tor Vergata", I-00133 Roma, Italy}
\altaffiltext{61}{Space Science Institute, Boulder, CO 80301, USA}
\altaffiltext{62}{Department of Physics, Royal Institute of Technology (KTH), AlbaNova, SE-106 91 Stockholm, Sweden}
\altaffiltext{63}{School of Pure and Applied Natural Sciences, University of Kalmar, SE-391 82 Kalmar, Sweden}
\altaffiltext{64}{Institut de F\'isica d'Altes Energies (IFAE), Edifici Cn, Universitat Aut\`onoma de Barcelona (UAB), E-08193 Bellaterra (Barcelona), Spain}
\altaffiltext{65}{INAF National Institute for Astrophysics, I-00136 Roma, Italy}
\altaffiltext{66}{Universit\`a di Siena and INFN Pisa, I-53100 Siena, Italy}
\altaffiltext{67}{Technische Universit\"at Dortmund, D-44221 Dortmund, Germany}
\altaffiltext{68}{Universidad Complutense, E-28040 Madrid, Spain}
\altaffiltext{69}{Instituto de Astrof\'isica de Canarias, E38205 - La Laguna (Tenerife), Spain}
\altaffiltext{70}{Departamento de Astrofisica, Universidad de La Laguna, E-38205 La Laguna, Tenerife, Spain}
\altaffiltext{71}{University of  {\L}\'od\'z, PL-90236 {\L}\'od\'z, Poland}
\altaffiltext{72}{Tuorla Observatory, University of Turku, FI-21500 Piikki\"o, Finland}
\altaffiltext{73}{Deutsches Elektronen Synchrotron DESY, D-15738 Zeuthen, Germany}
\altaffiltext{74}{ETH Zurich, CH-8093 Zurich, Switzerland}
\altaffiltext{75}{Max-Planck-Institut f\"ur Physik, D-80805 M\"unchen, Germany}
\altaffiltext{76}{Universitat de Barcelona (ICC/IEED), E-08028 Barcelona, Spain}
\altaffiltext{77}{Institut f\"ur Theoretische Physik and Astrophysik, Universit\"at W\"urzburg, D-97074 W\"urzburg, Germany}
\altaffiltext{78}{Supported by INFN Padova}
\altaffiltext{79}{Istituto Nazionale di Fisica Nucleare, Sezione di Trieste, and Universit\`a di Trieste, I-34127 Trieste, Italy}
\altaffiltext{80}{Centro de Investigaciones Energ\'eticas, Medioambientales y Tecnol\'ogicas (CIEMAT), Madrid, Spain}
\altaffiltext{81}{Instituto de Astrof\'isica de Andaluc\'ia, CSIC, E-18080 Granada, Spain}
\altaffiltext{82}{Croatian MAGIC Consortium, Institute R. Bo\v{s}kovi\'c, University of Rijeka and University of Split, HR-10000 Zagreb, Croatia}
\altaffiltext{83}{Universitat Aut\'onoma de Barcelona, E-08193 Bellaterra, Spain}
\altaffiltext{84}{Institute for Nuclear Research and Nuclear Energy, BG-1784 Sofia, Bulgaria}
\altaffiltext{85}{INAF Osservatorio Astronomico di Brera, I-23807 Merate, Italy}
\altaffiltext{86}{INAF Osservatorio Astronomico di Trieste, I-34143 Trieste, Italy}
\altaffiltext{87}{Dipartimento di Fisica ``Enrico Fermi", Universit\`a di Pisa, Pisa I-56127, Italy}
\altaffiltext{88}{Fred Lawrence Whipple Observatory, Harvard-Smithsonian Center for Astrophysics, Amado, AZ 85645, USA}
\altaffiltext{89}{Department of Physics and Astronomy, University of California, Los Angeles, CA 90095-1547, USA}
\altaffiltext{90}{Department of Physics and Astronomy and the Bartol Research Institute, University of Delaware, Newark, DE 19716, USA}
\altaffiltext{91}{School of Physics and Astronomy, University of Leeds, Leeds LS2 9JT, UK}
\altaffiltext{92}{Department of Physics, Washington University, St. Louis, MO 63130, USA}
\altaffiltext{93}{National University of Ireland Galway, University Road, Galway, Ireland}
\altaffiltext{94}{Adler Planetarium and Astronomy Museum, Chicago, IL 60605, USA}
\altaffiltext{95}{Department of Physics, Purdue University, West Lafayette, IN 47907, USA}
\altaffiltext{96}{Department of Physics and Astronomy, Barnard College, Columbia University, New York, NY 10027, USA}
\altaffiltext{97}{Department of Astronomy and Astrophysics, Pennsylvania State University, University Park, PA 16802, USA}
\altaffiltext{98}{Department of Physics and Astronomy, University of Utah, Salt Lake City, UT 84112, USA}
\altaffiltext{99}{Department of Physics, McGill University, Montreal, PQ, Canada H3A 2T8}
\altaffiltext{100}{Department of Physics, Pittsburg State University, Pittsburg, KS 66762, USA}
\altaffiltext{101}{Enrico Fermi Institute, University of Chicago, Chicago, IL 60637, USA}
\altaffiltext{102}{Department of Physics and Astronomy, University of Iowa, Iowa City, IA 52242, USA}
\altaffiltext{103}{Department of Physics and Astronomy, DePauw University, Greencastle, IN 46135-0037, USA}
\altaffiltext{104}{Department of Physics and Astronomy, Iowa State University, Ames, IA 50011-3160, USA}
\altaffiltext{105}{Department of Life and Physical Sciences, Galway-Mayo Institute of Technology, Galway, Ireland}
\altaffiltext{106}{Instituto de Astronom\'ia y Fisica del Espacio, Parbell\'on IAFE, Cdad. Universitaria, Buenos Aires, Argentina}
\altaffiltext{107}{Institut f\"ur Physik und Astronomie, Universit\"at Potsdam, 14476 Potsdam, Germany}
\altaffiltext{108}{Kavli Institute for Cosmological Physics, University of Chicago, Chicago, IL 60637, USA}
\altaffiltext{109}{Department of Applied Physics and Instrumentation, Cork Institute of Technology, Bishopstown, Cork, Ireland}
\altaffiltext{110}{Columbia Astrophysics Laboratory, Columbia University, New York, NY 10027, USA}
\altaffiltext{111}{Los Alamos National Laboratory, Los Alamos, NM 87545, USA}
\altaffiltext{112}{Deceased}
\altaffiltext{113}{Department of Astronomy, University of Michigan, Ann Arbor, MI 48109-1042, USA}
\altaffiltext{114}{Pulkovo Observatory, 196140 St. Petersburg, Russia}
\altaffiltext{115}{Osservatorio Astronomico della Regione Autonoma Valle d'Aosta, Italy}
\altaffiltext{116}{Graduate Institute of Astronomy, National Central University, Jhongli 32054, Taiwan}
\altaffiltext{117}{Astronomical Institute, St. Petersburg State University, St. Petersburg, Russia}
\altaffiltext{118}{Abastumani Observatory, Mt. Kanobili, 0301 Abastumani, Georgia}
\altaffiltext{119}{Aalto University Mets\"ahovi Radio Observatory, FIN-02540 Kylm\"al\"a, Finland}
%\altaffiltext{119}{Mets\"ahovi Radio Observatory, Helsinki University of Technology TKK, FIN-02540 Kylmala, Finland}
\altaffiltext{120}{Isaac Newton Institute of Chile, St. Petersburg Branch, St. Petersburg, Russia}
\altaffiltext{121}{Circolo Astrofili Talmassons, I-33030 Campoformido (UD), Italy}
\altaffiltext{122}{INAF, Osservatorio Astronomico di Torino, I-10025 Pino Torinese (TO), Italy}
\altaffiltext{123}{Instituto Nacional de Astrof\'isica, \'Optica y Electr\'onica, Tonantzintla, Puebla 72840, Mexico}
\altaffiltext{124}{INAF Istituto di Radioastronomia, Sezione di Noto, Contrada Renna Bassa, 96017 Noto (SR), Italy}
\altaffiltext{125}{Harvard-Smithsonian Center for Astrophysics, Cambridge, MA 02138, USA}
\altaffiltext{126}{Astro Space Center of the Lebedev Physical Institute, 117810 Moscow, Russia}
\altaffiltext{127}{Osservatorio Astrofisico di Catania, 95123 Catania, Italy}
\altaffiltext{128}{INAF Istituto di Radioastronomia, Stazione Radioastronomica di Medicina, I-40059 Medicina (Bologna), Italy}
\altaffiltext{129}{Department of Physics and Astronomy, Brigham Young University, Provo Utah 84602, USA}
\altaffiltext{130}{Department of Physics, Tokyo Institute of Technology, Meguro City, Tokyo 152-8551, Japan}
\altaffiltext{131}{Department of Physics and Astronomy, University of Leicester, Leicester, LE1 7RH, UK}
\altaffiltext{132}{Department of Physics and Astronomy, Whittier College, Whittier, CA, USA}
\altaffiltext{133}{Crimean Astrophysical Observatory, 98409 Nauchny, Crimea, Ukraine}
\altaffiltext{134}{Universitat de Val\`encia, 46010 Val\`encia, Spain}
\altaffiltext{135}{Department of Physics, University of Colorado, Denver, CO 80220, USA}
\altaffiltext{136}{Department of Physics and Astronomy, Pomona College, Claremont CA 91711-6312, USA}
\altaffiltext{137}{Instututo de Astronomia y Meteorologia, Dpto. de Fisica, CUCEI, Universidad de Guadalajara}

\begin{abstract}

We report on the $\gamma$-ray activity of the blazar Mrk\,501 during
the first 480 days of \Fermi operation. We find that the average LAT
$\gamma$-ray spectrum of Mrk\,501 can be well described by a single
power-law function with a photon index of $1.78 \pm 0.03$. While we
observe relatively mild flux variations with the \FermiLAT (within
less than a factor of 2), we detect remarkable spectral variability
where the hardest observed spectral index within the LAT energy range
is $1.52 \pm 0.14$, and the softest one is $2.51 \pm 0.20$. These
unexpected spectral changes do not correlate with the measured flux
variations above $0.3$\,GeV. In this paper, we also present the first
results from the 4.5-month-long multifrequency campaign (2009 March
15 -- August 1) on Mrk\,501, which included the VLBA, \Swiftc,
\RXTEc, MAGIC and VERITAS, the  F-GAMMA, GASP-WEBT, and other
collaborations and instruments which provided excellent time and
energy coverage of the source throughout the entire campaign.  The
extensive radio to TeV data set from this campaign provides us with
the most detailed spectral energy distribution yet collected for this
source during its relatively low activity. The average spectral energy
distribution of Mrk\,501 is well described by the standard one-zone
synchrotron self-Compton model. In the framework of this model, we
find that the dominant emission region is characterized by a size
$\lesssim 0.1$\,pc (comparable within a factor of few to the size of
the partially-resolved VLBA core at 15-43 GHz), and that the total jet power ($\simeq 10^{44}$\,erg\,s$^{-1}$) constitutes only a small fraction ($\sim 10^{-3}$) of the Eddington luminosity. The energy distribution of the freshly-accelerated radiating electrons required to fit the time-averaged data has a broken power-law form in the energy range $0.3$\,GeV$-10$\,TeV, with spectral indices 2.2 and 2.7 below and above the break energy of $20$\,GeV. We argue that such a form is consistent with a scenario in which the bulk of the energy dissipation within the dominant emission zone of Mrk\,501 is due to relativistic, proton-mediated shocks. We find that the ultrarelativistic electrons and mildly relativistic protons within the blazar zone, if comparable in number, are in approximate energy equipartition,  with their energy dominating the jet magnetic field energy by about two orders of magnitude.

\end{abstract}


\section{Introduction}
\label{Intro}

Blazars constitute a subclass of radio-loud active galactic nuclei (AGN), in which a jet of magnetized plasma assumed to emanate with relativistic bulk velocity from close to a central supermassive black hole points almost along the line of sight. The broadband emission spectra of these objects are dominated by non-thermal, strongly Doppler-boosted and variable radiation produced in the innermost part of the jet. Most of the identified extragalactic $\gamma$-ray sources detected with the EGRET instrument (Hartman et al. 1999) on board the {\em Compton} Gamma Ray Observatory belong to this category. Blazars include flat-spectrum radio quasars (FSRQs) and BL Lacertae objects (BL Lacs).
Even though blazars have been observed for several decades at different frequencies, the existing experimental data did not permit unambiguous identification of the physical mechanisms responsible for the  production of their high-energy ($\gamma$-ray) emission. Given the existing high-sensitivity detectors which allow detailed study of the low-energy (synchrotron) component of blazar sources (extending from radio up to hard X-rays), one of the reasons for the incomplete understanding of those objects was only moderate sensitivity of previous $\gamma$-ray instruments. This often precluded detailed cross-correlation studies between the low- and high-energy emission and did not provide enough constraints on the parameters of the theoretical models. Some of the open and fundamental questions regarding blazar sources are  (i) the content of their jets, (ii) the location and structure of their dominant emission zones, (iii) the origin of their variability, observed on timescales from minutes to tens of years, (iv) the role of external photon fields (including the extragalactic background light, EBL) in shaping their observed $\gamma$-ray spectra, and (v) the energy distribution and the dominant acceleration mechanism for the underlying radiating particles.

The Large Area Telescope (LAT) instrument \citep{Atwood2009} on board
the \Fermi Gamma-ray Space Telescope satellite provides a large improvement in the experimental
capability for performing $\gamma$-ray astronomy, and hence it is
shedding new light on the blazar phenomenon. In this paper, we report
on the \Fermi observations of the TeV-emitting high-frequency-peaked
--- or, according to a more recent classification
\citep{SEDFermiBlazars}, high-synchrotron-peaked (HSP) --- BL Lac
object Markarian\,501 (Mrk\,501; RA=16$^h$ 45$^m$ 52.22$^s$, Dec=
39$^\circ$ 45' 36.6" , J2000,  redshift $z = 0.034$), which is one of
the brightest extragalactic sources in the X-ray/TeV sky. Mrk 501 was
the second extragalactic object (after Markarian\,421) identified as a
very high energy (thereafter VHE) $\gamma$-ray emitter
\citep{Quinn1996,Bradbury1997}. After a phase of moderate emission
lasting for about a year following its discovery (1996), Mrk\,501 went into a state of surprisingly high activity and strong variability, becoming $>$\,10 times brighter than the Crab Nebula at energies $> 1$\,TeV, as reported by various instruments/groups \citep{Catanese1997,Samuelson1998,Aharonian1999a,Aharonian1999b,Aharonian1999c,Djannati1999}. In 1998-1999, the mean VHE $\gamma$-ray flux dropped by an order of magnitude, and the overall VHE spectrum softened significantly \citep{Piron2000,Aharonian2001}. In 2005, $\gamma$-ray flux variability on minute timescales was observed in the VHE band, thus establishing Mrk\,501 as one of the sources with the fastest $\gamma$-ray flux changes \citep{Albert2007}. During the 2005 VHE flux variations (when Mrk\,501 was 3--4 times dimmer than it was in 1997), significant spectral variability was detected as well, with a clear ``harder when brighter'' behavior. Those spectral variations are even more pronounced when compared with the spectrum measured during the low activity level recently reported in \citet{Anderhub2009}. 

The spectral energy distribution (SED) and the multifrequency correlations of Mrk\,501 have been intensively studied in the past \citep[e.g.,][]{pia98,Villata1999,Krawczynski2000,Sambruna2000,tav01,katar01,Ghisellini2002,gli06,Anderhub2009}, but the nature of this object is still far from being understood. The main reasons for this lack of knowledge are the sparse multifrequency data during long periods of time, and the moderate sensitivity available in the past to study the $\gamma$-ray emission of this source. Besides, most of the previous multifrequency campaigns were triggered by an enhanced flux level in some energy band, and hence much of our information about the source is biased towards ``high-activity'' states, where perhaps distinct physical processes play a dominant role. In addition, until now we knew very little about the GeV emission of Mrk\,501. The only detection reported at GeV energies before \Fermi was in \cite{Kataoka1999}, but the significance of this detection was too low to include Mrk\,501 in the 3rd EGRET catalog \citep{Hartman1999}. Moreover, Mrk\,501 was not detected by EGRET during the large X-ray and VHE $\gamma$-ray flare which lasted for several months in 1997 \citep{pia98}.

The large improvement in the performance provided by the \FermiLAT
compared with its predecessor, EGRET, provides us with a new
perspective for the study of blazars like Mrk\,501. However, it is
important to emphasize that blazars can vary their emitted power by
one or two orders of magnitude on short timescales, and that they emit
radiation over the entire observable electromagnetic spectrum (from
$\sim 10^{-6}$\,eV up to $\sim 10^{13}$\,eV). For this reason, the
information from \FermiLAT alone  is not enough to understand the
broadband emission of Mrk\,501, and hence simultaneous data in other
frequency ranges are required. In particular, the frequency ranges
where the low- and high-energy spectral components peak in the SED
representation are of major importance. In the case of Mrk\,501, those
peaks are typically located around $1$\,keV (low-energy bump) and
$100$\,GeV (high-energy bump), and hence simultaneous UV/X-ray and
GeV/TeV observations are essential for the proper reconstruction of
the overall SED of Mrk\,501. At TeV energies there has been a
substantial improvement in the instrumental capability as a result of
the deployment of a new generation of imaging atmospheric Cherenkov
telescopes (IACTs). In particular, for the study of Mrk\,501, the new
telescope systems MAGIC and VERITAS provide greater sensitivity, wider
energy range and improved energy resolution compared with the previous
generation of instruments. Simultaneous observations with \FermiLAT
and IACTs like MAGIC or VERITAS (potentially covering six decades in
energy, from $20$\,MeV to $20$\,TeV) can, for the first time,
significantly resolve both the rising and the falling segments of the
high-energy emission component of Mrk\,501, with the expected location
of the SED peak in the overlapping energy range between those
instruments. Because of the smaller collection area, and the self-veto
problem\footnote{The self-veto problem in EGRET is the degradation of
  the effective area at high energies ($>$5~GeV) due to backsplash of
  secondary particles from the calorimeter causing the anticoincidence
  system to veto the event. This problem is substantially reduced in
  LAT by using a segmented anticoincidence detector.}, the sensitivity
of EGRET to detect $\gamma$-rays with energies larger than 10 GeV was about two orders of
magnitude lower than that of \FermiLATc \footnote{This estimate
  includes the larger exposure from \FermiLAT due to the 4 times
  larger field of view.}. Besides, during the period of operation of EGRET, the sensitivity of  the previous generation of IACTs was only moderate, with relatively low sensitivity below $0.5$\,TeV. Therefore, the higher sensitivity and larger energy range of the newer $\gamma$-ray instruments has become a crucial tool for studying Mrk\,501, and the blazar phenomenon in general.

In order to exploit the performance of the \FermiLAT and the new
IACTs, as well as the capabilities of several existing instruments
observing at radio-to-X-ray frequencies, a multifrequency (from radio
to TeV photon energies) campaign was organized to monitor Mrk\,501
during a period of 4.5 months, from mid-March to August 2009.  
%The list of all the instruments participating in the campaign and the
%observing schedule can be found
%online\footnote{\url{https://confluence.slac.stanford.edu/display/GLAMCOG/Campaign+on+Mrk501+(March+2009+to+July+2009)}}. 
The scientific goal was to collect a very complete, simultaneous,
multifrequency data set that would allow current theoretical models of
broadband blazar emission to be tested. This, in turn, should help us
to understand the origin of high-energy emission of blazar sources and
the physical mechanisms responsible for the acceleration of radiating
particles in relativistic jets in general. In this paper, the only
reported result from the multifrequency observations is the overall
SED averaged over the duration of the observing campaign. A more
in-depth analysis of the multifrequency data set will be given in a
forthcoming paper. The scientific results from the data collected
during the two-day time interval 2009 March 23-25 (which includes extensive observations with the Suzaku X-ray satellite) will be reported in a separate paper \citep{Acciari2010}.
The paper is organized as follows. In \S2 we introduce the LAT instrument and describe the LAT data analysis. In \S3 we report on the flux/spectral variability of Mrk\,501 observed during the first 16 months of \FermiLAT operation, and compare it with the flux variability observed in X-rays by the all-sky instruments \RXTE \citep[]{RXTERef} All Sky Monitor (ASM) and the \Swift \citep[]{SwiftRef} Burst Alert Telescope (BAT). In \S4 we analyze the $\gamma$-ray spectrum of Mrk\,501 measured by \FermiLAT in the energy range $0.1-400$\,GeV. \S5 reports on the overall SED obtained during the 4.5-month-long multifrequency campaign organized in 2009. \S6 is devoted to SED modeling, the results of which are further discussed in \S7. Conclusions are presented in \S8.

\section{\FermiLAT Data Selection and Analysis}
\label{FermiData}

The \FermiLAT is an instrument to perform $\gamma$-ray astronomy above
$20$\,MeV. The instrument is an array of $4 \times 4$ identical
towers, each one consisting of a tracker (where the photons are
pair-converted) and a calorimeter (where the energies of the pair-converted photons are measured). The entire instrument is covered with an anticoincidence detector to reject charged-particle background. The LAT has a peak effective area of $0.8$\,m$^2$ for $1$\,GeV photons, an energy resolution typically better than $10\%$ and a field of view (FoV) of about $2.4$\,sr, with an angular resolution  ($68\%$ containment angle) better than $1^{\circ}$ for energies above $1$\,GeV. Further details on the  LAT can be found in \cite{Atwood2009}.

The LAT data reported in this paper were collected from 2008 August 5
 (MJD 54683) to 2009 November 27 (MJD 55162). During this time, the \FermiLAT instrument
operated mostly in survey mode. The analysis was performed with the
Fermi Science Tools software package version \texttt{v9r15p6}. Only
events with the highest probability of being photons --- those in the
``diffuse'' class --- were used. The LAT data were extracted from a
circular region of $10^{\circ}$ radius centered at the location of
Mrk\,501. The spectral fits were performed using photon energies
in the energy range $0.3-400$\,GeV. At photon energies above 0.3 GeV the effective area of the instrument is
relatively large ($>0.5$\,m$^{2}$) and the angular resolution
relatively good ($68\%$ containment angle smaller than
$2^{\circ}$). In particular, because of the better angular resolution,
the spectral fits using energies above $0.3$\,GeV (instead of
$0.1$\,GeV) are less sensitive to possible contamination from
unaccounted (perhaps transient), neighboring $\gamma$-ray sources and hence have smaller systematic errors, at the expense of reducing somewhat the number of photons from the source. In addition, a cut on zenith angle ($> 105^{\circ}$) was applied to reduce contamination from Earth-albedo $\gamma$-rays, which are produced by cosmic rays interacting with the upper atmosphere.

The background model used to extract the $\gamma$-ray signal includes
a Galactic diffuse emission component and an isotropic component. 
The model that we adopted for the Galactic component is
\texttt{gll\_iem\_v02.fit}\footnote{\url{http://fermi.gsfc.nasa.gov/ssc/data/access/lat/BackgroundModels.html}}.
The isotropic component, which is the sum
of the extragalactic diffuse emission and the residual
charged-particle background, is parametrized here with a single
power-law function.  To reduce systematic uncertainties in the
analysis, the photon index of the isotropic 
component and the normalization of both components in the background model were allowed to vary freely during the spectral point fitting. 
Owing to the relatively small size of the region
analyzed (radius 10$^\circ$) and the hardness of the spectrum of
Mrk\,501, the high-energy structure in the standard tabulated
isotropic background spectrum isotropic\_iem\_v02.txt does not dominate
the total counts at high energies. In addition we find that for this
region a power-law approximation to the isotropic background results
in somewhat smaller residuals for the overall model, possibly because
the isotropic term, with a free spectral index, compensates for an
inaccuracy in the model for the Galactic diffuse emission, which is
also approximately isotropic at the high Galactic latitude of Mrk\,501 ($b \sim 39^\circ$).  In any case, the resulting spectral fits for Mrk\,501 are not significantly different if isotropic\_iem\_v02.txt is used for the analysis. In addition, the model also includes five nearby sources from the 1FGL catalog \citep{1FGL}: 1FGL\,J1724.0+4002, 1FGL\,J1642.5+3947, 1FGL\,J1635.0+3808, 1FGL\,J1734.4+3859, and 1FGL\,J1709.6+4320. The spectra of those sources were also parameterized by a power-law functions, whose photon index values were fixed to the values from the 1FGL catalog, and only the normalization factors for the single sources were left as free parameters. The spectral analysis was performed with the post-launch instrument-response functions \texttt{P6\_V3\_DIFFUSE} using an unbinned maximum-likelihood method \citep[]{Mattox1996}. The systematic uncertainties on the flux were estimated as $10\%$ at $0.1$\,GeV, $5\%$ at $560$\,MeV and $20\%$ at $10$\,GeV and above\footnote{\url{http://fermi.gsfc.nasa.gov/ssc/data/analysis/LAT_caveats.html}}.

\section{Flux and Spectral Variability}
\label{LC}

\begin{figure}
  \centering
  \includegraphics[height=2.2in,width=4.0in]{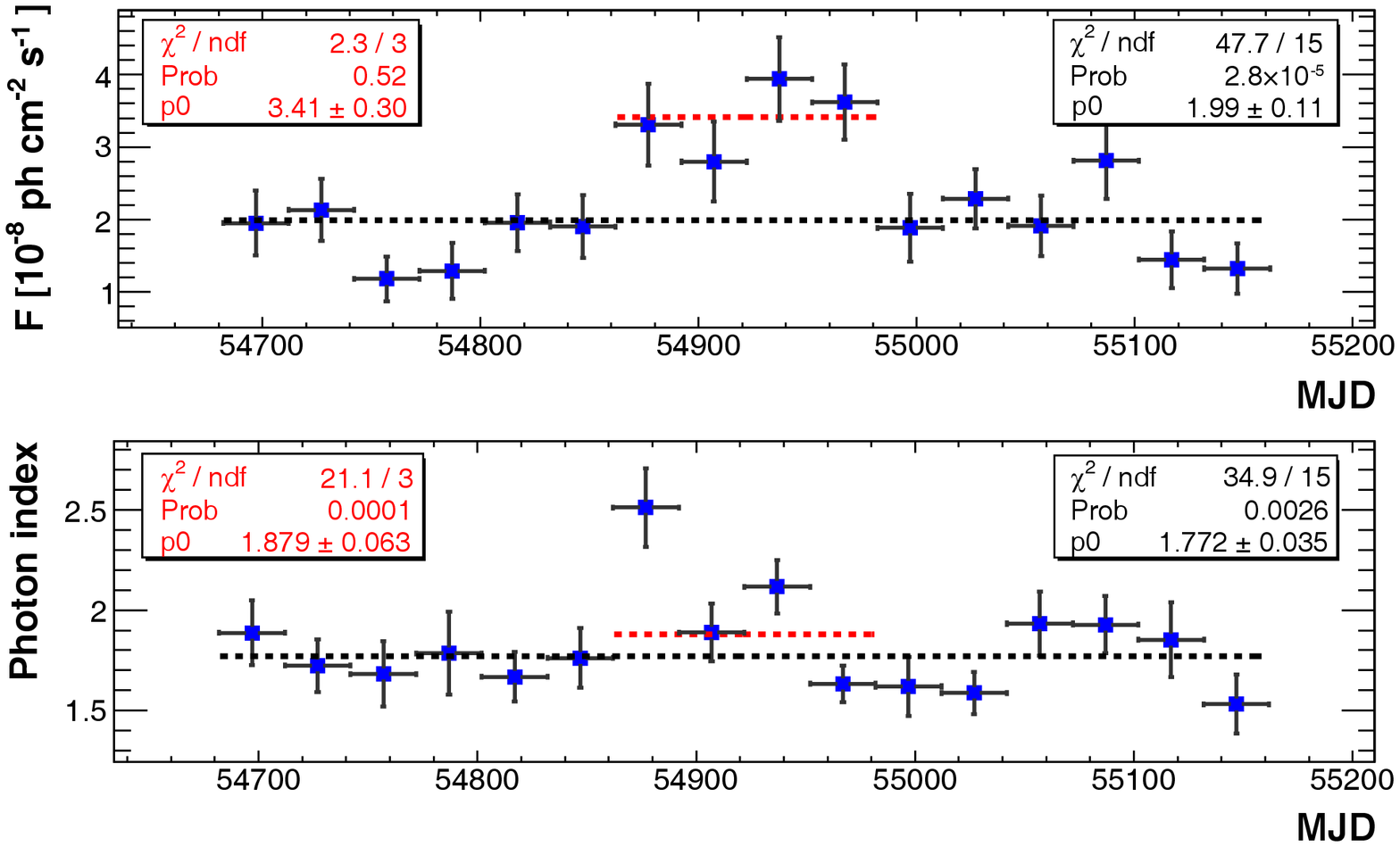}
  \includegraphics[height=2.2in,width=2.4in]{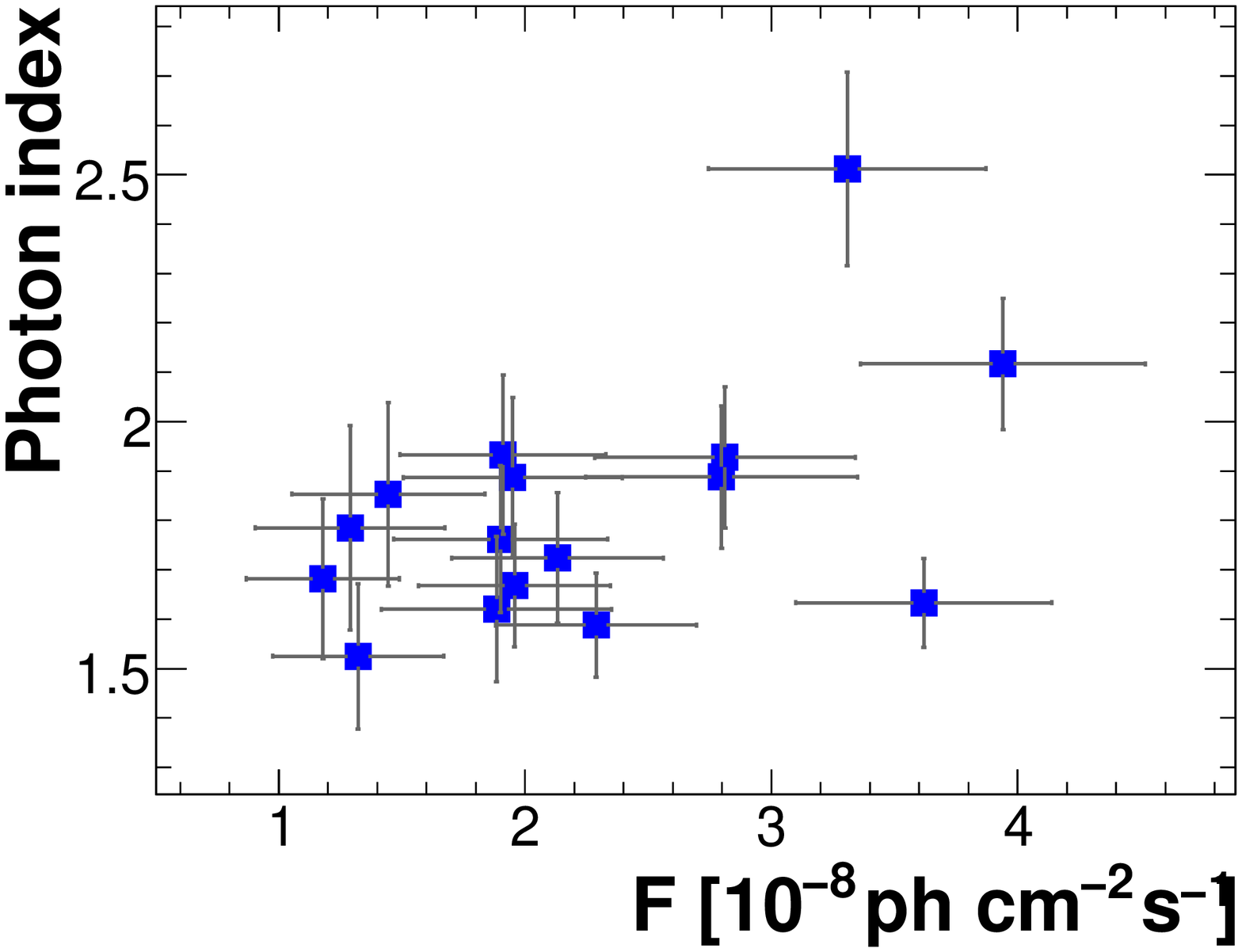}
  \caption{ {\bf Left:} \FermiLAT $\gamma$-ray flux in the energy
    range $0.3-400$\,GeV (top panel) and spectral photon index from a
    power-law fit (bottom panel) for Mrk\,501 for 30-day time intervals from
    2008 August 5 (MJD 54683) to 2009 November 27 (MJD 55162). Vertical bars denote $1 \sigma$ uncertainties and the horizontal bars denote the width of the time interval. The red dashed line and the red legend show the results from a constant fit to the time interval MJD 54862--54982, while the black dashed line and black legend show the results from a constant fit to the entire 480-day data set. {\bf Right:} Scatter plot of the photon index vs flux values.}
 \label{fig:LcAndScatter}
 \end{figure}

The high sensitivity and survey-mode operation of \FermiLAT permit
systematic, uninterrupted monitoring of Mrk\,501 in $\gamma$-rays,
regardless of the activity level of the source. The measured
$\gamma$-ray flux above $0.3$\,GeV and the photon index from a
power-law fit are shown in the left panel of
Figure\,\ref{fig:LcAndScatter}. The data spans the time from 2008 August 5
 (MJD 54683) to 2009 November 27 (MJD 55162), binned in time intervals of 30 days. The Test Statistic (TS) values\footnote{The Test Statistic value quantifies the probability of having a point $\gamma$-ray source at the location specified. It is roughly the square of the significance value: a TS of 25 would correspond to a signal of approximately 5 standard deviations \citep[]{Mattox1996}.} for the 16 time intervals are all in excess of 50 (i.e., $\sim 7$ standard deviations, hereafter $\sigma$), with three-quarters of them greater than 100 (i.e., $\sim 10\,\sigma$). During this 480-day period, Mrk\,501 did not show any outstanding flaring activity in the \FermiLAT energy range, but there appear to be flux and spectral variations on timescales of the order of 30 days. During the 120-day period MJD 54862--54982, the photon flux above $0.3$\,GeV was $(3.41 \pm 0.28) \times 10^{-8}$\,ph\,cm$^{-2}$\,s$^{-1}$, which is about twice as large as the averaged flux values before and after that time period, which are $(1.65 \pm 0.16) \times 10^{-8}$\,ph\,cm$^{-2}$\,s$^{-1}$ and $(1.84 \pm 0.17) \times 10^{-8}$\,ph\,cm$^{-2}$\,s$^{-1}$, respectively. Remarkably, the photon index changed from $2.51 \pm 0.20$ for the first 30-day interval of this ``enhanced-flux period'' to $1.63 \pm 0.09$ for the last 30-day interval. As shown in the red legend of the bottom plot in the left panel of Figure\,\ref{fig:LcAndScatter}, a constant fit to the photon index values of this 120-day period gives a null probability of $10^{-4}$, hence a deviation of $4\,\sigma$. A constant fit to the entire 16-month period gives  a null probability of $2.6 \times 10^{-3}$, hence spectral variability is detected for the entire data set at the level of $3\,\sigma$. It is worth stressing that the spectral variability in the 480-day time interval is entirely dominated by the spectral variability occurring during the 120-day time interval of MJD 54862--54982, with no significant spectral variability before or after this ``enhanced-flux period''. The right plot in Figure\,\ref{fig:LcAndScatter} does not show any clear correlation between the flux and the spectral variations. The discrete correlation function computed as prescribed in \citet{Edelson1988} gives $DCF=0.5 \pm 0.3$ for a time lag of zero.

Mrk\,501 is known for showing spectral variability at VHE $\gamma$-ray energies. During the large X-ray/$\gamma$-ray flare in 1997, Whipple and (especially) CAT observations showed evidence of spectral curvature and variability \citep{Samuelson1998,Djannati1999}. The spectral changes are larger when comparing the measurements from 1997 with the low states from 1998 and 1999, as reported by CAT and HEGRA \citep{Piron2000,Aharonian2001}. The MAGIC telescope, with lower energy threshold and higher sensitivity than the Whipple, HEGRA and CAT telescopes, observed remarkable spectral variability in 2005, when the $\gamma$-ray activity of Mrk\,501 was significantly smaller than that observed in 1997 \citep{Albert2007}. The spectral variability is even larger when comparing the MAGIC measurements from 2005 with those from 2006 when the source was in an even lower state \citep{Anderhub2009}. However, despite the measured spectral variability at VHE $\gamma$-ray energies, the outstanding spectral steepening at GeV energies observed during the time interval MJD 54862--54892 was not envisioned in any of the previous works in the literature; the modeled spectrum of Mrk\,501 at GeV energies was always assumed to be hard (photon indices $\sim 1.5-1.8$). This observational finding, further discussed in \S\ref{FermiSpectrum} and \S\ref{Discussion}, shows the importance of having a $\gamma$-ray instrument capable of long-term, uninterrupted, high-sensitivity monitoring of Mrk\,501 and other HSP BL Lacs, and it points to the important role \FermiLAT will play in improving our understanding of the physics behind the blazar phenomenon.

\begin{figure}[!t]
  \centering
  \includegraphics[width=6.5 in]{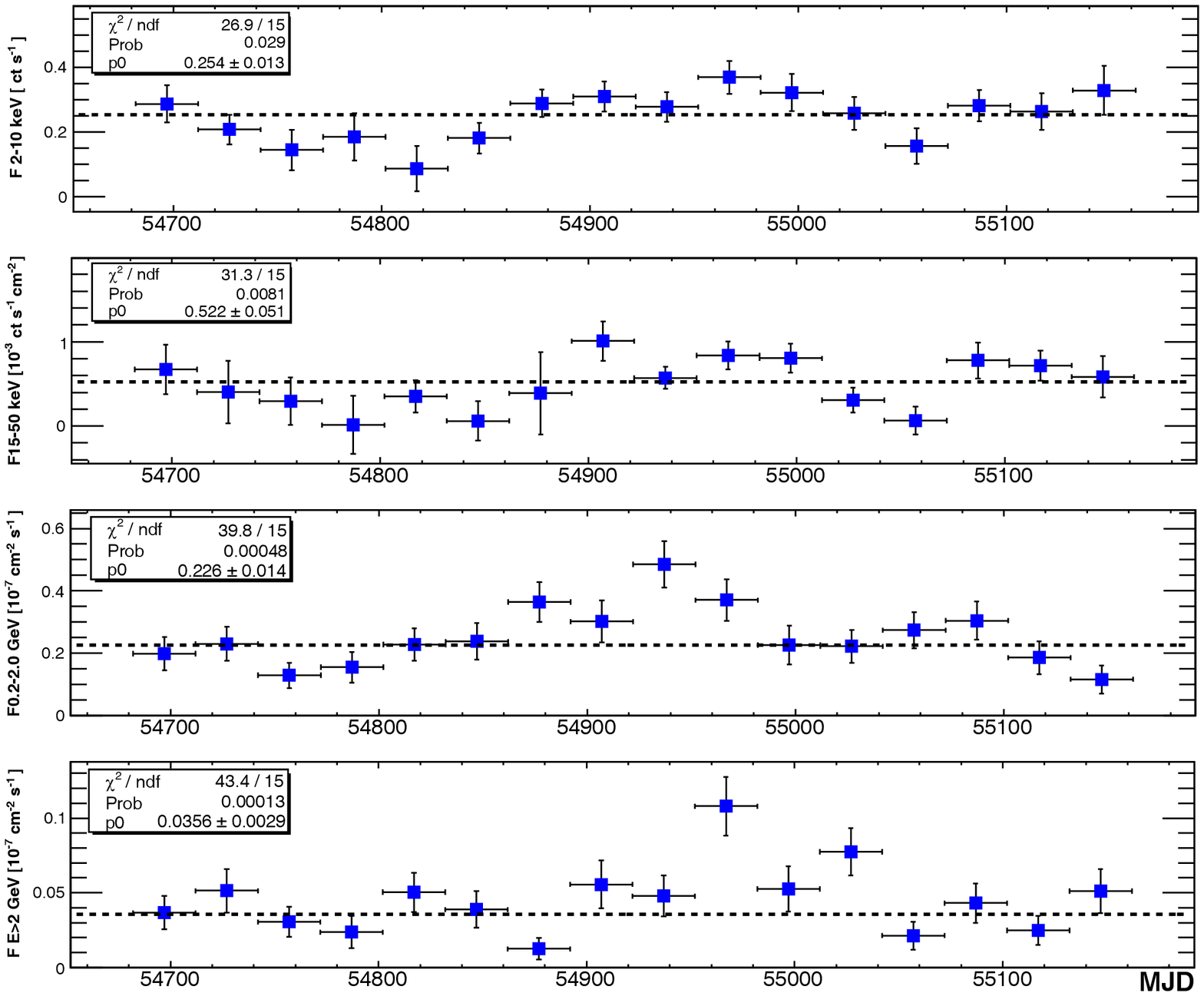}
  \caption{Multifrequency light curves of Mrk\,501 with 30-day time
    bins obtained with 3 all-sky-monitoring instruments: \RXTEc-ASM
    ($2-10$\,keV, first from the top); \Swiftc-BAT ($15-50$\,keV,
    second) and \FermiLAT for two different energy ranges
    ($0.2-2$\,GeV, third, and $>2$\,GeV, fourth). The light curves
    cover the period from2008 August 5 (MJD 54683) to 2009 November 27 (MJD 55162). Vertical bars denote $1\sigma$ uncertainties and horizontal bars show the width of the time interval. The horizontal dashed lines and the legends (for all the plots) show the results from a constant fit to the entire 480-day data set.}
 \label{fig:Lc30daysMW}
 \end{figure}

The \FermiLAT capability for continuous source monitoring is complemented at X-ray frequencies by \RXTEc-ASM and \Swiftc-BAT, the two all-sky instruments that can probe the X-ray activity of Mrk\,501 on a 30-day timescale. Figure\,\ref{fig:Lc30daysMW} shows the fluxes measured by ASM in the energy range $2-10$\,keV, by BAT in the energy range $15-50$\,keV, and by LAT in two different energy bands: $0.2-2$\,GeV (low-energy band) and $>2$\,GeV (high-energy band)\footnote{The fluxes depicted in the \FermiLAT light curves were computed fixing the photon index to 1.78 (average index during the first 480 days of Fermi operation) and fitting only the normalization factor of the power-law function.}. The data from \RXTEc-ASM were obtained from the ASM web page\footnote{\url{http://xte.mit.edu/ASM_lc.html}}. The data were filtered according to the prescription provided there, and the weighted average over all of the dwells\footnote{A dwell is a scan/rotation of the ASM Scanning Shadow Cameras lasting 90 seconds.} was determined for the 30-day time intervals defined for the Fermi data. The data from \Swiftc-BAT were gathered from the BAT web page\footnote{\url{http://swift.gsfc.nasa.gov/docs/swift/results/transients/}}. We retrieved the daily averaged BAT values and made the weighted average over all the days from the 30-day time intervals defined for the Fermi data. The X-ray intensity from Mrk\,501, averaged over the 16 months, is $0.25 \pm 0.01$\,ct\,s$^{-1}$ per Scanning Shadow Camera (\texttt{SSC}) in ASM, and $(0.52 \pm 0.05) \times 10^{-3}$\,ct\,s$^{-1}$\,cm$^{-2}$ in BAT (close to the BAT 30-day detection limit). This X-ray activity is compatible with that recorded in recent years, but quite different from the activity of the source during 1997, when the ASM flux was above $1$\,ct\,s$^{-1}$ per \texttt{SSC} during most of the year, with a peak well above $2$\,ct\,s$^{-1}$ around June 1997.

As noted previously (\S\ref{Intro}), Mrk\,501 is not in the 3rd EGRET
catalog, although there was a marginally significant EGRET detection
during the $\gamma$-ray outburst (with no clear X-ray counterpart) in
1996 \citep{Kataoka1999}. At that time, the source was detected at a
level of $4.0\,\sigma$ at energies above $0.1$\,GeV and at
$5.2\,\sigma$ above $0.5$\,GeV. The flux from the EGRET 1996 flare
above $0.5$\,GeV was $(6 \pm 2) \times
10^{-8}$\,ph\,cm$^{-2}$\,s$^{-1}$, which is about five times higher
than the average flux observed by Fermi from 2008 August 5 (MJD 54683) to 2009 November 27 (MJD 55162), namely $(1.39 \pm 0.07) \times 10^{-8}$\,ph\,cm$^{-2}$\,s$^{-1}$  (also above photon energy $0.5$\,GeV). The \FermiLAT flux measured during the 120 days with the ``enhanced'' $\gamma$-ray activity (MJD 54862--54982) is $(2.03 \pm 0.18) \times 10^{-8}$\,ph\,cm$^{-2}$\,s$^{-1}$ (above photon energy $0.5$\,GeV), about a factor of three lower than that detected by EGRET in 1996. 

In spite of the relatively low activity, the ASM and BAT fluxes show some flux variations and a positive correlation between the fluxes measured by these two instruments. The discrete correlation function for the ASM/BAT data points shown in Figure\,\ref{fig:Lc30daysMW} is $DCF=0.73 \pm 0.17$ for a time lag of zero. On the other hand, the X-ray ASM/BAT fluxes are not significantly correlated with the $\gamma$-ray LAT fluxes. We found, for a time lag of zero, $DCF=0.32 \pm 0.22$ for the ASM/LAT ($<2$\,GeV) and $DCF=0.43 \pm 0.30$ for the ASM/LAT ($>2$\,GeV) flux data points shown in Figure\,\ref{fig:Lc30daysMW}. It is also interesting to note that the largest flux variations occur at the highest Fermi energies ($>2$\,GeV), where the $\gamma$-ray flux increased by one order of magnitude during the 120-day interval MJD 54862--54892. This trend is consistent with the photon index hardening revealed by the spectral analysis reported above (see Figure\,\ref{fig:LcAndScatter}).

We followed the description given in \cite{Vaughan2003} to quantify the flux variability by means of the fractional variability parameter, $F_{var}$, as a function of photon energy. In order to account for the individual flux measurement errors ($\sigma_{err,\,i}$), we used the ``excess variance'' as an estimator of the intrinsic source variance \citep{Nandra1997,Edelson2002}. This is the variance after subtracting the expected contribution from the measurement errors. For a given energy range, $F_{var}$\ is calculated as
\begin{equation}
F_{var} = \sqrt{\frac{S^2 - \langle \sigma_{err}^2 \rangle}{\langle F_{\gamma} \rangle^2}}
\end{equation}
where $\langle F_{\gamma} \rangle$ is the mean photon flux, $S$ is the standard deviation of the $N$ flux points, and $\langle\sigma_{err}^2\rangle$ is the average mean square error, all determined for a given energy bin.

\begin{figure}[!t]
  \centering
  \includegraphics[width=5.0in]{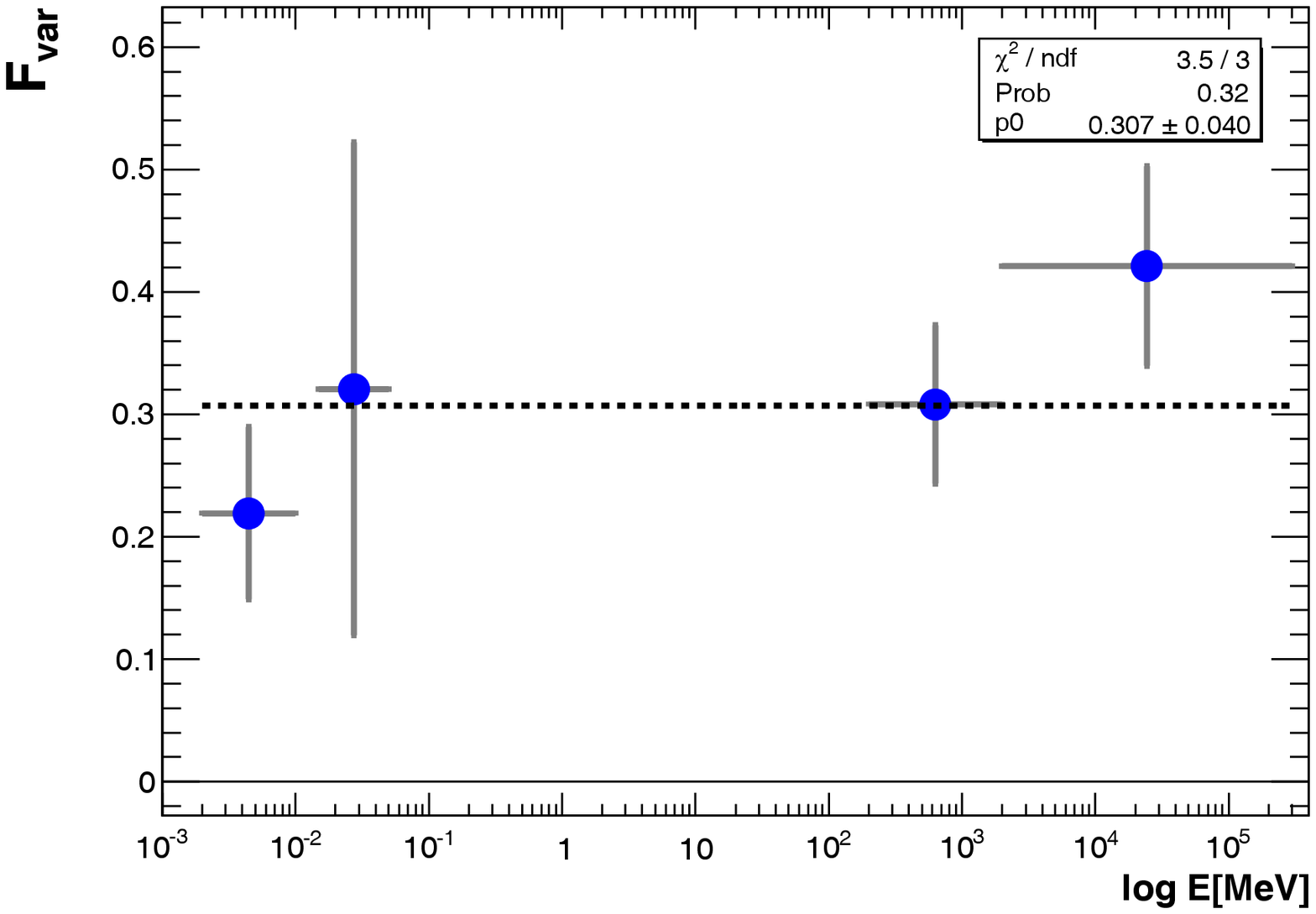}
  \caption{Fractional variability parameter for 16 months data (2008
    August 5 --- 2009 November 27) from 3 all-sky-monitoring instruments: \RXTEc-ASM ($2-10$\,keV); \Swiftc-BAT ($15-50$\,keV) and \FermiLAT (two energy ranges $0.2-2$\,GeV and $2-300$\,GeV). The fractional variability was computed according to \cite{Vaughan2003} using the light curves from Figure\,\ref{fig:Lc30daysMW}. Vertical bars denote $1\sigma$ uncertainties and horizontal bars indicate the width of each energy bin. The horizontal dashed line and the legend show the results from a constant fit.}
 \label{fig:nva}
 \end{figure}

Figure\,\ref{fig:nva} shows the $F_{var}$ values derived for the four different energy ranges and the time window covered by the light curves shown in Figure\,\ref{fig:Lc30daysMW}. The source is variable at all energies. The uncertainty in the variability quantification for the \Swiftc-BAT energies is large due to the fact that Mrk\,501 is a relatively weak X-ray source, and is therefore difficult to detect above $15$\,keV in exposure times as short as 30 days. On the contrary, the variability at the \RXTEc-ASM and, especially, \FermiLAT energies, is significant ($> 3\,\sigma$ level). The amplitude variability in the two X-ray bands is compatible within errors, and the same holds for the variability in the two $\gamma$-ray bands. As shown in Figure\,\ref{fig:nva}, for the hypothesis of a constant  $F_{var}$ over the four energy bands one obtains $\chi^2$= 3.5 for 3 degrees of freedom (probability of 0.32), implying that the energy-dependent variability is not statistically significant. It is worth noticing that the limited sensitivity of ASM and (particularly) BAT instruments to detect Mrk\,501 in 30-day time intervals, as well as the relatively stable X-ray emission of Mrk\,501 during the analyzed observations, precludes any detailed X-ray/$\gamma$-ray variability and correlation analysis.

\section{Spectral Analysis up to 400\,GeV} 
\label{FermiSpectrum}

\begin{figure}[!t]
 \centering
 \includegraphics[width=5.0 in]{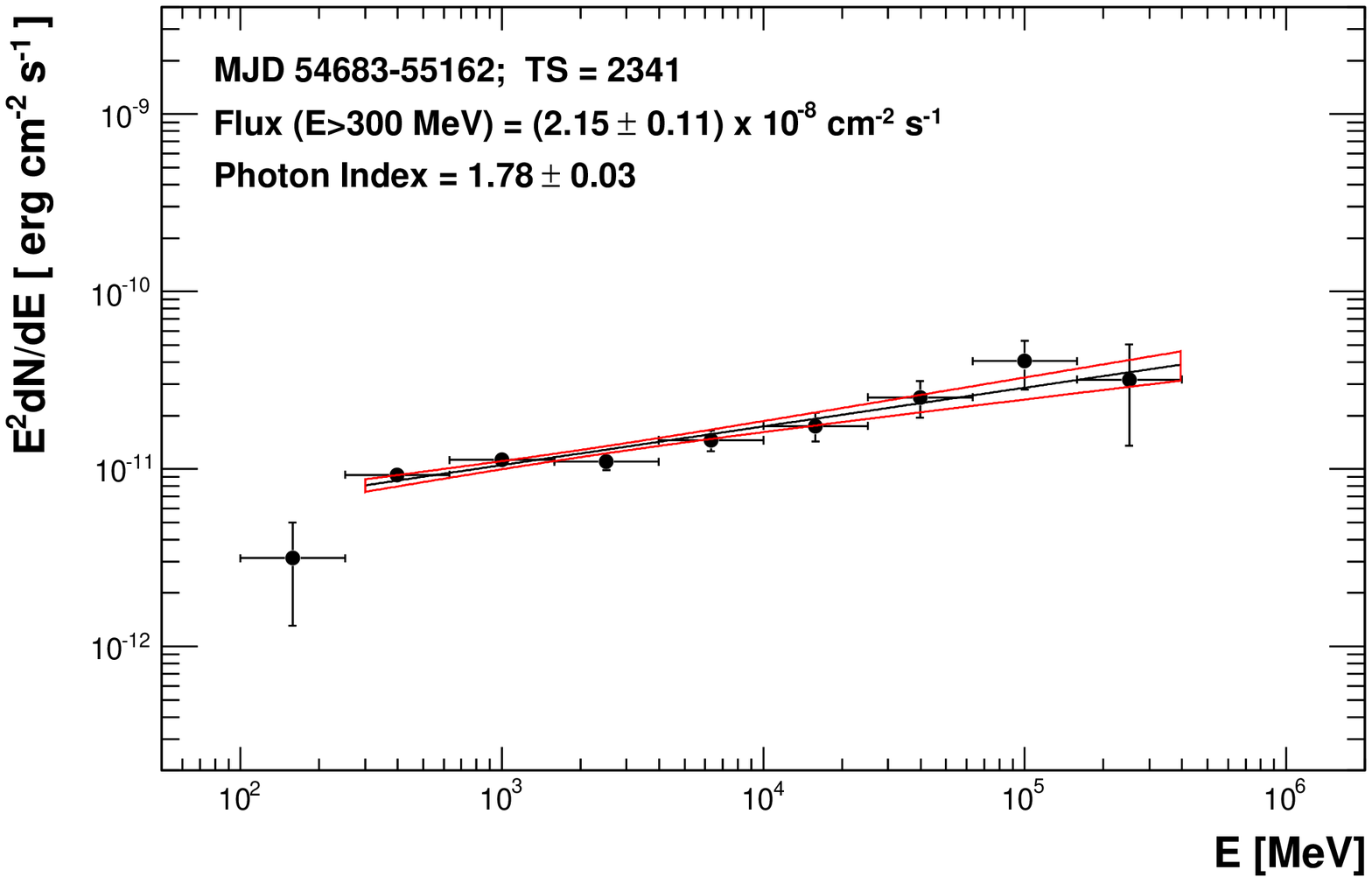}
 \caption{Spectral energy distribution for Mrk\,501 from \FermiLAT
   during the period from 2008 August 5 (MJD 54683) to 2009 November 27 (MJD 55162). The black line depicts the result of the unbinned likelihood power-law fit, the red contour is the $68\%$ uncertainty of the fit, and the black data points show the energy fluxes in differential energy ranges.  The legend reports the results from the unbinned likelihood power-law fit in the energy range $0.3-400$\,GeV.}
 \label{fig:SED}
 \end{figure}

The large effective area of the \FermiLAT instrument permits photon energy reconstruction over many orders of magnitude. As a result, the spectrum of Mrk\,501 could be resolved within the energy range $0.1-400$\,GeV, as shown in Figure\,\ref{fig:SED}. This is the first time the spectrum of Mrk\,501 has been studied with high accuracy over this energy range. The fluxes were computed using the analysis procedures described in \S\ref{FermiData}. The black line in Figure\,\ref{fig:SED} is the result of an unbinned likelihood fit with a single power-law function in the energy range $0.3-400$\,GeV\footnote{The unbinned likelihood fit was performed on photon energies above 0.3\,GeV in order to reduce systematics. See  \S\ref{FermiData} for further details.}, and the red contour is the $68\%$ uncertainty of the fit. The data are consistent with a pure power-law function with a photon index of $1.78 \pm 0.03$. The black data points result from the analysis in differential energy ranges\footnote{Because the analysis was carried out in small energy ranges, it was decided to fix the spectral index at $1.78$ (the value obtained from fitting the entire energy range) and fit only the normalization factor. We repeated the same procedure fixing the photon indices to 1.5 and 2.0 and found no significant change. Therefore, the results from the differential energy analysis are not sensitive to the photon index used in the analysis.} ($\log\Delta E=0.4$). The points are well within $1-2\sigma$ from the fit to the overall spectrum (black line), which confirms that the entire Fermi spectrum is consistent with a pure power-law function. Note, however, that, due to the low photon count, the error bars for the highest energy data points are rather large. The predicted (by the model for Mrk\,501) number of photons detected by LAT in the energy bins $60-160$\,GeV and $160-400$\,GeV are only $11$ and $3$, respectively. Therefore, even though the signal significance in the highest-energy bins are very high due to the very low background (the TS values for the two highest-energy ranges is 162 and 61, respectively), the large statistical uncertainties could hide a potential turnover in the spectrum of Mrk\,501 around $100$\,GeV photon energies. As we know from past observations, the VHE spectrum is significantly softer than the one observed by Fermi \citep[e.g.,][]{Aharonian2001,Anderhub2009}, and hence the spectrum of Mrk\,501 must have a break around the highest \FermiLAT energies. 

\begin{figure}[!t]
  \centering
  \includegraphics[width=3.2 in]{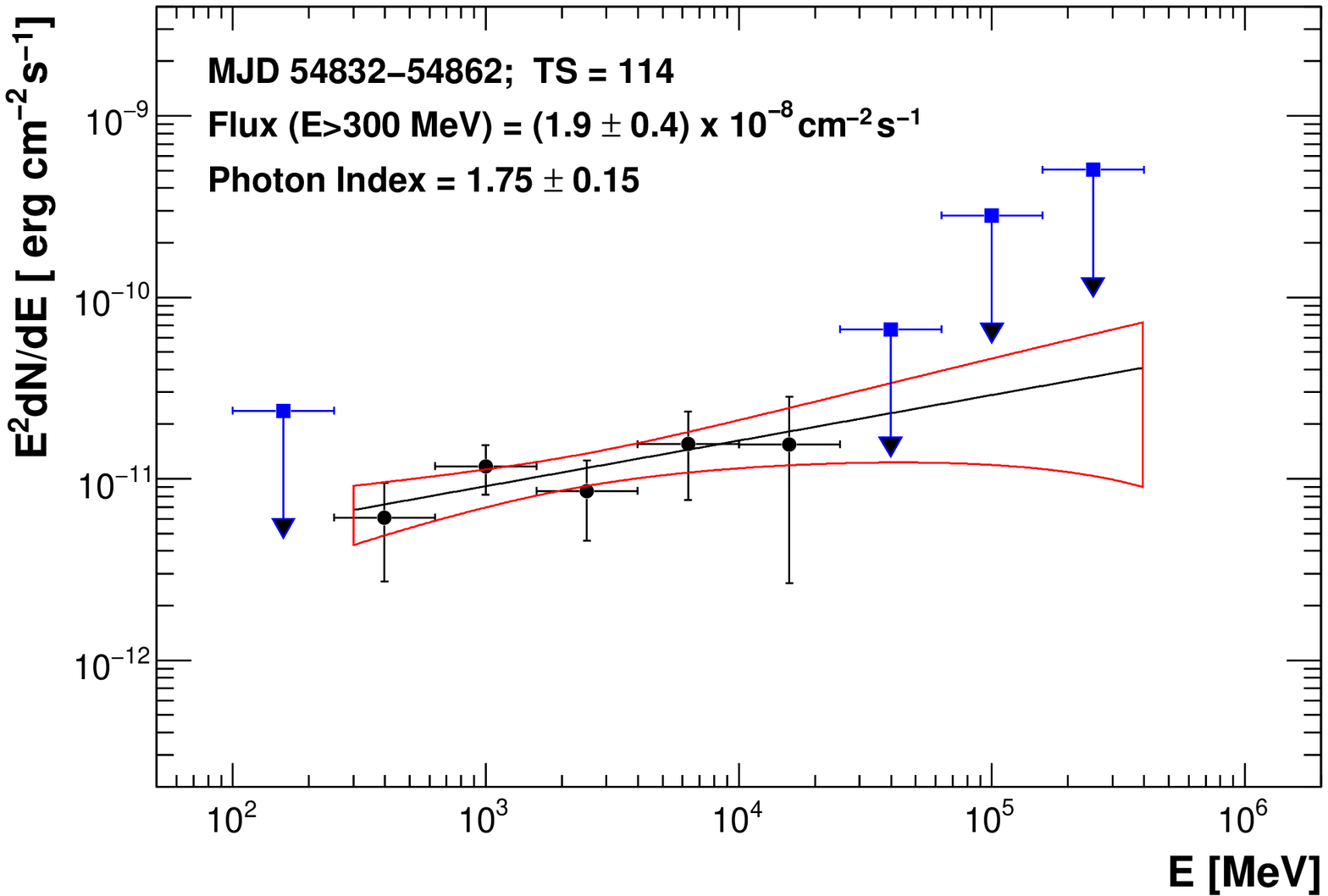}
  \includegraphics[width=3.2 in]{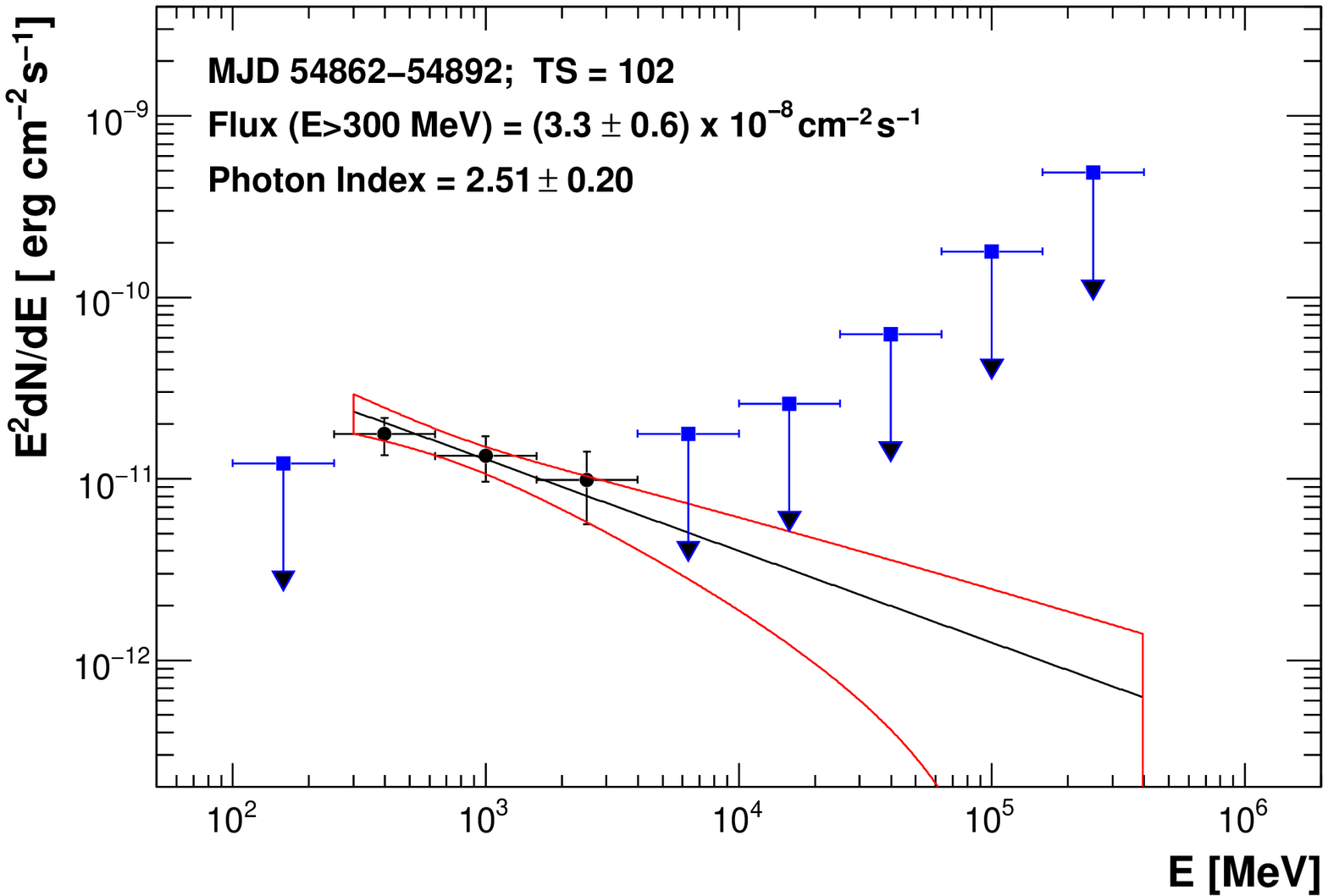}
  \includegraphics[width=3.2 in]{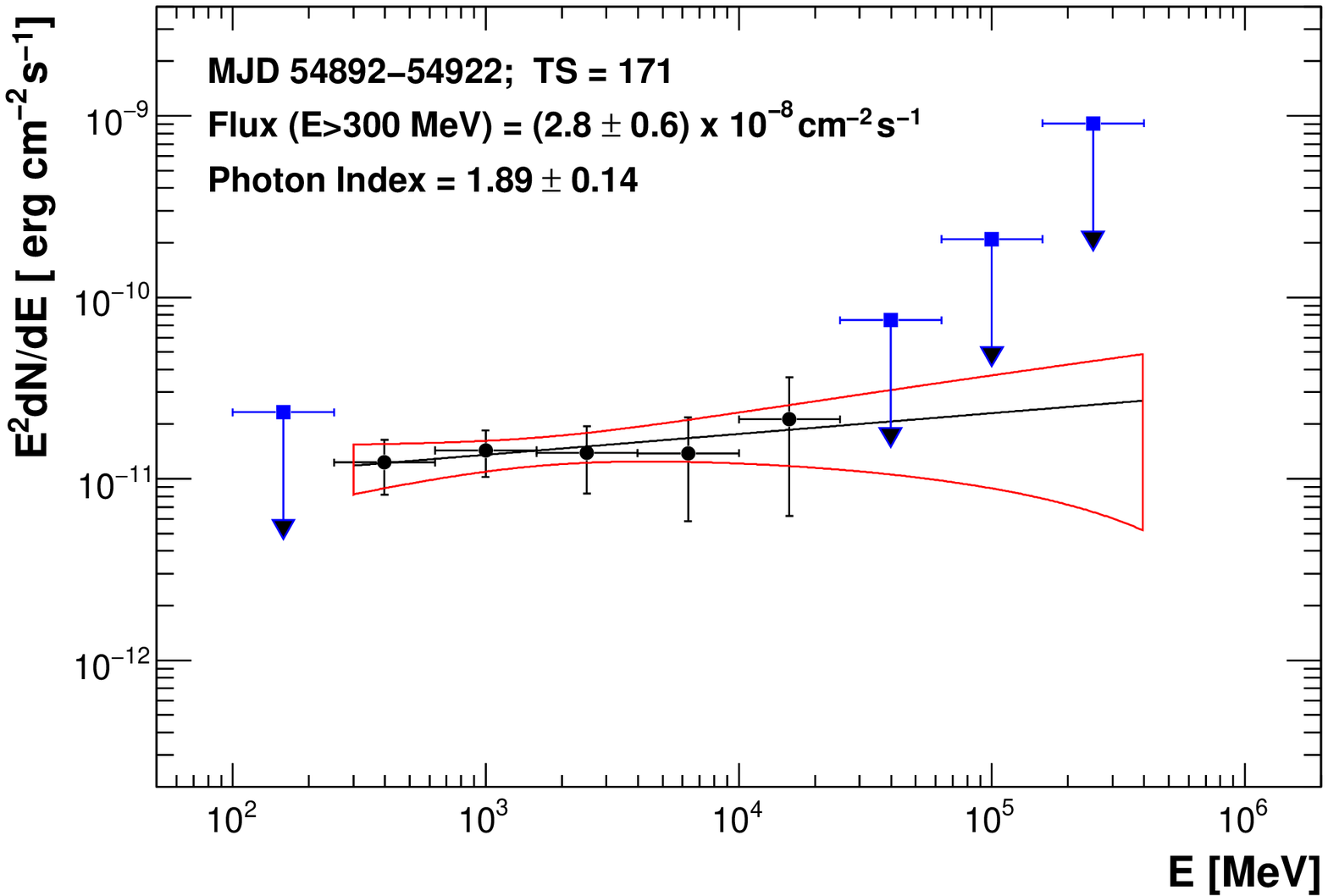}
  \includegraphics[width=3.2 in]{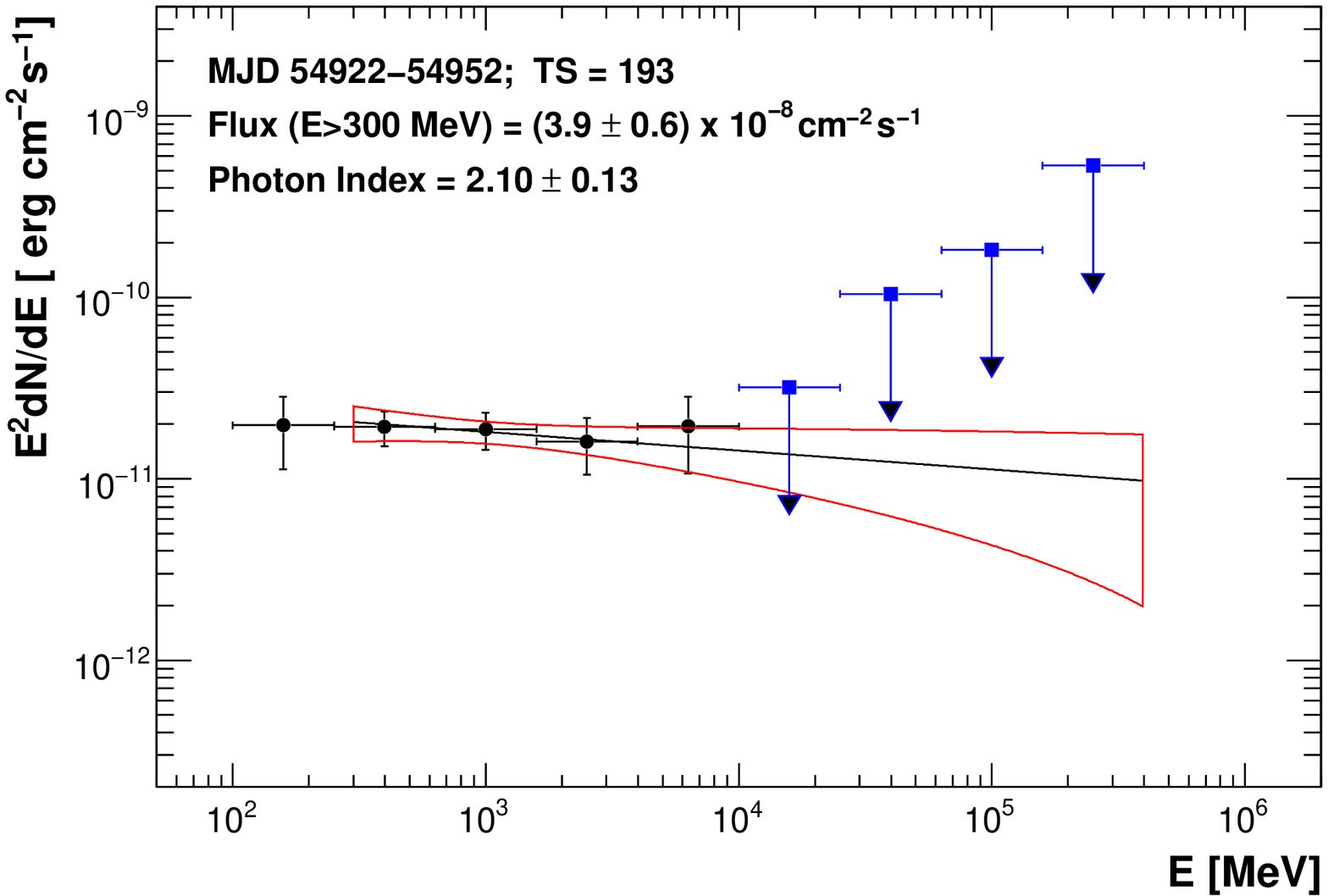}
  \includegraphics[width=3.2 in]{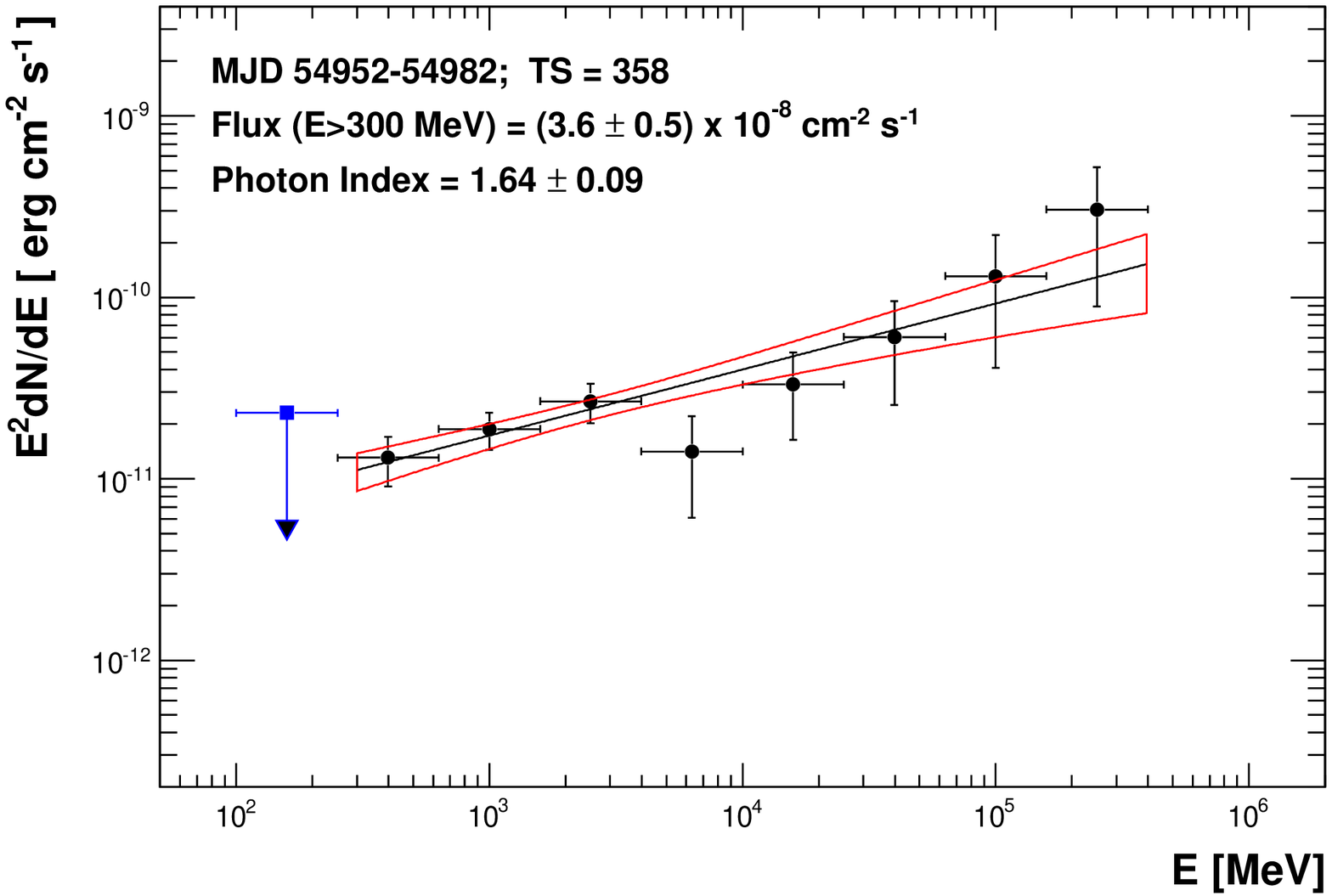}
  \includegraphics[width=3.2 in]{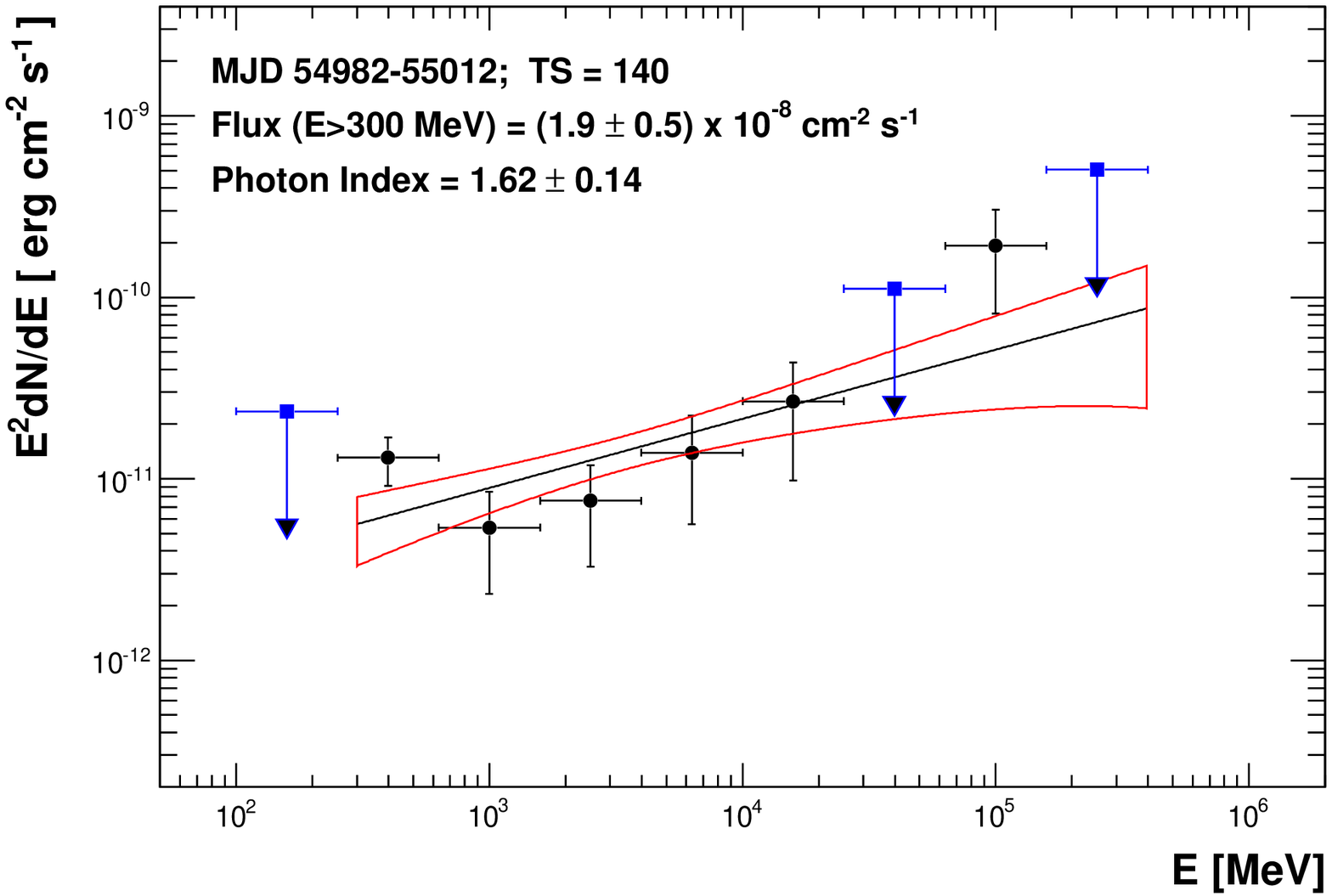}
  \caption{Spectral energy distribution for Mrk\,501 from \FermiLAT
    for six 30-day time intervals: MJD 54832--54862 (top left), MJD
    54862--54892 (top right), MJD 54892--54922 (middle left),  MJD
    54922--54952 (middle right), MJD 54952--54982 (bottom left) and
    MJD 54982--55012 (bottom right). In all the panels, the black line
    depicts the result of the unbinned likelihood power-law fit, the
    red contour denotes the $68\%$ uncertainty of the fit, and the
    black data points show the energy fluxes computed for differential
    energy ranges. The blue arrows denote 95\% upper limits, which
    were computed for the differential energy ranges with a signal of
    $TS<4$ or less than two photons predicted by the analysis model
    for Mrk\,501. The legend reports the results from the unbinned  
    likelihood power-law fit in the energy range $0.3-400$\,GeV. }
  \label{fig:SED_30day}
\end{figure}

In \S\ref{LC} we reported remarkable spectral variability during the 120-day time interval MJD 54862--54982, when Mrk\,501 was characterized by a photon flux (at $> 0.3$\,GeV) twice as large as during the rest of the exposure. In order to understand better the behaviour of the source during that time, we produced SED plots (analogous to that of Figure\,\ref{fig:SED}) for each of the 30-day time intervals from the period with the enhanced flux level. These are shown in Figure\,\ref{fig:SED_30day}, together with the SED plots from the 30-day time intervals before and after this 120-day epoch, which are representative of the average source behaviour during the other 360 days. The variability of the SED data points below a few GeV is rather mild (factor of two), but above a few GeV the spectra vary substantially (factor of ten). The $\gamma$-ray signal at the highest energies is suppressed during MJD 54862--54982, while it increases by a large factor during MJD 54952--54982, where the analysis model for Mrk\,501 predicts 2.0 photons in the energy range $160-400$\,GeV. It is worth stressing that for the SED from Figure\,\ref{fig:SED}, which corresponds to the total exposure of 480 days, the analysis model for Mrk\,501 predicts only 3.2 photons in the highest energy bin. Hence the time interval MJD 54952--54982 holds almost all the signal detected by LAT in the energy range $160-400$\,GeV during 16 months. The situation changes somewhat for the lower energy bin $60-160$\,GeV, for which the analysis model for Mrk\,501 predicts 2.4 photons for the time interval MJD 54952--54982, while it does predict 11.3 photons for the entire 16-month time interval. Fortunately, the 30-day time interval characterized by hard spectrum is covered by the 4.5-month campaign that we organized, and hence simultaneous multifrequency observations (radio to TeV) are available for this particular period, as discussed further below. 

\section{Broadband Spectral Energy Distribution of Mrk\,501} 
\label{MWSED}

As mentioned in \S\ref{Intro}, we organized a multifrequency campaign (from radio to TeV photon energies) to monitor Mrk\,501 during a time period of 4.5 months. The observing campaign started on March 15, 2009 (MJD 54905) and finished on August 01, 2009 (MJD 55044). The observing goal for this campaign was to sample the broadband emission of Mrk\,501 every 5 days, which was largely accomplished whenever the weather and/or technical limitations allowed. The underlying scientific goal has already been outlined in \S1. A detailed analysis of the multifrequency variability and correlations, as well as the evolution of the overall spectral energy distribution  with time, will be reported in a forthcoming paper. In this section of the manuscript, we describe the source coverage during the campaign and the data analysis for several of the participating instruments, and we report on the averaged SED resulting from the campaign. The modeling of these data and the physical implications are given in \S\ref{SEDModel} and \S\ref{Discussion} below, respectively. 

\subsection{Details of the Campaign: Participating Instruments  and Temporal Coverage }
\label{CampaignDetails}

\begin{figure}[!t]
  \centering
  \includegraphics[width=6.5in]{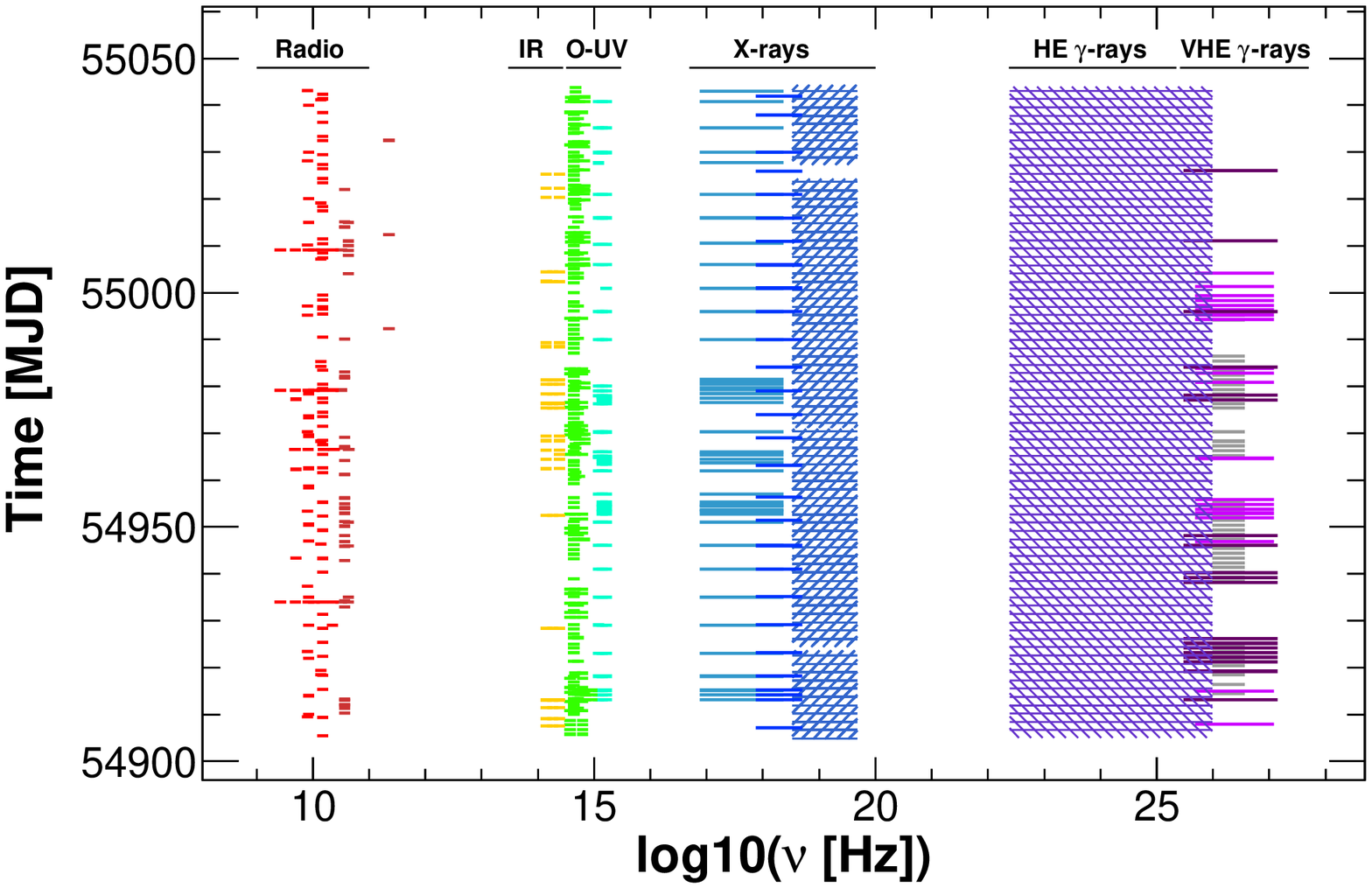}
   \caption{Time and energy coverage during the multifrequency campaign. For the sake of clarity, the minimum observing time displayed in the plot was set to half a day. }
  \label{fig:TimeEnergyCoverage}
\end{figure}

The list of all the instruments that participated in the campaign is given in Table\,\ref{TableWithInstruments}, and the scheduled observations can be found online\footnote{\url{https://confluence.slac.stanford.edu/display/GLAMCOG/Campaign+on+Mrk501+(March+2009+to+July+2009)}}. In some cases the planned observations could not be performed due to bad observing conditions, while in some other occasions the observations were performed but the data could not be properly analyzed due to technical problems or rapidly changing weather conditions. In order to quantify the actual time and energy coverage during the campaign on Mrk\,501, Figure\,\ref{fig:TimeEnergyCoverage} shows the exposure time as a function of the energy range for the instruments/observations used to produce the SED shown in Figure\,\ref{fig:MWSED}. Apart from the unprecedented energy coverage (including, for the first time, the GeV energy range from \FermiLATc), the source was sampled quite uniformly with the various instruments participating in the campaign and, consequently, it is reasonable to consider the SED constructed below as the actual average (typical) SED of Mrk\,501 during the time interval covered by this multifrequency campaign. The largest non-uniformity in the sampling of the source comes from the Cherenkov Telescopes, which are the instruments most sensitive to weather conditions. Moreover, while there are many radio/optical instruments spread all over the globe, there are only three Cherenkov Telescope observatories in the northern hemisphere we could utilize (MAGIC, VERITAS, Whipple). Hence, the impact of observing conditions was more important to the coverage at the VHE $\gamma$-ray energies. 

We note that  Figure\,\ref{fig:TimeEnergyCoverage}  shows the MAGIC, VERITAS and Whipple coverage at VHE $\gamma$-ray energies, but only the MAGIC and VERITAS observations were used to produce the spectra shown in Figure\,\ref{fig:MWSED}.  The more extensive (120 hr), but less sensitive, Whipple data (shown as grey boxes in Figure\,\ref{fig:TimeEnergyCoverage}) were primarily taken to determine the light curve \citep{Pichel2009} and a re-optimization was required to derive the spectrum which will be reported elsewhere.

In the following paragraphs we briefly discuss the procedures used in
the data analysis of the instruments participating in the campaign. The analysis of the \FermiLAT data was described in \S\ref{FermiData} and the results obtained will be described in detail in \S\ref{FermiSED_InMW}.

\subsubsection{Radio Instruments}

Radio data were taken for this campaign from single-dish telescopes,
one mm-interferometer, and one Very Long Baseline Interferometry
(VLBI) array, at frequencies between $2.6$\,GHz and $225$\,GHz (see
Table\,\ref{TableWithInstruments}). The single-dish telescopes were
the Effelsberg 100\,m radio telescope, the 32\,m Medicina radio
telescope, the 14\,m Mets\"ahovi radio telescope, the 32\,m Noto radio
telescope, the Owens Valley Radio Observatory (OVRO) 40\,m telescope,
the 26\,m University of Michigan Radio Astronomy Observatory (UMRAO)
and the 600 meter ring radio telescope RATAN-600. The
mm-interferometer is the Sub-millimeter Array (SMA). The NRAO Very
Long Baseline Array (VLBA) was used for the VLBI observations. For the
single-dish instruments and SMA, Mrk\,501 
is point-like and unresolved at all observing frequencies. Consequently, the single-dish measurements denote the total flux density of the source integrated over the whole source extension. Details of the observing strategy and data reduction can be found in \citet[F-GAMMA project]{Fuhrmann2008,Angelakis2008}, \citet[Mets\"ahovi]{Terasranta1998}, \citet[UMRAO]{Aller1985}, \citet[Medicina and Noto]{Venturi2001},  \citet[RATAN-600]{Kovalev1999} and \citet[in preparation, OVRO]{Richards2010}.

In the case of the VLBA, the data were obtained at various frequencies
from $5$\,GHz to $43$\,GHz through various programs (BP143, BK150 and
MOJAVE). The data were reduced following standard procedures for data
reduction and calibration \citep[see, for example,][for a description
of the MOJAVE program which provided the $15$\,GHz
data]{Lister2009}. Since the VLBA angular resolution is smaller than
the radio source extension, measurements were performed for the most
compact core region, as well as for the total radio structure at
parsec scales. The VLBA core size was determined with
two-dimensional circular or elliptical Gaussian fits to the measured
visibilities. The FWHM size of the core was estimeted to be in the range 
0.14--0.18 mas at the highest observing frqeuencies, 15--43 GHz.
Both the total and the core radio flux densities from the VLBA data are depicted in Figure\,\ref{fig:MWSED}.

\subsubsection{Optical and Near-IR Instruments}

The coverage at optical frequencies was obtained through various telescopes around the globe, and this decreased the sensitivity to weather/technical difficulties and provided good overall coverage of the source, as depicted in Figure\,\ref{fig:TimeEnergyCoverage}. Many of the observations were performed within the GASP-WEBT program \citep[e.g.,][]{Villata2008, Villata2009}; that is the case for the data collected by the telescopes at Abastumani, Lulin, Roque de los Muchachos (KVA), St. Petersburg, Talmassons, and Valle d'Aosta observatories ($R$ band), and also for Campo Imperatore (near-infrared frequencies, $JHK$ bands). In addition, the telescopes GRT, ROVOR and MitSume provided data with various optical filters, while OAGH and WIRO provided data at near-infrared wavelengths. See Table\,\ref{TableWithInstruments} for further details. 

All the instruments used the calibration stars reported in \citet{Villata1998}, and the Galactic extinction was corrected with the coefficients given in \citet{schlegel98}. On the other hand, the flux from the host galaxy, which in the $R$ band accounts for about two-thirds of the overall measured optical flux \citep[]{Nilsson2007}, was not subtracted. As can be seen from Figure\,\ref{fig:MWSED}, the host galaxy contribution shows up as an additional (narrow) bump in the SED with the peak located at infrared frequencies and the flux decreasing rapidly with increasing frequency. At frequencies above $10^{15}$\,Hz, the blazar emission again dominates the radiative output of Mrk\,501. 

\subsubsection{\Swiftc-UVOT} 

The \Swiftc-Ultra-Violet/Optical Telescope \citep[UVOT;][]{Roming2005} data used in this analysis include all the observations performed during     
the time interval MJD 54905 and 55044, which amounts to 41 single pointing observations    
that were requested to provide UV coverage during the Mrk\,501 multifrequency campaign.  
The UVOT telescope cycled through each of six optical and ultraviolet passbands (V, B,     
U, UVW1, UVM2, UVW2). Photometry was computed using a $5$\,arcsec  source 
region around Mrk\,501 using a custom UVOT pipeline that obtains similar photometric 
results to the public pipeline \citep{Poole2008}. The custom pipeline also allows for separate, observation-by-observation corrections for astrometric mis-alignments \citep[][in preparation]{AcciariMrk4212008}. 
A visual inspection was also performed on 
each of the observations to ensure proper data quality selection and correction. 
 The flux measurements obtained have been corrected for Galactic extinction $E_{B-V} = 0.019$\,mag \citep[]{schlegel98} in each spectral band \citep[]{Fitzpatrick99}.

\subsubsection{\Swiftc-XRT} 

All the \Swiftc-X-ray Telescope \citep[XRT;][]{Burrows2005} Windowed Timing observations carried out from MJD 54905 to 55044 were used for the analysis: this amounts to a total of 41 observations performed within this dedicated multi-instrument effort to study Mrk\,501. The XRT data set was first processed with the \texttt{XRTDAS} software package (v.2.5.0) developed at the ASI Science Data Center (ASDC) and distributed by HEASARC within the \texttt{HEASoft} package (v.6.7). Event files were calibrated and cleaned with standard filtering criteria with the \texttt{xrtpipeline} task using the latest calibration files available in the Swift CALDB. The individual XRT event files were then merged together using the \texttt{XSELECT} package and the average spectrum was extracted from the summed event file. Events for the spectral analysis were selected within a circle of 20-pixel ($\sim 47$\,arcsec) radius centered at the source position and enclosing about $95\%$ of the point-spread function (PSF) of the instrument. The background was extracted from a nearby circular region of 40-pixel radius. The source spectrum was binned to ensure a minimum of 20 counts per bin to utilize the $\chi^{2}$ minimization fitting technique. The ancillary response files were generated with the \texttt{xrtmkarf} task applying corrections for the PSF losses and CCD defects using the cumulative exposure map. The latest response matrices (v.011) available in the \Swift CALDB were used.

The XRT average spectrum in the $0.3-10$\,keV energy band was fitted using the \texttt{XSPEC} package. We adopted a log-parabolic model for the photon flux spectral density \citep{Massaro2004a,Massaro2004b} of the form $\log [\mathcal{F}(E)] = \log K - a \, \log [E/{\rm keV}] - b \, \log^2 [E/{\rm keV}]$, with an absorption hydrogen-equivalent column density fixed to the Galactic value in the direction of the source, namely $1.56 \times 10^{20}$\,cm$^{-2}$ \citep{Kalberla2005}. This model provided a good description of the observed spectrum, with the exception of the $1.4-2.3$\,keV energy band where spectral fit residuals were present. These residuals are due to known XRT calibration uncertainties (SWIFT-XRT-CALDB-12)\footnote{\url{http://heasarc.gsfc.nasa.gov/docs/heasarc/caldb/swift/docs/xrt/SWIFT-XRT-CALDB-09_v12.pdf}} and hence we decided to exclude the $1.4-2.3$\,keV energy band from the analysis. In addition, we had to apply a small energy offset ($\sim 40$\,eV) to the observed energy spectrum. The origin of this correction is likely to be CCD charge traps generated by radiation and high-energy proton damage (SWIFT-XRT-CALDB-12), which affects mostly the lowest energies (first one or two bins) in the spectrum. The resulting spectral fit gave the following parameters: $K = (3.41 \pm 0.03) \times 10^{-2}$\,ph\,cm$^{-2}$\,s$^{-1}$\,keV$^{-1}$, $a=1.96 \pm  0.04$, and $b= 0.308  \pm  0.010$. The XRT SED data shown in Figure~\ref{fig:MWSED} were corrected for the Galactic absorption and then binned into 10 energy intervals. 

\subsubsection{\RXTEc-PCA}

The {\it Rossi}-X-ray Timing Explorer \citep[\RXTEc;][]{RXTERef} satellite performed 29 pointing observations of Mrk\,501 during the time interval  MJD 54905 and 55044. These observations amount to a total exposure of 52\,ks, which was requested through a dedicated Cycle 13 proposal to provide X-ray coverage for our campaign. We did not find a significant signal in the \RXTEc-HEXTE data and hence we only report on the data from \RXTEc-PCA, which is the main pointing instrument on board \RXTEc. The data analysis was performed using \texttt{FTOOLS} v6.5 and following the procedures and filtering criteria recommended by the \RXTE Guest Observer Facility\footnote{\url{http://www.universe.nasa.gov/xrays/programs/rxte/pca/doc/bkg/bkg-2007-saa/}} after September 2007. In particular, the observations were filtered following the conservative procedures for faint sources\footnote{The average net count rate from Mrk\,501 was about 7\,ct/s/pcu (in the energy range $3-20$\,keV) with flux variations typically much smaller than a factor of two.}: Earth elevation angle greater than $10^\circ$, pointing offset less than $0.02^{\circ}$, time since the peak of the last SAA (South Atlantic Anomaly) passage greater than 30 minutes, and electron contamination less than $0.1$. For further details on the analysis of faint sources with RXTE, see the online Cook Book\footnote{\url{http://heasarc.gsfc.nasa.gov/docs/xte/recipes/cook_book.html}}. In the data analysis, in order to increase the quality of the signal, only the first xenon layer of PCU2 was used. We used the package \texttt{pcabackest} to model the background and the package \texttt{saextrct} to produce spectra for the source and background files and the script\footnote{The CALDB files are located at \url{http://heasarc.gsfc.nasa.gov/FTP/caldb}}  \texttt{pcarsp} to produce the response matrix.

The PCA average spectrum in the $3-28$\,keV energy band was fitted using the XSPEC package with a single power-law function $\log [\mathcal{F}(E)] = \log K - a \, \log [E/{\rm keV}]$ with a constant neutral hydrogen column density $N_{\rm H}$ fixed at the Galactic value in the direction of the source, namely $1.56 \times 10^{20}$\,cm$^{-2}$ \citep{Kalberla2005}. However, since the PCA bandpass starts at $3$\,keV, the value used for $N_{\rm H}$ does not significantly affect our results. The resulting spectral fit provided a good representation of the data for the following parameters: $K = (4.34 \pm 0.11) \times 10^{-2}$\,ph\,cm$^{-2}$\,s$^{-1}$\,keV$^{-1}$, and $a=2.28 \pm  0.02$. The PCA average spectrum obtained using 23 energy bins is shown in Figure~\ref{fig:MWSED}.

\subsubsection{\Swiftc-BAT} 

The \Swiftc-Burst Alert Telescope \citep[BAT;][]{Barthelmy2005} analysis results presented in this paper were derived with all the  available data during the time interval MJD 54905 and 55044. The spectrum was extracted following the recipes presented in \citet[][]{ajello08,ajello09b}. This spectrum is constructed by weighted averaging of the source spectra extracted from short exposures (e.g., $300$\,s) and is representative of the averaged source emission over the time range spanned by the observations. These spectra are accurate to the mCrab level and the reader is referred to \cite{ajello09a} for more details. The \Swiftc-BAT spectrum is consistent with a power-law function with normalization parameter $K = 0.24 \pm 0.16$\,ph\,cm$^{-2}$\,s$^{-1}$\,keV$^{-1}$ and photon index $\mathrm{a = 2.8 \pm 0.4}$.

\subsubsection{MAGIC}

MAGIC is a system of two 17\,m-diameter IACTs for very high energy $\gamma$-ray astronomy located on the Canary Island of La Palma, at an altitude of 2200\,m above sea level. At the time of the observation, MAGIC-II, the new second telescope of the current array system, was still in its commissioning phase so that Mrk\,501 was observed in stand-alone mode by MAGIC-I, which is in scientific operation since 2004 \citep{Albert2008}. The MAGIC telescope monitored the VHE activity of  Mrk\,501 in the framework of the organized multifrequency campaign. The observations were performed in the so-called ``wobble'' mode \citep{Daum1997}. In order to have a low energy threshold, only observations at zenith angles less than $35^{\circ}$ were used in this analysis. Bad weather and a shut-down for a scheduled hardware system upgrade during the period MJD 54948--54960 (April 27 -- May 13) significantly reduced the actual amount of observing time compared to what had initially been scheduled for this campaign. The data were analyzed following the prescription given in \cite{Albert2008} and \cite{Aliu2009}. The data surviving the quality cuts amount to a total of 16.2 hours. The preliminary reconstructed photon fluxes for the individual observations gave an average activity of about $30\%$ the flux of the Crab Nebula, with small (typically much less than a factor of two) flux variations. The derived spectrum was unfolded to correct for the effects of the limited energy resolution of the detector and of possible bias \citep{Albert2007b}. The resulting spectrum was fitted satisfactorily with a single power-law function of the form $\log [\mathcal{F}(E)] = \log K - a \, \log [E/{\rm TeV}]$, giving normalization parameter $K = (0.90\pm0.05) \times 10^{-11}$ph\,cm$^{-2}$\,s$^{-1}$\,TeV$^{-1}$ and photon index $a = 2.51 \pm 0.05$.  

\subsubsection{VERITAS}

VERITAS is a state-of-the-art TeV $\gamma$-ray observatory consisting of four 12\,m-diameter IACTs. VERITAS is located at the basecamp of the F.L. Whipple Observatory in southern Arizona, USA, at an altitude of 1250\,m above sea level, and the system has been fully operational since fall 2007 \citep{Acciari2010a}. VERITAS observed Mrk\,501 as part of the long-term monitoring campaign between March and June of 2009.  The observations were performed in ``wobble'' mode \citep{Daum1997} at relatively low zenith angle ($<40^{\circ}$). These data were analyzed following the prescription reported in \citet{veritas}. After removal of data runs with poor observing conditions, a total of 9.7 hours of good quality data was obtained between MJD 54907 and MJD 55004. Due to the long-term nature of these observations, several factors had to be taken into account when analyzing the data. The initial portion of the campaign includes data taken under standard 4-telescope operating conditions. Two nights of data were taken with only two operational telescopes due to technical difficulties. For the latter portion of the campaign, data were taken over several nights with three operational telescopes because one of the telescopes was being relocated as part of an upgrade to the array \citep{t1move}. The effective collection areas for the array in these three configurations were calculated using Monte Carlo simulations of extensive air showers passed through the analysis chain with detector configurations corresponding to the respective data-taking conditions.

An initial analysis of the VHE activity showed an increase in the flux by a factor of about five during MJD  54953--54956. Because of the large difference in the VHE flux, we decided to analyze this 3-day data set (corresponding to a ``flaring'' state of Mrk\,501) separately from the rest of the collected data (``non-flaring''). The ``flaring'' epoch consists of $2.4$\,h of data taken during MJD 54953--54956. The ``non-flaring'' epoch consists of $7.3$\,h of data taken during the remaining portion of the campaign. The spectra from these two data sets were each fitted with a single power-law function of the form $\log [\mathcal{F}(E)] = \log K - a \, \log [E/{\rm TeV}]$. The resulting fit parameter values are $K = (4.17\pm0.24) \times 10^{-11}$\,ph\,cm$^{-2}$\,s$^{-1}$\,TeV$^{-1}$ with $a = 2.26 \pm 0.06$ for the ``flaring'' state, and $K = (0.88 \pm 0.06) \times 10^{-11}$\,ph\,cm$^{-2}$\,s$^{-1}$\,TeV$^{-1}$ with photon index $a = 2.48 \pm 0.07$ for the ``non-flaring'' state. 

\begin{deluxetable}{lll}
\rotate
\tabletypesize{\scriptsize}
\tablecolumns{3} 
\tablewidth{0pc}
\tablecaption{List of instruments participating in the multifrequency campaign and used in the construction of the SED in Figure\,\ref{fig:MWSED}}
\tablehead{ 
\colhead{Instrument/observatory}                   &\colhead{Energy range covered}  &\colhead{Web page} 
}  
\startdata 

MAGIC                 & 0.12-5.8\,TeV               & \url{http://wwwmagic.mppmu.mpg.de/} \\
VERITAS              & 0.20-5.0\,TeV               & \url{http://veritas.sao.arizona.edu/} \\          
Whipple$^{a}$              &0.4-1.5\,TeV                  & \url{http://veritas.sao.arizona.edu/content/blogsection/6/40/} \\           
\FermiLAT                & 0.1-400\,GeV               & \url{http://www-glast.stanford.edu/index.html} \\
\Swiftc-BAT                & 14-195\,keV               & \url{http://heasarc.gsfc.nasa.gov/docs/swift/swiftsc.html/} \\
\RXTEc-PCA                & 3-28\,keV               & \url{http://heasarc.gsfc.nasa.gov/docs/xte/rxte.html} \\
\Swiftc-XRT                & 0.3-9.6\,keV               & \url{http://heasarc.gsfc.nasa.gov/docs/swift/swiftsc.html} \\
\Swiftc-UVOT                & V, B, U, UVW1, UVM2, UVW2            & \url{http://heasarc.gsfc.nasa.gov/docs/swift/swiftsc.html} \\
Abastumani {\scriptsize (through GASP-WEBT program)}              & R       band        & \url{http://www.oato.inaf.it/blazars/webt/} \\
Lulin {\scriptsize (through GASP-WEBT program)}              & R       band        & \url{http://www.oato.inaf.it/blazars/webt/} \\
Roque de los Muchachos (KVA) {\scriptsize (through GASP-WEBT program)}              & R       band        & \url{http://www.oato.inaf.it/blazars/webt/} \\
St. Petersburg {\scriptsize (through GASP-WEBT program)}              & R       band        & \url{http://www.oato.inaf.it/blazars/webt/} \\
Talmassons {\scriptsize (through GASP-WEBT program)}              & R       band        & \url{http://www.oato.inaf.it/blazars/webt/} \\
Valle d'Aosta {\scriptsize (through GASP-WEBT program)}              & R       band        & \url{http://www.oato.inaf.it/blazars/webt/} \\
GRT                & V, R, B bands               & \url{http://asd.gsfc.nasa.gov/Takanori.Sakamoto/GRT/index.html} \\
MitSume                & g, Rc, Ic bands               & \url{http://www.hp.phys.titech.ac.jp/mitsume/index.html} \\
ROVOR                &   B, R, V, I bands            & \url{http://rovor.byu.edu/} \\
Campo Imperatore {\scriptsize (through GASP-WEBT program)}                &  H, J, K bands             & \url{http://www.oato.inaf.it/blazars/webt/} \\
OAGH    &  H, J, K bands             & \url{http://astro.inaoep.mx/en/observatories/oagh/} \\
WIRO                & J, K bands            & \url{http://physics.uwyo.edu/~chip/wiro/wiro.html} \\
SMA      &   225 GHz  &  \url{http://sma1.sma.hawaii.edu/} \\
VLBA  &  4.8, 8.3, 15.4, 23.8, 43.2 GHz               & \url{http://www.vlba.nrao.edu/} \\
Noto & 8.4, 43 GHz & \url{http://www.noto.ira.inaf.it/} \\
Mets\"ahovi {\scriptsize (through GASP-WEBT program)}                 & 37 GHz               & \url{http://www.metsahovi.fi/} \\
VLBA  {\scriptsize (through MOJAVE program)}                  &  15 GHz               & \url{http://www.physics.purdue.edu/MOJAVE/} \\
OVRO                & 15 GHz              & \url{http://www.astro.caltech.edu/ovroblazars} \\
Medicina & 8.4, 22.3 GHz & \url{http://www.med.ira.inaf.it/index_EN.htm} \\
UMRAO {\scriptsize (through GASP-WEBT program)}                & 4.8, 8.0, 14.5 GHz               & \url{http://www.oato.inaf.it/blazars/webt/} \\
RATAN-600  &  2.3, 4.8, 7.7, 11.1, 22.2 GHz & \url{http://w0.sao.ru/ratan/} \\
Effelsberg {\scriptsize (through F-GAMMA program)}               & 2.6, 4.6, 7.8, 10.3, 13.6, 21.7, 31 GHz           & \url{http://www.mpifr-bonn.mpg.de/div/effelsberg/index_e.html/} \\

\enddata
\tablecomments{The energy range shown in column two is the actual energy range covered during the Mrk\,501 observations, and not the instrument's nominal energy range, which might only be achievable for bright sources and excellent observing conditions. }
\tablecomments{$(a)$ The Whipple spectra were not included in Figure\,\ref{fig:MWSED}. See text for further comments.}
\label{TableWithInstruments}
\end{deluxetable}

\subsection{\FermiLAT Spectra During the Campaign}
\label{FermiSED_InMW}

\begin{figure}[!t]
  \centering
   \includegraphics[height=2.2 in]{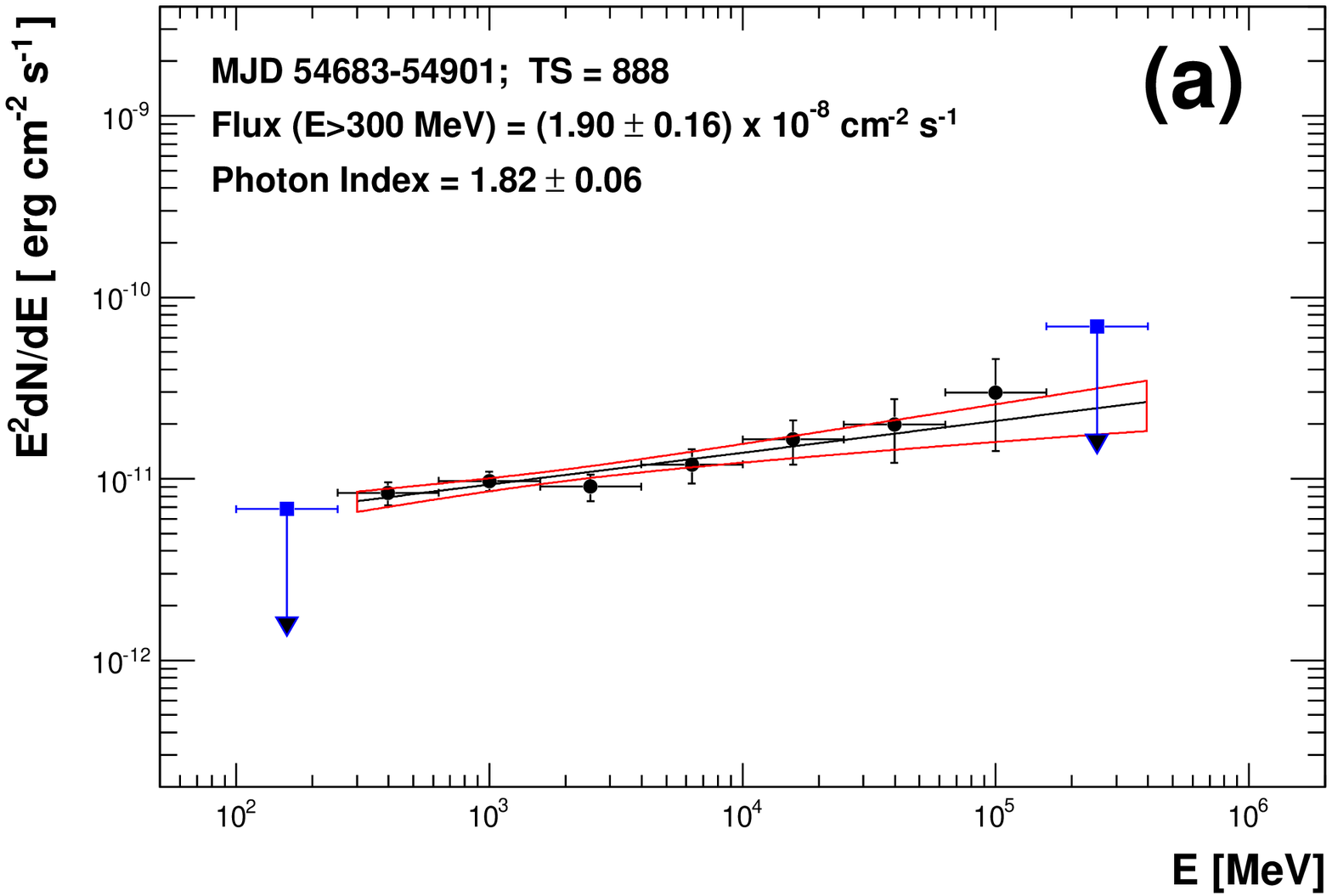}
  \includegraphics[height=2.2 in]{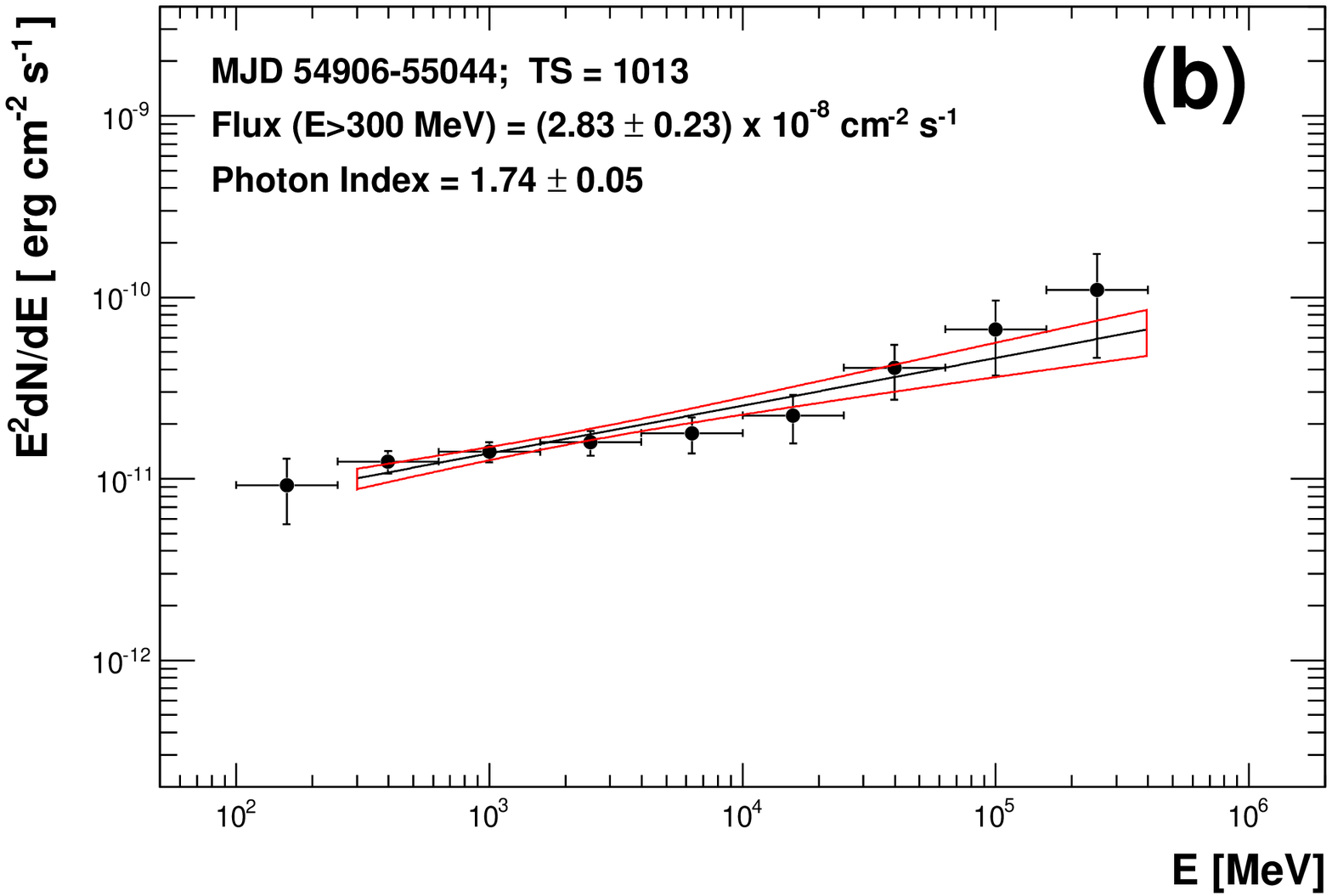}
  \includegraphics[height=2.2 in]{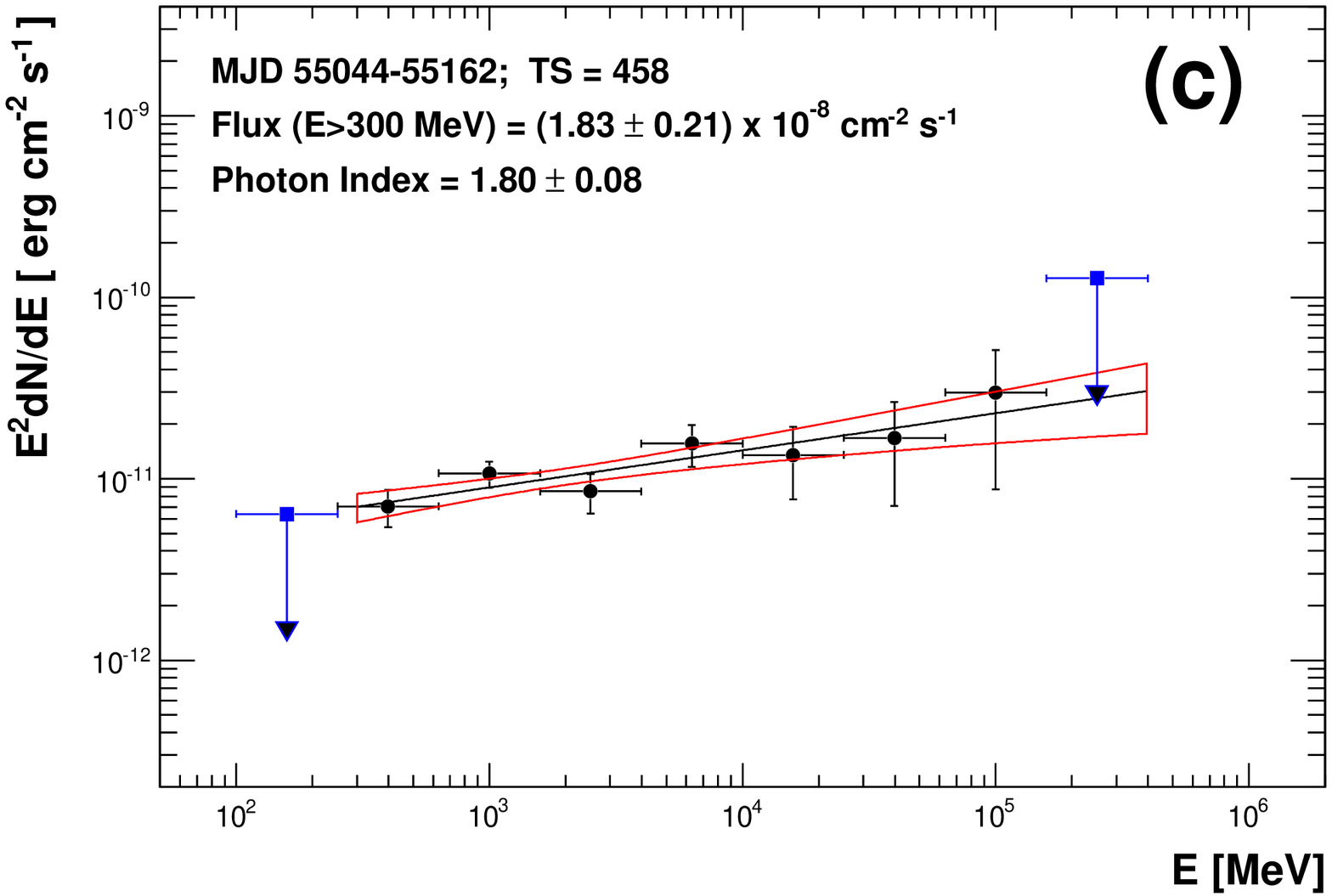}
  \caption{Spectral energy distribution for Mrk\,501 from \FermiLAT
    for several time intervals of interest. The panel {\it (a)} shows
    the SED for the time period before the multifrequency campaign
    (MJD 54683--54901), the panel {\it (b)} for the time interval
    corresponding to the multifrequency campaign (MJD 54905--55044)
    and the panel {\it (c)} for the period after the campaign (MJD
    55044--55162). In all panels,  the black line depicts the result
    of the unbinned likelihood power-law fit, the red contours denote
    the $68\%$ uncertainty of the power-law fit and blue arrows denote
    upper limits at 95\% confidence level, which were computed for the
    differential energy ranges with a signal of $TS<4$ or less than two photons predicted by the analysis model
    for Mrk\,501. The legend reports the results from the unbinned likelihood power-law fit in the energy range $0.3-400$\,GeV.}
  \label{fig:SED_LongIntervals}
\end{figure}

The Mrk\,501 spectrum measured by \FermiLATc, integrated during the time interval of the multifrequency campaign, is shown in the panel {\it (b)} of Figure\,\ref{fig:SED_LongIntervals}. The spectrum can be described by a power-law function with photon index $1.74 \pm 0.05$. The flux data points resulting from the analysis in differential energy ranges are within $1-2\,\sigma$ of the power-law fit result; this is an independent indication that a single power-law function is a good representation of the spectrum during the multifrequency campaign. On the other hand, the shape of the spectrum depicted by the differential energy flux data points suggests the possibility of a concave spectrum. As it was discussed in \S\ref{LC} and \S\ref{FermiSpectrum} (see Figures\,\ref{fig:LcAndScatter} and \ref{fig:SED_30day}), Mrk\,501 showed substantial spectral variability during the time period covered by the multifrequency campaign, with some 30-day time intervals characterized by relatively soft spectra (photon index $\sim 2$ for the 30-day intervals MJD 54892--54922 and MJD 54922--54952) and others by relatively hard spectra (photon index $\sim 1.6$ for the 30-day intervals MJD 54952--54982, MJD 54982--55012 and MJD 55012--55042). The panel {\it (b} of Figure\,\ref{fig:SED_LongIntervals} presents the average spectrum over those time intervals, and hence it would not be surprising to see two slopes (instead of one) in the spectrum. In order to evaluate this possibility, a broken power-law fit was applied, yielding indices of $1.86 \pm 0.08$ and $1.44 \pm 0.14$ below and above a break energy of $10 \pm 3$\,GeV, respectively. The likelihood ratio of the broken power law and the power law is $2.2$. Given that the broken power law has two additional degrees of freedom, this indicates that the broken power law is not statistically preferred over the single power law function. 

For comparison purposes we also computed the spectra for time
intervals before and after the multifrequency campaign (MJD
54683--54901 and MJD 55044--55162)\footnote{Technical problems
  prevented the scientific operation of the \FermiLAT instrument
  during the interval MJD 54901--54905.}. These two spectra, shown in
the panel {\it (a)}   and {\it (ac}  of Figure\,\ref{fig:SED_LongIntervals}, can both be described satisfactorily by single power-law functions with photon indices $1.82 \pm 0.06$ and $1.80 \pm 0.08$. Note that the two spectra are perfectly compatible with each other, which is consistent with the relatively small flux/spectral variability shown in Figures\,\ref{fig:LcAndScatter} and \ref{fig:Lc30daysMW} for those time periods. 

\subsection{The Average Broadband SED During the Campaign}
\label{MWSEDDataResults}

\begin{figure}[!t]
  \centering
  \includegraphics[width=6.5 in]{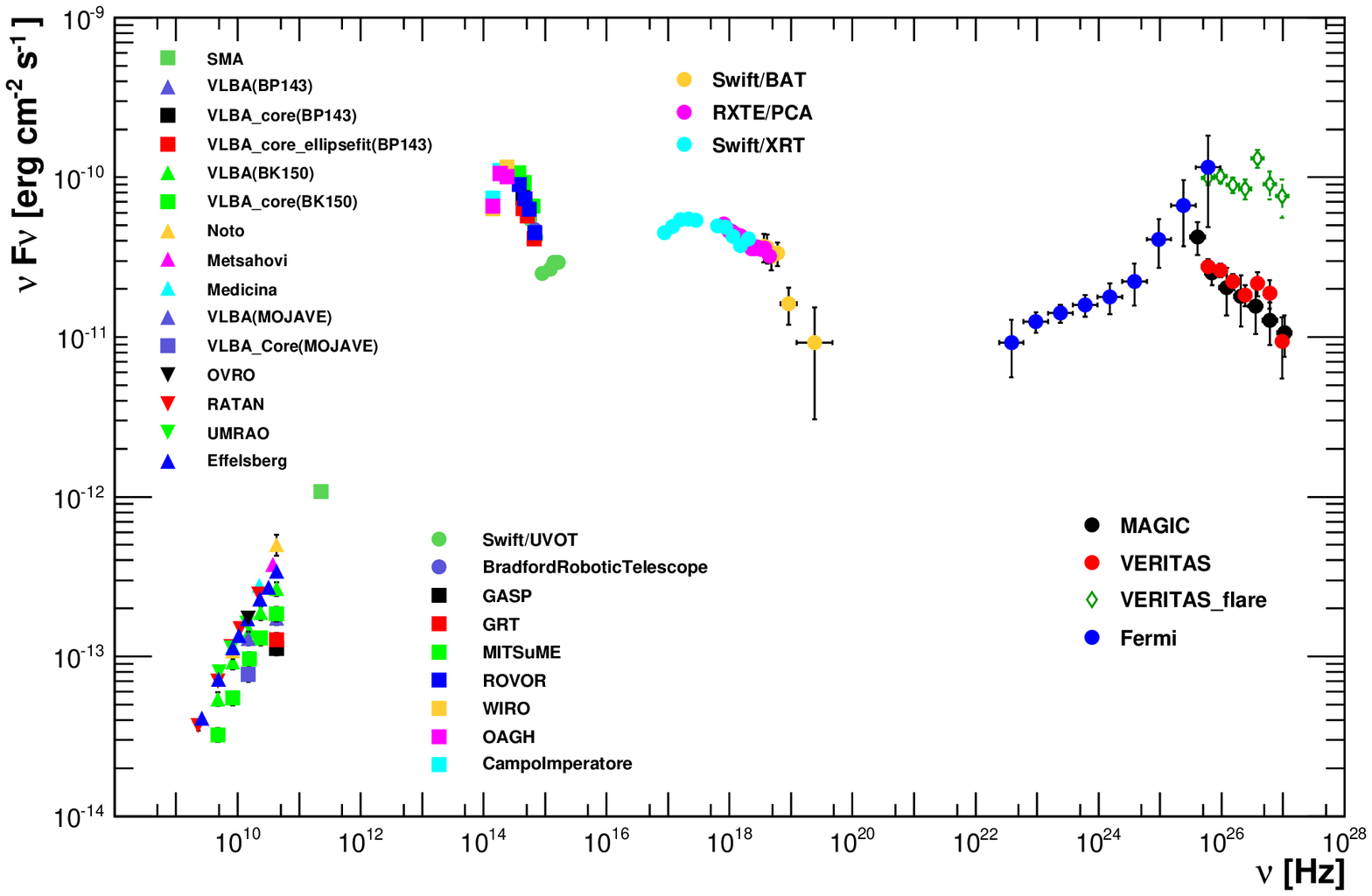}
  \caption{Spectral energy distribution for Mrk\,501 averaged over all
    observations taken during the multifrequency campaign performed
    between  2009 March 15 (MJD 54905)  and 2009 August 1 (MJD 55044). The legend reports the correspondence between the instruments and the measured fluxes. Further details about the instruments are given in \S\ref{CampaignDetails}. The optical and X-ray data have been corrected for Galactic extinction, but the host galaxy (which is clearly visible at the IR/optical frequencies) has not been subtracted. The TeV data from MAGIC and VERITAS have been corrected for the absorption in the extragalactic background light using the model reported in \cite{fran08}. The VERITAS data from the time interval MJD 54952.9--54955.9 were removed from the data set used to compute the average spectrum, and are depicted separately in the SED plot (in green diamonds). See text for further details.}
  \label{fig:MWSED}
 \end{figure}

The average broadband SED of Mrk\,501 resulting from our 4.5-month-long multifrequency campaign is shown in  Figure\,\ref{fig:MWSED}. The TeV data from MAGIC and VERITAS have been corrected for the absorption in the EBL using the particular EBL model by \cite{fran08}. The corrections given by the other low-EBL-level models \citep{kneiske04,gilmore09,finke10} are very similar for the low redshift of Mrk\,501 ($z =0.034$). The attenuation factor at a photon energy of $6$\,TeV (the highest energy detected from Mrk\,501 during this campaign) is in the range $e^{-\tau_{\gamma\gamma}} \simeq 0.4-0.5$, and smaller at lower energies.

During the campaign, as already noted above, the source did not show large flux variations like those recorded by EGRET in 1996, or those measured by X-ray and TeV instruments in 1997. Nevertheless, significant flux and spectral variations at $\gamma$-ray energies occurred in the time interval MJD 54905--55044. The largest flux variation during the campaign was observed at TeV energies during the time interval MJD 54952.9--54955.9, when VERITAS measured a flux about five times higher than the average one during the campaign. Because of the remarkable difference with respect to the rest of the analyzed exposure, these observations were excluded from the data set used to compute the average VERITAS spectrum for the campaign; the three-day ``flaring-state'' spectrum (2.4 hours of observation) is presented separately in Figure\,\ref{fig:MWSED}. Such a remarkable flux enhancement was not observed in the other energy ranges and hence Figure\,\ref{fig:MWSED} shows only the averaged spectra for the other instruments\footnote{The MAGIC telescope did not operate during the time interval MJD 54948--54965 due to a drive system upgrade.}. 

\begin{figure}[!t]
  \centering
 \includegraphics[width=5.0 in]{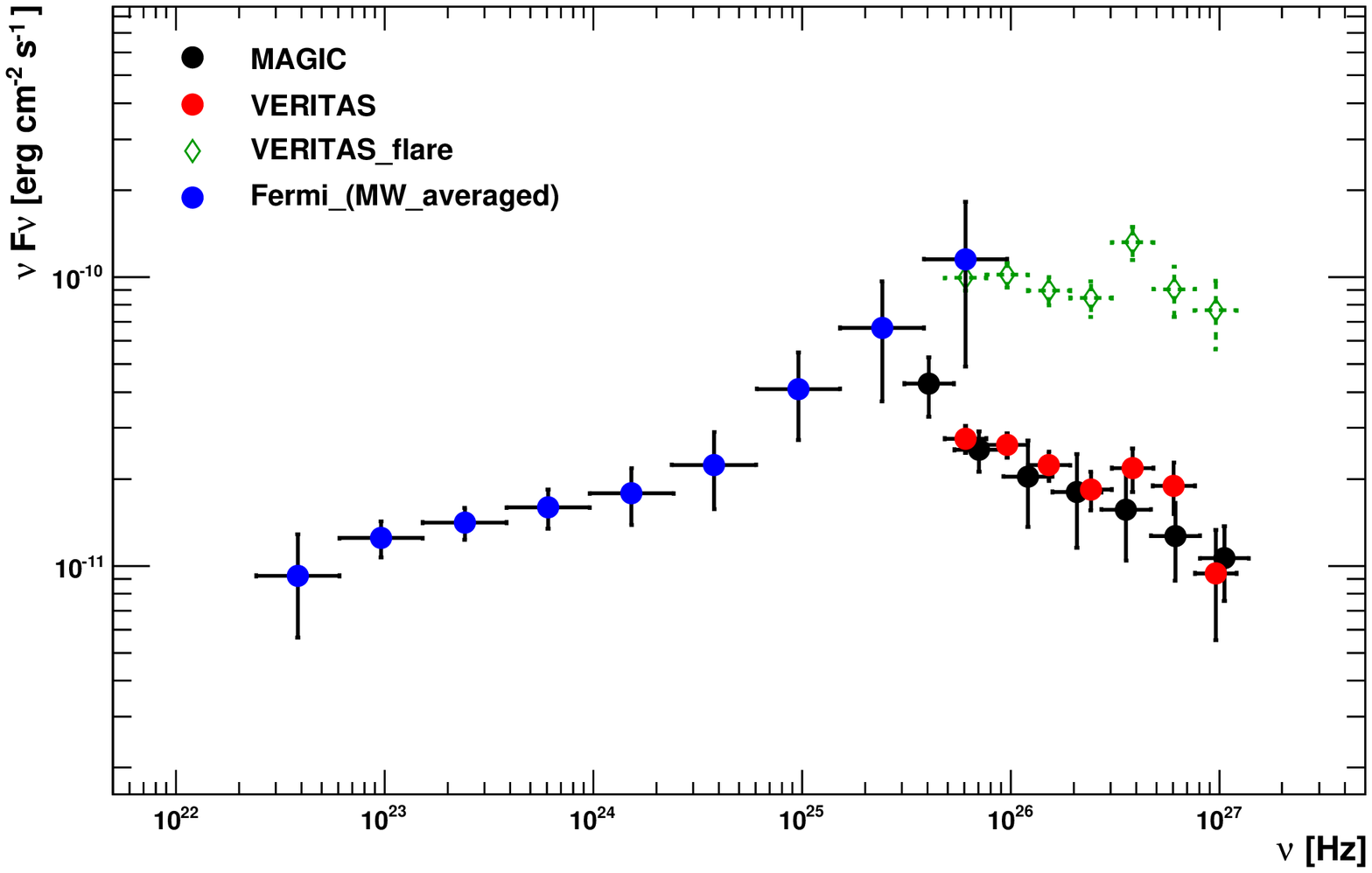}
 \includegraphics[width=5.0 in]{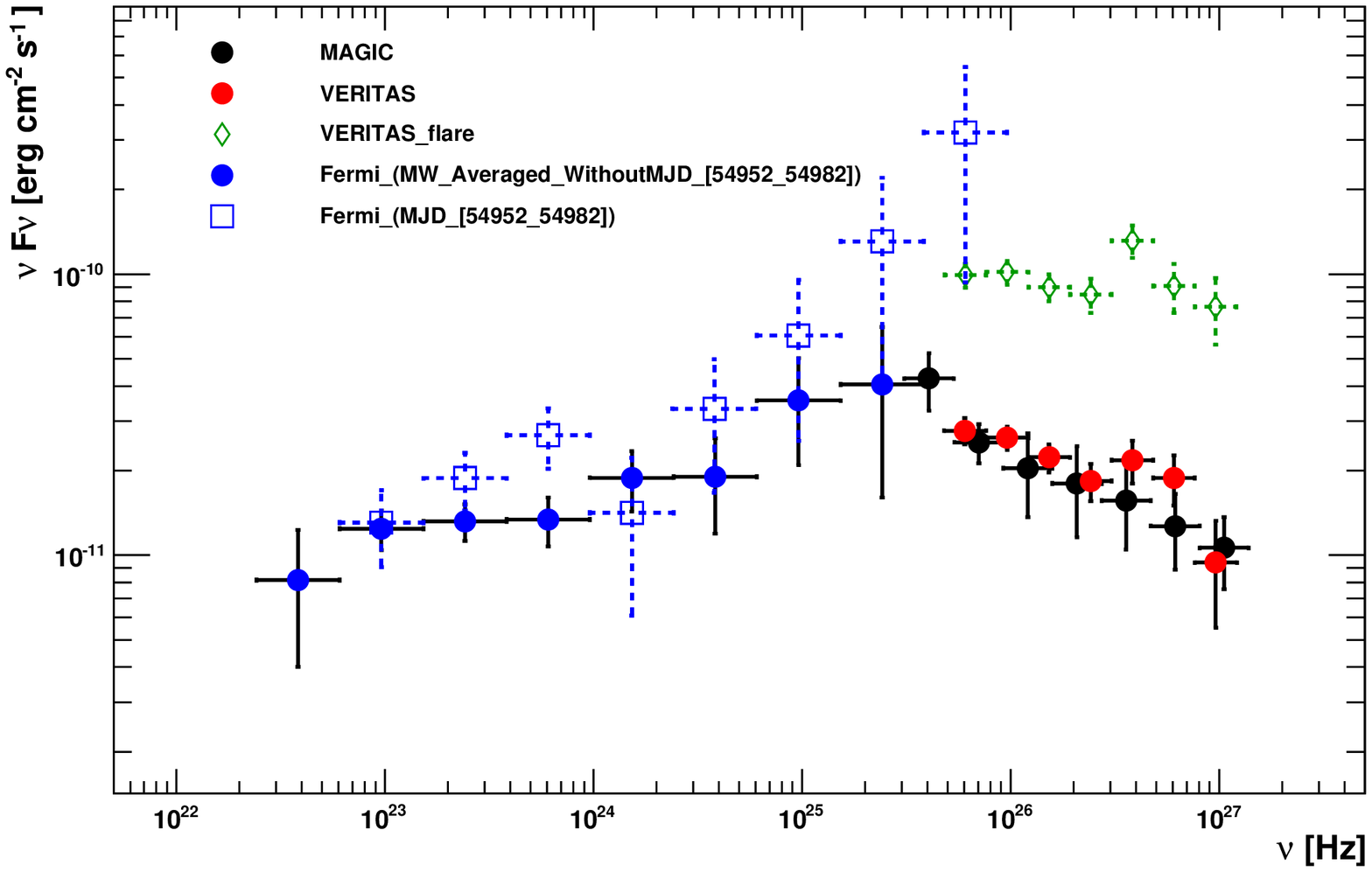}
  \caption{{\bf Top panel:} Enlargement of the $\gamma$-ray energy range from
    Figure\,\ref{fig:MWSED}.
% together with the \FermiLAT spectrum  during the 30-day time
% interval MJD 54862--54892 (open violet squares). This 30-day time
% interval, which corresponds to the softest spectrum (photon index
% $2.5 \pm 0.2$) measured by\FermiLATc, was not part of the
% multifrequency campaign. 
{\bf Bottom panel:} Same SED as in the top panel, but with the \FermiLAT data from the multifrequency campaign split in two data sets: MJD 54952--54982 (open blue squares) and the rest (filled blue circles).}
  \label{fig:MWSEDGamma}
 \end{figure}

The top panel in Figure\,\ref{fig:MWSEDGamma} shows a zoom of the
high-energy bump depicted in Figure\,\ref{fig:MWSED}.  
%For comparison purposes, in this figure we have included the
%\FermiLAT spectrum from the 30-day time interval MJD 54862--54892,
%when Mrk\,501 showed the largest photon index ($2.5 \pm 0.2$). 
 The last two energy bins from \Fermi ($60-160$ and $160-400$\,GeV) are systematically above (1-2$\sigma$) the measured/extrapolated spectrum from MAGIC and VERITAS. Even though this mismatch is not statistically significant, we believe that the spectral variability observed during the 4.5 month long campaign (see \S\ref{FermiSpectrum} and \S\ref{FermiSED_InMW}) could be the origin of such a difference. Because \FermiLAT operates in a survey mode, Mrk\,501 is constantly monitored at GeV energies\footnote{During every three hours of \Fermi operation, Mrk\,501 is in the LAT field of view for about 0.5 hour.}, while this is not the case for the other instruments which typically sampled the source during $\leq$1 hour every 5 days approximately. Moreover, because of bad weather or moonlight conditions, the monitoring at the TeV energies with Cherenkov telescopes was even less regular than that at lower frequencies. Therefore, \FermiLAT may have measured high activity that was missed by the other instruments. Indeed, the 2.4-hour high-flux spectrum from VERITAS depicted in Figure\,\ref{fig:MWSED} (which was obtained during the 3-day interval MJD 54952.9--54955.9)  demonstrates that, during the multifrequency campaign, there were time periods with substantially (factor of five) higher TeV activity. It is possible that the highest-energy LAT observations ($\geq$50 GeV) include high TeV flux states which occurred while the IACTs were not observing. 

If the flaring activity occurred only at the highest photon energies, then the computed \FermiLAT flux ($>$0.3\,GeV) would not change very much and the effect might only be visible in the measured power-law photon index. This seems to be the case in the presented data set. As was shown in Figure\,\ref{fig:SED_30day},  the 30-day intervals  MJD 54922--54952 and MJD 54952--54982 have photon fluxes above $0.3$\,GeV of $(3.9 \pm 0.6) \times 10^{-8}$\,ph\,cm$^{-2}$\,s$^{-1}$ and $(3.6 \pm 0.5) \times 10^{-8}$\,ph\,cm$^{-2}$s$^{-1}$, while their photon indices are $2.10 \pm 0.13$ and $1.63 \pm 0.09$, respectively. Therefore, the spectral information (together with the enhanced photon flux) indicates the presence of flaring activity at the highest $\gamma$-ray energies during the second 30-day time period. Besides the factor $\sim 5$ VHE flux enhancement recorded by VERITAS and Whipple at the beginning of the time interval MJD 54952--54982,  MAGIC and Whipple also recorded a factor $\sim 2$ VHE flux enhancement at the end of this 30-day time interval \citep[see preliminary fluxes reported in ][]{PanequeComo2009,Pichel2009}. This flux enhancement was measured for the time interval MJD 54975--54977, but there were no VHE measurements during the period MJD 54970.5--54975.0. Thus, the average \FermiLAT spectrum could have been affected by elevated VHE activity during the 30-day time interval MJD 54952--54982, which was only partly covered by the IACTs participating in the campaign.

For illustrative purposes, in the bottom panel of Figure\,\ref{fig:MWSEDGamma} we show separately the \FermiLAT spectra for the 30-day time interval MJD 54952--54982 (high photon flux and hard spectrum), and for the rest of the campaign. It is interesting to note that the \FermiLAT spectrum without the 30-day time interval MJD 54952--54982 (blue data points in the bottom panel of Figure\,\ref{fig:MWSEDGamma}) agrees perfectly with the VHE spectrum measured by IACTs. We also want to point out that the power-law fit to the \FermiLAT spectrum without the 30-day interval MJD 54952--54982 gave a photon flux above $0.3$\,GeV of $(2.62 \pm 0.25) \times 10^{-8}$\,ph\,cm$^{-2}$\,s$^{-1}$ with a photon index of $1.78 \pm 0.07$, which is statistically compatible with the results for the power-law fit to the \FermiLAT data from the entire campaign (see  panel {\it (b)} in Figure\,\ref{fig:SED_LongIntervals}). As discussed above, the flaring activity occurred mostly at the highest energies, where the (relatively) low photon count has little impact on the overall power-law fit performed above $0.3$\,GeV.

This is the most complete quasi-simultaneous SED ever collected for Mrk\,501, or for any other TeV-emitting BL Lac \citep[see also][in preparation]{Mrk421FermiView}. At the highest energies, the combination of Fermi and MAGIC/VERITAS data allows us to measure, for the first time, the high-energy bump without any spectral gap. The low-energy spectral component is also very well characterized with \Swiftc-UVOT, \Swiftc-XRT and \RXTEc-PCA data, covering the peak of the synchrotron continuum. The only (large) region of the SED with no available data corresponds to the photon energy range $200$\,keV\,$-100$\,MeV, where the sensitivity of current instruments is not good enough to detect Mrk\,501. It is worth stressing that the excellent agreement in the overlapping energies among the various instruments (which had somewhat different time coverage) indicates that the collected data are representative of the truly average SED during the multi-instrument campaign.

\section{Modeling the Spectral Energy Distribution of Mrk\,501} 
\label{SEDModel}

The simultaneous broadband data set resulting from the multifrequency
campaign reported above offers an unprecedented opportunity to model
the emission of an archetypal TeV blazar in a more robust way than in
the past. It is widely believed that the radio-to-$\gamma$-ray emission of the BL Lac class of AGN is produced predominantly via the synchrotron and synchrotron self-Compton (SSC) processes, and hence the homogeneous one-zone approximation of the SSC scenario is the simplest model to consider. Here we therefore adopt the `standard' one-zone SSC model, which has had moderate success in accounting for the spectral and temporal properties of the TeV-emitting BL Lacs analyzed so far \citep[e.g.,][and references therein]{fin08,ghi09SEDModel}. We also note that one-zone SSC analyses have been widely applied before to the particular case of Mrk\,501 \citep[e.g.,][]{bednarek1999,katar01,tav01,kin02,Albert2007}. However, it is important to stress that the modeling results from the previous works related almost exclusively to the high-activity state of Mrk\,501. In the more recent work by \cite{Anderhub2009} the source was studied also during its low-activity state, yet the simultaneous observations used in the modeling covered only the X-ray and TeV photon energies. In this paper we study Mrk\,501 during a relatively low activity state, and the modeling is applied to a more complete broadband SED extending from radio to TeV energies, including the previously unavailable GeV data from Fermi. This constitutes a substantial difference with respect to previous works. The resulting constraints on the physical parameters of the source, together with several limitations of the applied scenario, are discussed further down in the next sections.

We want to note that modeling of the average blazar SED based on a 
scenario assuming steady-state homogenous emission zone could be an
over-simplification of the problem. The blazar emission may be produced 
in an inhomogeneous region, involving stratification of the emitting 
plasma both along and accross a relativistic outflow. In such a case, 
the observed radiative output of a blazar could be due to a complex 
superposition of different emission zones characterized by very different 
parameters and emission properties. Some first attempts to approach this 
problem in a more quantitative way have been already discussed in the 
literature
\citep[e.g.][]{Ghisellini2005,Katar2008,Graff2008,Giannios2009}. 
The main drawback of the proposed models, however, is 
the increased number of free parameters (over the simplest homogeneous 
one-zone scenario), what reduces considerably the predictive power of 
the modeling. That is particularly problematic if a ``limited'' (in a 
time and energy coverage) dataset is considered in the modeling. Only a 
truly simultaneous multifrequency dataset covering a large 
fraction of the available electromagnetic spectrum and a wide range of 
timescales --- like the one collected during this and future campaigns which will 
be further exploited in forthcoming publications --- will enable to
test such more sophisticated 
and possibly more realistic blazar emission models in a time-deoendent manner.

\subsection{SSC Modeling}
\label{SSCModel}

Let us assume that the emitting region is a homogeneous and roughly spherically symmetric moving blob, with radius $R$ and comoving volume $V' \simeq (4 \pi/ 3) \, R^3$. For this, we evaluate the comoving synchrotron and synchrotron self-Compton emissivities, $\nu'\!j'_{\nu'}$, assuming isotropic distributions of ultrarelativistic electrons and synchrotron photons in the rest frame of the emitting region. Thus, we use the exact synchrotron and inverse-Compton kernels (the latter one valid in both Thomson and Klein-Nishina regimes), as given in \cite{cru86} and \cite{blu70}, respectively. The intrinsic monochromatic synchrotron and SSC luminosities are then $\nu'\!L'_{\nu'} = 4 \pi \, V' \, \nu'\!j'_{\nu'}$, while the observed monochromatic flux densities (measured in erg\,cm$^{-2}$\,s$^{-1}$) can be found as
\begin{equation}
\nu F_{\nu} = {\delta^4 \over 4 \pi \, d_L^2} \, [\nu'\!L'_{\nu'}]_{\nu' = \nu \, (1+z)/\delta} \simeq {4 \pi \, \delta^4 R^3 \over 3 \, d_L^2} \, [\nu'\!j'_{\nu'}]_{\nu' = \nu \, (1+z)/\delta} \, ,
\end{equation}
\noindent
where $\delta$ is the jet Doppler factor, $z = 0.034$ is the source redshift, and $d_L = 142$\,Mpc is the luminosity distance to Mrk\,501. In order to evaluate the comoving emissivities $\nu'\!j'_{\nu'}$, the electron energy distribution $n'_e(\gamma)$ has to be specified. For this, we assume a general power-law form between the minimum and maximum electron energies, $\gamma_{min}$ and $\gamma_{max}$, allowing for multiple spectral breaks in between, as well as for an exponential cut-off above $\gamma_{max}$. In fact, the broadband data set for Mrk\,501 requires two different electron break energies, and hence we take the electron energy distribution in a form
\begin{equation}
n'_e(\gamma) \propto \left\{ \begin{array}{ccc} \gamma^{-s_1} & {\rm for} & \gamma_{min} \leq \gamma < \gamma_{br,\,1} \\
\gamma^{-s_2} & {\rm for} & \gamma_{br,\,1} \leq \gamma < \gamma_{br,\,2} \\
\gamma^{-s_3} \, \exp\left[-\gamma/\gamma_{max}\right] & {\rm for} & \gamma_{br,\,2} \leq \gamma
\end{array} \right. \quad ,
\end{equation}
\noindent
with the normalization expressed in terms of the equipartition parameter (the ratio of the comoving electron and magnetic field energy densities), namely
\begin{equation}
\eta_e \equiv {U'_e \over U'_B} = {\int \, \gamma \, m_e c^2 \, n'_e(\gamma) \, d\gamma \over B^2 / 8 \pi} \, .
\end{equation}
\noindent
The measured SED is hardly compatible with a simpler form of the electron distribution with only one break and an exponential cutoff. However, some smoothly curved spectral shape might perhaps be an alternative representation of the electron spectrum \citep[e.g.,][]{sta08,Tramacere2009}.

The model adopted is thus characterized by four main free parameters ($B$, $R$, $\delta$, and $\eta_e$), plus seven additional ones related to the electron energy distribution ($\gamma_{min}$, $\gamma_{br,\,1}$, $\gamma_{br,\,2}$, $\gamma_{max}$, $s_1$, $s_2$, and $s_3$). These seven additional parameters are determined by the spectral shape of the non-thermal emission continuum probed by the observations, predominantly by the spectral shape of the synchrotron bump (rather than the inverse-Compton bump), and depend only slightly on the particular choice of the magnetic field $B$ and the Doppler factor $\delta$ within the allowed range\footnote{For example, for a given critical (break) synchrotron frequency in the observed SED, the corresponding electron break Lorentz factor scales as $\gamma_{br} \propto 1/\sqrt{B \, \delta}$.}. There is a substantial degeneracy regarding the four main free parameters: the average emission spectrum of Mrk\,501 may be fitted by different combinations of $B$, $R$, $\delta$, and $\eta_e$ with little variation in the shape of the electron energy distribution. Note that, for example, $[\nu F_{\nu}]_{syn} \propto R^3 \, \eta_e$, but at the same time $[\nu F_{\nu}]_{ssc} \propto R^4 \, \eta_e^2$. 
We can attempt to reduce this degeneracy by assuming that the observed main variability timescale is related to the size of the emission region and its Doppler factor according to the formula
\begin{equation}
t_{var} \simeq {(1+z) \, R \over c \, \delta} \, .
\end{equation}
\noindent
The multifrequency data collected during the 4.5-month campaign (see \S\ref{MWSED}) allows us to study the variability of Mrk\,501 on timescales from months to a few days. We found that, during this time period, the multifrequency activity varied typically on a timescale of $5-10$ days, with the exception of a few particular epochs when the source became very active in VHE $\gamma$-rays, and flux variations with timescales of a day or shorter were found at TeV energies. Nevertheless, it is important to stress that several authors concluded in the past that the dominant emission site of Mrk\,501 is characterized by variability timescales longer than one day \citep[see][for a comprehensive study of the Mrk\,501 variability in X-rays]{kat01}, and that the power in the intraday flickering of this source is small, in agreement with the results of our campaign. Nevertheless, one should keep in mind that this object is known for showing sporadic but extreme changes in its activity that can give flux variations on timescales as short as a few minutes \citep{Albert2007}. In this work we aim to model the average/typical behaviour of Mrk\,501 (corresponding to the 4.5-month campaign) rather than specific/short periods with outstanding activity, and hence we constrained the minimum (typical) variability timescale $t_{var}$ in the model to the range $1-5$ days.

Even with $t_{var}$ fixed as discussed above, the reconstructed SED of Mrk\,501 may be fitted by different combinations of $B$, $R$, $\delta$, and $\eta_e$. Such a degeneracy between the main model parameters is an inevitable feature of the SSC modeling of blazars \citep[e.g.,][]{Kataoka1999}, and it is therefore necessary to impose additional constraints on the physical parameters of the dominant emission zone. Here we argue that such constraints follow from the requirement for the electron energy distribution to be in agreement with the one resulting from the simplest prescription of the energy evolution of the radiating electrons within the emission region, as discussed below.

The idea of separating the sites for the particle acceleration and emission processes is commonly invoked in modeling different astrophysical sources of high-energy radiation, and blazar jets in particular. Such a procedure is not always justified, because interactions of ultrarelativistic particles with the magnetic field (leading to particle diffusion and convection in momentum space) are generally accompanied by particle radiative losses (and vice versa). On the other hand, if the characteristic timescale for energy gains is much shorter than the timescales for radiative cooling ($t'_{rad}$) or escape ($t'_{esc}$) from the system, the particle acceleration processes may be indeed approximated as being `instantaneous', and may be modeled by a single injection term ${\dot Q}(\gamma)$ in the simplified version of the kinetic equation 
\begin{equation}
{\partial n'_e(\gamma) \over \partial t} = - {\partial \over \partial \gamma}
\left[{\gamma\, n_e(\gamma) \over t'_{rad}(\gamma)}\right]  - {n_e(\gamma) \over t'_{esc}} +
{\dot Q}(\gamma)
\label{eq:kin}
\end{equation}
describing a very particular scenario for the energy evolution of the radiating ultrarelativistic electrons.

It is widely believed that the above equation is a good approximation for the energy evolution of particles undergoing diffusive (`first-order Fermi') shock acceleration, and cooling radiatively in the downstream region of the shock. In such a case, the term ${\dot Q}(\gamma)$ specifies the energy spectrum and the injection rate of the electrons freshly accelerated at the shock front and not affected by radiative losses, while the escape term corresponds to the energy-independent dynamical timescale for the advection of the radiating particles from the downstream region of a given size $R$, namely $t'_{esc} \simeq t'_{dyn} \simeq R/c$. The steady-state electron energy distribution is then very roughly $n'_e(\gamma) \sim t'_{dyn} \, {\dot Q}(\gamma)$ below the critical energy for which $t'_{rad}(\gamma) = t'_{dyn}$, and $n'_e(\gamma) \sim t'_{rad}(\gamma) \, {\dot Q}(\gamma)$ above this energy. Note that in the case of a power-law injection ${\dot Q}(\gamma) \propto \gamma^{-s}$ and a homogeneous emission region with dominant radiative losses of the synchrotron type, $t'_{rad}(\gamma) \propto \gamma^{-1}$, the injected electron spectrum is expected to steepen by $\Delta s = 1$ above the critical `cooling break' energy. This provides us with the additional constraint on the free model parameters for Mrk\,501: namely, we require that the position of the second break in the electron energy distribution needed to fit the reconstructed SED, $\gamma_{br\,2}$, should correspond to the location of the cooling break for a given chosen set of the model free parameters.

\begin{figure}[!t]
  \centering
  \includegraphics[scale=1.5]{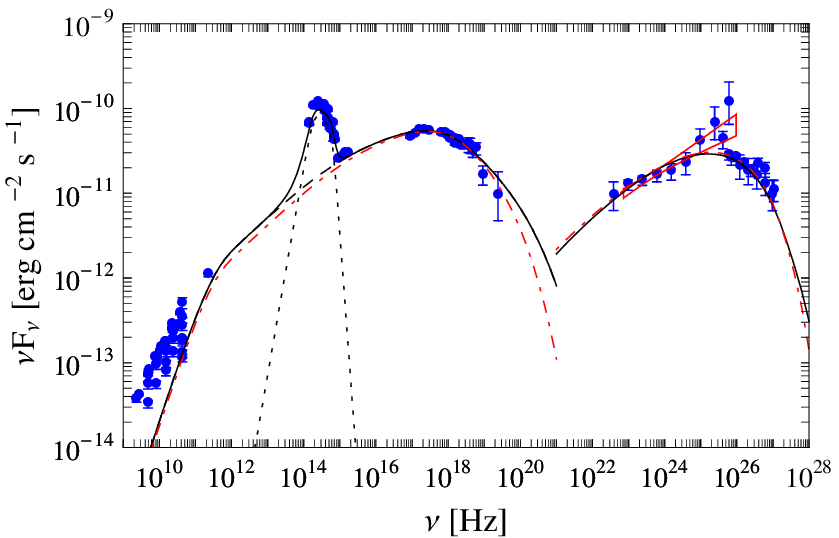}
  \caption{The SSC model fits to the broadband emission spectrum of
    Mrk\,501, averaged over all the observations made during the
    multifrequency campaign performed between  2009 March 15 (MJD
    54905)  and 2009 August 1 (MJD 55044). The red bow-tie in the
    figure corresponds to the 68\% containment of the power-law fit to
    the average \FermiLAT spectrum (photon index $1.74 \pm 0.05$). The
    dotted black curve denotes the fit to the starlight emission of
    the host galaxy assuming a template of a luminous elliptical as
    given in \cite{silva98}. The details of the model are given in
    \S\ref{SEDModel}. The black curves correspond to the main set of
    the model parameters considered (variability timescale $t_{var}
    \simeq 4$\,days), while the red dot-dashed curves to the
    alternative set of the model parameters with the emission region
    size decreased by an order of magnitude ($t_{var} \simeq
    0.35$\,day). See text for further discussion.}
  \label{fit1}
 \end{figure}

Figure\,\ref{fit1} (black curves) shows the resulting SSC model fit (summarized in Table~2) to the averaged broadband emission spectrum of Mrk\,501, which was obtained for the following parameters: $B = 0.015$\,G, $R=1.3 \times 10^{17}$\,cm, $\delta = 12$, $\eta_e = 56$, $\gamma_{min} = 600$, $\gamma_{br,\,1} = 4 \times 10^4$, $\gamma_{br,\,2} = 9 \times 10^5$, $\gamma_{max} = 1.5 \times 10^7$, $s_1 = 2.2$, $s_2 = 2.7$, and $s_3 = 3.65$. The overall good agreement of the model with the data is further discussed in \S\ref{SEDdata}. Here, we note that, for these model parameters, synchrotron self-absorption effects are important only below $1$\,GHz, where we do not have observations\footnote{The turnover frequency related to the synchrotron self-absorption may be evaluated using the formulae given in \cite{ghi91} and the parameter values from our SSC model fit as $\nu'_{ssa} \simeq 60$\,MHz, which in the observer frame reads $\nu_{ssa} = \delta \, \nu'_{ssa} / (1+z) \simeq 0.7$\,GHz.}. We also emphasize that with all the aforementioned constraints and for a given spectral shape of the synchrotron continuum (including all the data points aimed to be fitted by the model, as discussed below), and thus for a fixed spectral shape of the electron energy distribution (modulo critical electron Lorentz factors scaling as $\propto 1/\sqrt{B\,\delta}$), the allowed range for the free parameters of the model is relatively narrow. Namely, for the variability timescale between $1$ and $5$ days, the main model parameters may change within the ranges $R \simeq (0.35-1.45)\times 10^{17}$\,cm, $\delta \simeq 11-14$, and $B \simeq 0.01-0.04$\,G. The parameter $\eta_e$ depends predominantly on the minimum Lorentz factor of the radiating electrons. Hence, it is determined uniquely as $\eta_e \simeq 50$ with the sub-mm flux included in the fitted dataset. Only with a different prescription for the spectral shape of the electron energy distribution could the main free parameters of the model be significantly different from those given above. 

Despite the absence of any fast variability during this multifrequency 
campaign (apart from the already discussed isolated 3-day-long flare), 
Mrk\,501 is known for the extremely rapid flux changes at the highest 
observed photon energies \citep[e.g.][]{Albert2007}. Hence, it is 
interesting to check whether any shorter than few-day-long variability 
timescales can be accommodated in the framework of the simplest SSC model 
applied here for the collected dataset. In order to do that, we decreased 
the minimum variability time scale by one order of magnitude (from 4 days 
to 0.4 days), and tried to model the data. A satisfactory fit 
could be obtained with those modified parameters, but only when we relaxed the 
requirement for the electron energy distribution to be in agreement with 
the one following from the steady-state solution to Equation \ref{eq:kin}, 
and in particular the resulting constraint for the second break in the 
electron spectrum to be equal the cooling break. This ``alternative'' 
model fit is shown in Figure\,\ref{fit1} (red curves) together with the 
``best'' model fit discussed above. The resulting model parameters for the 
``alternative'' fit are $B = 0.03$\,G, $R=2 \times 10^{16}$\,cm, $\delta = 
22$, $\eta_e = 130$, $\gamma_{min} = 300$, $\gamma_{br,\,1} = 3 \times 
10^4$, $\gamma_{br,\,2} = 5 \times 10^5$, $\gamma_{max} = 3 \times 10^6$, 
$s_1 = 2.2$, $s_2 = 2.7$, and $s_3 = 3.5$. This particular parameter set 
--- which should be considered as an illustrative one, only --- would be 
therefore consistent with a minimum variability timescale of $0.36$\,days, 
but at the price of much larger departures from the energy equipartition 
($\eta_e > 100$). The other source parameters, on the other hand, would 
change only slightly (see Table~2). Because of the mismatch (by factor
$\sim$3) between the location of the cooling break and the second
break in the electron distribution, 
we consider this 
``alternative'' fit less consistent with the hypothesis of steady-state 
homogeneous one-zone SSC scenario, which is the framework we chose to 
model the broad-band SED of Mrk\,501 emerging from the campaign.

\subsection{Notes on the Spectral Data Points}
\label{SEDdata}

The low-frequency radio observations performed with single-dish
instruments have a relatively large contamination from non-blazar
emission due to the underlying extended jet component, and hence they
only provide upper limits for the radio flux density of the blazar emission
zone. On the other hand, the flux measurements by the interferometric
instruments (such as VLBA), especially the ones corresponding to the
core, provide us with the radio flux density from a region that is
not much larger than the blazar emission region. 
%Even though it has been suggested that the VLBA core size determinations
%should be treated as actual measurements \citep{Kovalev2005}, the core sizes
%obtained are very close to the VLBA resolution and hence for this
%paper we adopted the conservative approach of treating this size as an upper limit to
%the actual size of the radio core. Therefore,  the term ``unresolved core'' will be used in
%the rest of the paper to refer to the VLBA core. 

The radio flux densities from interferometric observations (from the VLBA 
core) are expected to be close upper limits to the radio continuum of
the blazar emission component. The estimated size of the partially-resolved
VLBA core of Mrk\,501 at 15 GHz and 43 GHz is $\simeq$ 0.14--0.18\,mas\,$\simeq$ 2.9--3.7 $\times 10^{17}$\,cm
(with the appropriate conversion scale $0.67$\,pc/mas). 
The VLBA size estimation is the FWHM of a Gaussian 
representing the brightness distribution of the blob, which could be
approximated as 0.9 times the radius of a corresponding spherical blob \citep{Marscher1983}.
That implies that the size of the VLBA core is only a factor 2--3
larger than the emission region in our SSC model fit
($R=1.3 \times 10^{17}$\,cm). Therefore, it is reasonable to assume
that the radio flux density from the VLBA core is indeed dominated
by the radio flux density of the blazar emission. Forthcoming multi-band
correlation studies (in particular VLBA and SMA radio with the $\gamma$-rays
from \FermiLATc) will shed light on this particular subject. Interestingly, the magnetic field claimed for the partially-resolved radio core of Mrk\,501 (which has a size of $\lesssim 0.2$\,mas) and its sub-mas jet, namely $B \simeq (10-30)$\,mG \citep{gir04,gir08}, is in very good agreement with the value emerging from our model fits (15 mG), assuring self-consistency of the approach adopted.

In addition to this, in the modeling we also aimed at matching the sub-millimeter flux of Mrk\,501, given at the observed frequency of $225$\,GHz, assuming that it represents the low-frequency tail of the optically-thin synchrotron blazar component. One should emphasize in this context that it is not clear if the blazar emission zone is in general located deep within the millimeter photosphere, or not. However, the broadband variability of luminous blazars of the FSRQ type indicates that there is a significant overlap of the blazar zone with a region where the jet becomes optically thin at millimeter wavelengths \citep[as discussed by][for the particular case of the blazar 3C\,454.3]{sik08}. We have assumed that the same holds for BL Lac objects. 

The IR/optical flux measurements in the range $\sim (1-10) \times 10^{14}$\,Hz represent the starlight of the host galaxy and hence they should be excluded when fitting the non-thermal emission of Mrk\,501. We modelled these data points with the template spectrum of an elliptical galaxy instead \citep[including only the dominant stellar component due to the evolved red giants, as discussed in][]{silva98}, obtaining a very good match (see the dotted line in Figure\,\ref{fit1}) for the bolometric starlight luminosity $L_{star} \simeq 3 \times 10^{44}$\,erg\,s$^{-1}$. Such a luminosity is in fact expected for the elliptical host of a BL Lac object. The model spectrum of the galaxy falls off very rapidly above $5 \times 10^{14}$\,Hz, while the three UV data points (above $10^{15}$\,Hz) indicate a prominent, flat power-law UV excess over the starlight emission. Therefore, it is reasonable to assume that the observed UV fluxes correspond to the synchrotron (blazar-type) emission of Mrk\,501 and, consequently, we used them in our model fit. However, many elliptical galaxies do reveal in their spectra the so-called `UV upturn', or `UV excess', whose origin is not well known, but which is presumably related to the starlight continuum (most likely due to young stars from the residual star-forming activity within the central region of a galaxy) rather than to non-thermal (jet-related) emission processes \citep[see, e.g.,][]{cod79,atl09}. Hence, it is possible that the UV data points provided here include some additional contamination from the stellar emission, and as such might be considered as upper limits for the synchrotron radiation of the Mrk\,501 jet.

The observed X-ray spectrum of Mrk\,501 agrees very well with the SSC model fit, except for a small but statistically significant discrepancy between the model curve and the first two data points provided by \Swiftc-XRT, which correspond to the energy range $0.3-0.6$\,keV. As pointed out in \S\ref{CampaignDetails}, the \Swiftc-XRT data had to be corrected for a residual energy offset which affects the lowest energies. The correction for this effect could introduce some systematic differences with respect to the actual fluxes detected at those energies. These low-energy X-ray data points might be also influenced by intrinsic absorption of the X-ray photons within the gaseous environment of Mrk\,501 nucleus, as suggested by the earlier studies with the ASCA satellite \citep[see][]{Kataoka1999}. As a result, the small discrepancy between the data and the model curve within the range $0.3-0.6$\,keV can be ignored in the modelling.

The agreement between the applied SSC model and the $\gamma$-ray data is also very good. In particular, the model returns the $\gamma$-ray photon index 
%$\Gamma_{\rm 0.3-30\,GeV} \simeq 1.78$, 
$1.78$ in the energy range 0.3-30\,GeV, 
which can be compared with the one resulting from the power-law fit to the \FermiLAT data above 0.3\,GeV, namely $1.74 \pm 0.05$. 
%$\Gamma_{\rm >0.3\,GeV} \simeq 1.74 \pm 0.05$. 
However, the last two energy bins from Fermi ($60-160$ and $160-400$\,GeV) are systematically above ($2 \sigma$) the model curves, as well as above the averaged spectrum reported by MAGIC and VERITAS. A possible reason for mismatch between the average \FermiLAT spectrum and the one from MAGIC/VERITAS was discussed in \S\ref{MWSEDDataResults}.

\section{Discussion}
\label{Discussion}

In this section we discuss some of the implications of the model results presented above. After a brief analysis of the global parameters of the source resulting from the SSC fits (\S\ref{SEDfit}), the discussion focuses on two topics. Firstly (\S\ref{EED}), we show that the characteristics of the electron energy distribution emerging from our modeling can be used to constrain the physical processes responsible for the particle acceleration in Mrk\,501, processes which may also be at work in other BL Lac type objects. Secondly (\S\ref{SEDvariability}), we examine the broadband variability of Mrk\,501 in the framework of the model.

\subsection{Main Characteristics of the Blazar Emission Zone in Mrk\,501}
\label{SEDfit}

The values for the emission region size $R = 1.3 \times 10^{17}$\,cm and the jet Doppler factor $\delta = 12$ emerging from our SSC model fit give a minimum (typical) variability timescale of $t_{var} \simeq (1+z)\, R / c \, \delta \sim 4$\,days, which is consistent with the variability observed during the campaign and with previous studies of the X-ray activity of Mrk\,501 \citep{kat01}. At this point, it is necessary to determine whether an emission region characterized by these values of $R$ and $\delta$ is optically thin to internal two-photon pair creation $\gamma\gamma \to e^+e^-$ for the highest TeV energies observed during the campaign. We now affirm pair transparency due to insufficient density of soft target photons. 

Since Mrk\,501 is a cosmologically local object, pair conversion in the EBL is not expected to prevent its multi-TeV photons from reaching the Earth, although the impact of this process is not negligible, as mentioned in \S\ref{MWSED}. Therefore, dealing with a nearby source allows us to focus mostly on the intrinsic absorption processes, rather than on the cosmological, EBL-related, attenuation of the $\gamma$-ray emission. Moreover, because of the absence (or weakness) of accretion-disk-related circumnuclear photon fields in BL Lac objects like Mrk\,501, we only need to consider photon-photon pair production involving photon fields internal to the jet emission site. The analysis is therefore simpler than in the case of FSRQs, where the attenuation of high-energy $\gamma$-ray fluxes is dominated by interactions with photon fields external to the jet --- such as those provided by the broad line regions or tori --- for which the exact spatial distribution is still under debate.

Pair-creation optical depths can be estimated as follows. Using the $\delta$-function approximation for the photon-photon annihilation cross-section \citep{zdz85}, $\sigma_{\gamma \gamma}(\varepsilon'_0, \varepsilon'_\gamma) \simeq 0.2 \, \sigma_T \, \varepsilon'_{0} \, \delta[\varepsilon'_{0} - (2 m_e^2 c^4/\varepsilon'_{\gamma})]$, the corresponding optical depth for a $\gamma$-ray photon with observed energy $\varepsilon_{\gamma} = \delta \, \varepsilon'_{\gamma} / (1+z)$ interacting with a jet-originating soft photon with observed energy
\begin{equation}
\varepsilon_0 = {\delta \, \varepsilon'_0 \over 1+z} \simeq {2 \, \delta^2 m_e^2 c^4 \over \varepsilon_{\gamma} \, (1+z)^2} \simeq 50 \, \left({\delta \over 10}\right)^2 \left({ \varepsilon_{\gamma} \over {\rm TeV}}\right)^{-1}\,{\rm eV} \, ,
\end{equation}
may be found as
\begin{equation}
\tau_{\gamma \gamma} \simeq \int^R ds \, \int_{m_ec^2/\varepsilon_0} d\varepsilon_0 \, n'_0(\varepsilon_0) \, \sigma_{\gamma \gamma}(\varepsilon_0, \varepsilon_\gamma) \sim 0.2 \, \sigma_{\rm T} \, R \, \varepsilon'_0 \, n'_0(\varepsilon_0) \, ,
\end{equation}
where $n'_0(\varepsilon'_0)$ is the differential comoving number density of soft photons. Noting that ${\varepsilon'}^2_0 \, n'_0(\varepsilon'_0) = L'_0 / 4 \pi \, R^2 c$, where $L'_0$ is the intrinsic monochromatic luminosity at photon energy $\varepsilon'_0$, we obtain
\begin{equation}
\tau_{\gamma \gamma} \simeq {\sigma_{\rm T} \, d_L^2 F_0 \, \varepsilon_{\gamma} (1+z) \over 10 \, R \, m_e^2 c^5 \, \delta^5} \simeq 0.001 \, \left({ \varepsilon_{\gamma} \over {\rm TeV}}\right) \, \left({F_0 \over 10^{-11}\,{\rm erg/cm^2/s}}\right) \, \left({R \over 10^{17}\,{\rm cm}}\right)^{-1} \left({\delta \over 10}\right)^{-5} \, ,
\end{equation}
where $F_0 = L_0 / 4 \pi d_L^2$ is the observed monochromatic flux energy density as measured at the observed photon energy $\varepsilon_0$. Thus, for $5$\,TeV $\gamma$-rays and the model parameters discussed (implying the observed $\varepsilon_0 = 15$\,eV flux of Mrk\,501 roughly $F_0 \simeq 3.2 \times 10^{-11}$\,erg\,s$^{-1}$\,cm$^{-2}$), one has $\tau_{\gamma \gamma}(5\,{\rm TeV}) \simeq 0.005$. Therefore, the values of $R$ and $\delta$ from our SSC model fit do not need to be adjusted to take into account the influence of spectral modifications due to pair attenuation. Note that such opacity effects, studied extensively in the context of $\gamma$-ray bursts, generally yield a broken power law for the spectral form, with the position and magnitude of the break fixed by the pair-production kinematics \citep[e.g.,][and references therein]{baring06}. The broad-band continuum of Mrk\,501, and in particular its relatively flat spectrum VHE $\gamma$-ray segment, is inconsistent with such expected break. This deduction is in agreement with the above derived transparency of the emitting region for high energy $\gamma$-ray photons.

\begin{figure}[!t]
  \centering
  \includegraphics[scale=1.5]{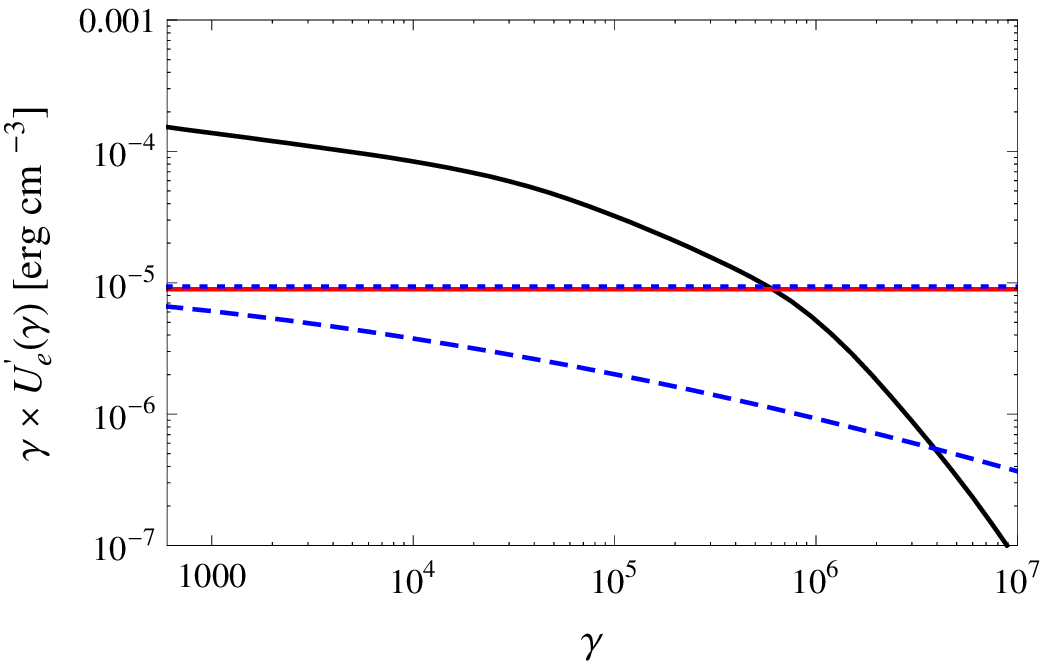}
  \caption{The jet comoving energy density of ultrarelativistic electrons per 
logarithmic energy bin, $\gamma U'_e(\gamma)$, as a function of the 
electron Lorentz factor $\gamma$ (solid black curve). For comparison, the 
comoving energy density of the magnetic field $U'_B$ (solid red line) and 
synchrotron photons $U'_{syn}$ (dotted blue line) are shown. The dashed 
blue curve denotes the comoving energy density of synchrotron photons 
which are inverse-Compton upscattered in the Thomson regime, $U'_{syn/T}$, 
for a given electron Lorentz factor $\gamma$ (see equation \ref{eq:USyncThomson}).}
  \label{U}
 \end{figure}

Next we evaluated the `monoenergetic' comoving energy density of ultrarelativistic electrons for a given electron Lorentz factor,
\begin{equation}
\gamma \, U'_e(\gamma) \equiv \gamma^2 m_e c^2 \, n'_e(\gamma) \, ,
\end{equation}
\noindent
and this is shown in Figure\,\ref{U} (solid black line). The total electron energy density is then $U'_e = \int U'_e(\gamma) \, d\gamma \simeq 5 \times 10^{-4}$\,erg\,cm$^{-3}$. As shown, most of the energy is stored in the lowest energy particles ($\gamma_{min} \simeq 600$). For comparison, the comoving energy density of the magnetic field and that of the synchrotron photons are plotted in the figure as well (horizontal solid red line and dotted blue line, respectively). These two quantities are approximately equal, namely $U'_B = B^2 / 8 \pi \simeq 0.9 \times 10^{-5}$\,erg\,cm$^{-3}$ and
\begin{equation}
U'_{syn} = {4 \pi \, R \over 3 \, c} \, \int j'_{\nu',\,syn} \, d\nu' \simeq 0.9 \times 10^{-5} \, {\rm erg\,cm^{-3}} \, .
\label{eq:USync}
\end{equation}
\noindent
In Figure\,\ref{U} we also plot the comoving energy density of synchrotron photons which are inverse-Compton upscattered in the Thomson regime for a given electron Lorentz factor $\gamma$,
\begin{equation}
U'_{syn/T}(\gamma) = {4 \pi \, R \over 3 \, c} \, \int^{\nu'_{KN}\!(\gamma)} j'_{\nu',\,syn} \, d\nu'
\label{eq:USyncThomson}
\end{equation}
\noindent
(dashed blue line), where $\nu'_{KN}\!(\gamma) \equiv m_e c^2 / 4 \gamma \, h$. Because of the well-known suppression of the inverse-Compton scattering rate in the Klein-Nishina regime, the scattering in the Thomson regime dominates the inverse-Compton energy losses\footnote{The inverse-Compton cross-section goes as $\sigma_{ic} \simeq \sigma_T$ for $\nu' < \nu'_{KN}\!(\gamma)$, and roughly as $\sigma_{ic} \sim \sigma_T \, (\nu' / \nu'_{KN})^{-1} \, \ln\!\left[\nu' / \nu'_{KN}\right]$ for $\nu' > \nu'_{KN}\!(\gamma)$ \citep[e.g.,][]{cop90}.}. Hence, one may conclude that even though the total energy density of the synchrotron photons in the jet rest frame is comparable to the comoving energy density of the magnetic field ($U'_{syn} \simeq U'_B$), the dominant radiative cooling for all the electrons is due to synchrotron emission, since $U'_{syn/T} < U'_B$ for any $\gamma$. 

The timescale for synchrotron cooling may be evaluated as
\begin{equation}
t'_{syn} \simeq {3 m_e c \over 4 \sigma_T \gamma \, U'_B} \simeq 4 \, \left({\gamma \over 10^7}\right)^{-1} \, {\rm day} \, .
\end{equation}
\noindent
Hence, $t'_{rad} \simeq t'_{syn}$ equals the dynamical timescale of the emitting region, $t'_{dyn} \simeq R/c$, for electron Lorentz factor $\gamma \simeq 8 \times 10^5$, i.e., close to the second electron break energy $\gamma_{br,\,2}$. Also the difference between the spectral indices below and above the break energy $\gamma_{br,\,2}$ determined by our modeling, namely $\Delta s_{3/2} = s_3 - s_2 = 0.95$, is very close to the `classical' synchrotron cooling break $\Delta s = 1$ expected for a uniform emission region, as discussed in \S\ref{SSCModel}. This agreement, which justifies at some level the assumed homogeneity of the emission zone, was in fact the additional constraint imposed on the model to break the degeneracy between the main free parameters. Note that in such a case the first break in the electron energy distribution around electron energy $\gamma_{br,\,1} = 4 \times 10^4$ is related to the nature of the underlying particle acceleration process. We come back to this issue in \S\ref{EED},

Another interesting result from our model fit comes from the evaluation of the mean energy of the electrons responsible for the observed non-thermal emission of Mrk\,501. In particular, the mean electron Lorentz factor is 
\begin{equation}
\langle \gamma \rangle \equiv {\int \gamma \, n'_e(\gamma) \, d\gamma \over \int n'_e(\gamma) \, d\gamma} \simeq 2400 
\end{equation}
\noindent
This value, which is determined predominantly by the minimum electron energy $\gamma_{min} = 600$ and by the position of the first break in the electron energy distribution, is comparable to the proton-to-electron mass ratio $m_p/m_e$. In other words, the mean energy of ultrarelativistic electrons within the blazar emission zone of Mrk\,501 is comparable to the energy of non-relativistic/mildly-relativistic (cold) protons. This topic will be discussed further in \S\ref{EED} as well.

The analysis presented allows us also to access the global energetics of the Mrk\,501 jet. In particular, with the given energy densities $U'_e$ and $U'_B$, we evaluate the total kinetic powers of the jet stored in ultrarelativistic electrons and magnetic field as
\begin{equation}
L_e = \pi R^2 c \Gamma^2 \, U'_e \simeq 10^{44} \, {\rm erg\,s^{-1}} \, ,
\label{eq:Le}
\end{equation}
\noindent
and
\begin{equation}
L_B = \pi R^2 c \Gamma^2 \, U'_B \simeq 2 \times 10^{42} \, {\rm erg\,s^{-1}} \, ,
\end{equation}
\noindent
respectively. In the above expressions, we have assumed that the emission region analyzed occupies the whole cross-sectional area of the outflow, and that the jet propagates at sufficiently small viewing angle that the bulk Lorentz factor of the jet equals the jet Doppler factor emerging from our modeling, $\Gamma = \delta$. These assumptions are justified in the framework of the one-zone homogeneous SSC scenario. Moreover, assuming one electron-proton pair per electron-positron pair within the emission region \citep[see][]{cel08}, or equivalently $N'_p \simeq N'_e/3$ where the total comoving number density of the jet leptons is
\begin{equation}
N'_e = \int n'_e(\gamma) \, d\gamma \simeq 0.26 \, {\rm cm^{-3}} \, ,
\end{equation}
\noindent
we obtain the comoving energy density of the jet protons $U'_p = \langle \gamma_p \rangle \, m_p c^2 N'_e/3$, and hence the proton kinetic flux $L_p = \pi R^2 c \Gamma^2 \, U'_p \simeq 0.3 \, \langle \gamma_p \rangle \, 10^{44}$\,erg\,s$^{-1}$. This is comparable to the kinetic power carried out by the leptons for mean proton Lorentz factor $\langle \gamma_p \rangle \simeq 4$ (see eq.~\ref{eq:Le}). It means that, within the dominant emission zone of Mrk\,501 (at least during non-flaring activity), ultrarelativistic electrons and mildly-relativistic protons, if comparable in number, are in approximate energy equipartition, and their energy dominates that of the jet magnetic field by two orders of magnitude. It is important to compare this result with the case of powerful blazars of the FSRQ type, for which the relatively low mean energy of the radiating electrons, $\langle \gamma \rangle \ll 10^3$, assures dynamical domination of cold protons even for a smaller proton content $N'_p/N'_e \lesssim 0.1$ \citep[see the discussion in][and references therein]{sik09}. 

Assuming $\langle \gamma_p \rangle \sim 1$ for simplicity, we find that the implied total jet power $L_j = L_e + L_p + L_B \simeq 1.4 \times 10^{44}\,{\rm erg\,s^{-1}}$ constitutes only a small fraction of the Eddington luminosity $L_{Edd} = 4 \pi \, G M_{BH} \, m_p c/\sigma_T \simeq (1.1-4.4) \times 10^{47}\,{\rm erg\,s^{-1}}$ for the Mrk\,501 black hole mass $M_{BH} \simeq (0.9-3.5) \times 10^9 M_{\odot}$ \citep{bar02}. In particular, our model implies $L_j/L_{Edd} \sim 10^{-3}$ in Mrk\,501. In this context, detailed investigation of the emission-line radiative output from the Mrk\,501 nucleus by \cite{bar02} allowed for an estimate of the bolometric, accretion-related luminosity as $L_{disk} \simeq 2.4 \times 10^{43}$\,erg\,s$^{-1}$, or $L_{disk}/L_{Edd} \sim 10^{-4}$. Such a relatively low luminosity is not surprising for BL Lacs, which are believed to accrete at low, sub-Eddington rates \citep[e.g.,][]{ghi09}. For low-accretion-rate AGN (i.e., those for which $L_{disk}/L_{Edd} < 10^{-2}$) the expected radiative efficiency of the accretion disk is $\eta_{disk} \equiv L_{disk}/L_{acc} < 0.1$ \citep{nar94,bla99}. Therefore, the jet power estimated here for Mrk\,501 is comfortably smaller than the available power of the accreting matter $L_{acc}$. 

Finally, we note that the total emitted radiative power is
\begin{equation}
L_{em} \simeq \Gamma^2 \, (L'_{syn} + L'_{ssc}) = 4 \pi R^2 c \, \Gamma^2 \, (U'_{syn} + U'_{ssc}) \sim 10^{43} \, {\rm erg\,s^{-1}} \, ,
\end{equation}
\noindent
where $U'_{syn}$ is given in Equation \ref{eq:USync} and the comoving energy density of $\gamma$-ray photons, $U'_{ssc}$, was evaluated in an analogous way as $\simeq 1.7 \times 10^{-6} \, {\rm erg\,cm^{-3}}$. This implies that the jet/blazar radiative efficiency was at the level of a few percent ($L_{em} / L_j \simeq 0.07$) during the period covered by the multifrequency campaign. Such a relatively low radiative efficiency is a common characteristic of blazar jets in general, typically claimed to be at the level of $1\%-10\%$ \citep[see][]{cel08,sik09}. On the other hand, the isotropic synchrotron and SSC luminosities of Mrk\,501 corresponding to the observed average flux levels are $L_{syn} = \delta^4 L'_{syn} \simeq 10^{45}$\,erg\,s$^{-1}$ and $L_{ssc} = \delta^4 L'_{ssc} \simeq 2 \times 10^{44}$\,erg\,s$^{-1}$, respectively.

\subsection{Electron Energy Distribution}
\label{EED}

The results of the SSC modeling presented in the previous sections indicate that the energy spectrum of freshly accelerated electrons within the blazar emission zone of Mrk\,501 is of the form $\propto E_e^{-2.2}$ between electron energy $E_{e,\,min} = \gamma_{min} m_e c^2 \sim 0.3$\,GeV and $E_{e,\,br} = \gamma_{br,\,1} m_e c^2 \sim 20$\,GeV, steepening to $\propto E_e^{-2.7}$ above $20$\,GeV, such that the mean electron energy is $\langle E_e \rangle \equiv \langle \gamma \rangle \, m_e c^2 \sim 1$\,GeV. At this point, a natural question arises: is this electron distribution consistent with the particle spectrum expected for a diffusive shock acceleration process? Note in this context that the formation of a strong shock in the innermost parts of Mrk\,501 might be expected around the location of the large bend (change in the position angle by $90^{\circ}$) observed in the outflow within the first few parsecs (projected) from the core \citep{edw02,pin09}. This distance scale could possibly be reconciled with the expected distance of the blazar emission zone from the center for the model parameters discussed, $r \sim R/\theta_j \sim 0.5$\,pc, for jet opening angle $\theta_j \simeq 1/\Gamma \ll 1$.

In order to address this question, let us first discuss the minimum electron energy implied by the modeling, $E_{e,\,min} \sim 0.3$\,GeV. In principle, electrons with lower energies may be present within the emission zone, although their energy distribution has to be very flat (possibly even inverted), in order not to overproduce the synchrotron radio photons and to account for the measured \FermiLAT $\gamma$-ray continuum. Therefore, the constrained minimum electron energy marks the injection threshold for the main acceleration mechanism, meaning that only electrons with energies larger than $E_{e,\,min}$ are picked up by this process to form the higher-energy (broken) power-law tail. Interestingly, the energy dissipation mechanisms operating at the shock fronts do introduce a particular characteristic (injection) energy scale, below which the particles are not freely able to cross the shock front. This energy scale depends on the shock microphysics, in particular on the thickness of the shock front. The shock thickness, in turn, is determined by the operating inertial length, or the diffusive mean free path of the radiating particles, or both. Such a scale depends critically on the constituents of the shocked plasma. For pure pair plasmas, only the electron thermal scale enters, and this sets $E_{e,\,min} \sim \Gamma m_e c^2$. In contrast, if there are approximately equal numbers of electrons and protons, the shock thickness can be relatively large. Diffusive shock acceleration can then operate on electrons only above a relatively high energy, establishing $E_{e,\,min} \sim \epsilon \, \Gamma m_p c^2$. Here, $\epsilon$ represents some efficiency of the equilibration in the shock layer between shocked thermal protons and their electron counterparts, perhaps resulting from electrostatic potentials induced by charge separation of species of different masses \citep{barsum07}. Our multifrequency model fits suggest that $\epsilon\sim 0.025$, providing an important blazar shock diagnostic. 

At even lower electron energies, other energization processes must play a dominant role \citep[e.g.,][]{hos92}, resulting in formation of very flat electron spectra \citep[see the related discussion in][]{sik09}. Which of these energy dissipation mechanisms are the most relevant, as well as how flat the particle spectra could be, are subjects of ongoing debates.  Different models and numerical simulations presented in the literature indicate a wide possible range for the lowest-energy particle spectral indices (below $E_{e,\,min}$), from $s_{inj} \simeq 1.0-1.5$ down to $s_{inj} \simeq -2$, depending on the particular shock conditions and parameters \citep{ama06,sir09}.

All in all, we argue that the relatively high minimum energy of the radiating electrons implied by the SSC modeling of the Mrk\,501 broadband spectrum and the overall character of the electron energy evolution in this source are consistent with a proton-mediated shock scenario. In addition, the fact that the mean energy of the ultrarelativistic electrons is of the order of the proton rest energy, $\langle E_e \rangle \sim m_p c^2$, can be reconciled with such a model. Moreover, the constrained power-law slope of the electrons with energies $E_{e,\,min} \leq E_e \leq E_{e,\,br}$, namely $s_1 = 2.2$, seems to suggest a dominant role for diffusive shock acceleration above the injection energy $E_{e,\,min}$, as this value of the spectral index is often claimed in the literature for particles undergoing first-order Fermi acceleration at relativistic shock \citep{bed98,kir00,ach01}. The caveat here is that this result for the `universal' particle spectral index holds only for particular conditions \citep[namely, for ultrarelativistic shock velocities with highly turbulent conditions downstream of the shock: see the discussion in][]{ost02}, whereas in general a variety of particle spectra may result from the relativistic first-order Fermi mechanism, depending on the local magnetic field and turbulence parameters at the shock front, and the speed of the upstream flow \citep{kh89,nie04,lem06,sir09,bar10}. Nevertheless, the evidence for ultrarelativistic electrons with spectral index $s_1 = 2.2$ in the jet of Mrk\,501 may be considered as an indication that the plasma conditions within the blazar emission zone allow for efficient diffusive shock acceleration (at least in this source), as described by the simplest asymptotic test-particle models, though only in a relatively narrow electron energy range. 

If the relativistic shock acceleration plays a dominant role in the blazar Mrk 501 (as argued above), the observations and the model results impose important constraints on this mechanism, many aspects of which are still not well understood. Firstly, this process must be very efficient in the sense that all the electrons pre-accelerated/preheated to the energy $E_{e,\,min} \sim 0.3$\,GeV are picked up by the acceleration process so that a single electron component is formed above the injection threshold and there is no Maxwellian-like population of particles around $E_{e,\,min}$ outnumbering the higher energy ones from the power-law tail\footnote{Note that due to low accretion rates and thus low luminosities of the accretion disks in BL Lacs, the number of non-relativistic/mildly-relativistic electrons (Lorentz factors $\gamma \sim 1$) cannot be constrained by analyzing `bulk-Compton' spectral features in the observed SED of Mrk\,501, in contrast to the situation in FSRQs \citep[see][]{sik00}.}. The second constraint is due to the presence of the spectral break $\Delta s_{2/1} = s_2-s_1 \simeq 0.5$ around electron energies $E_{e,\,br} \sim 20$\,GeV. As discussed in the previous section, this break cannot be simply a result of cooling or internal pair-attenuation effects, and hence must be accounted for by the acceleration mechanism. The discussion regarding the origin of this break --- which may reflect variations in the global field orientation or turbulence levels sampled by particles of different energy --- is beyond the scope of this paper. However, the presence of high-energy breaks in the electron energy distribution (intrinsic to the particle spectrum rather than forming due to cooling or absorption effects) seems to be a common property of relativistic jet sources observed by \FermiLATc, such as the FSRQ objects 3C\,454.3 and AO~0235+164 \citep{abd09b,abdoSpectralProperties}.

\subsection{Variability}
\label{SEDvariability}

In \S\ref{LC}--\S\ref{FermiSpectrum}, we reported on the $\gamma$-ray flux and spectral variability of Mrk\,501 as measured by the \FermiLAT instrument during the first 16 months of operation. In this section, we discuss whether the observed spectral evolution can be accounted for by our simple one-zone SSC model. 

\begin{figure}[!t]
  \centering
  \includegraphics[scale=1.25]{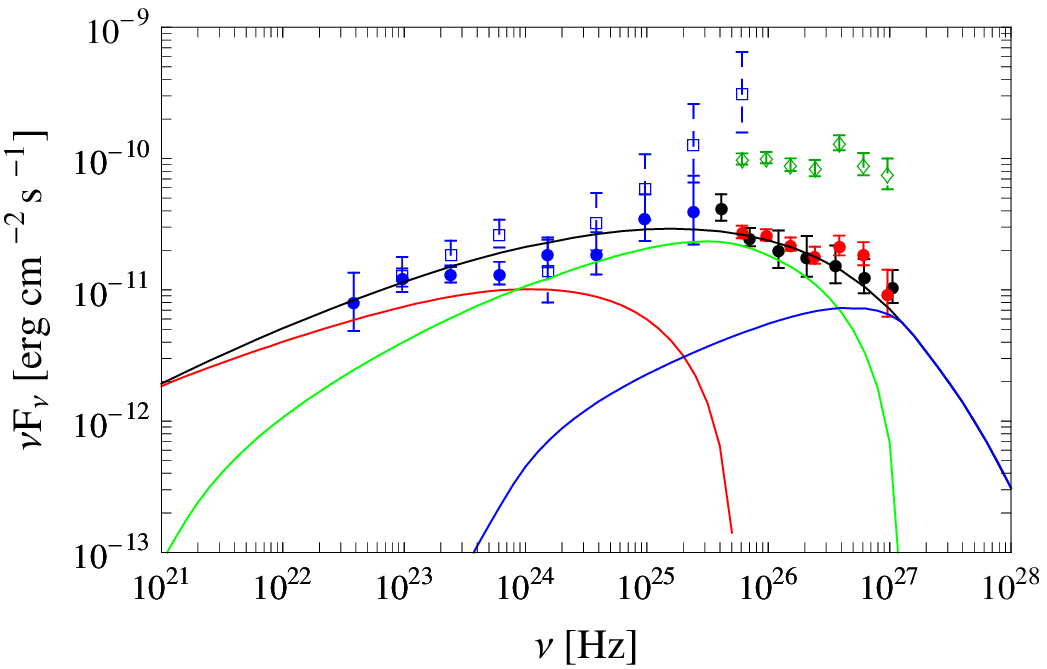}
  \includegraphics[scale=1.25]{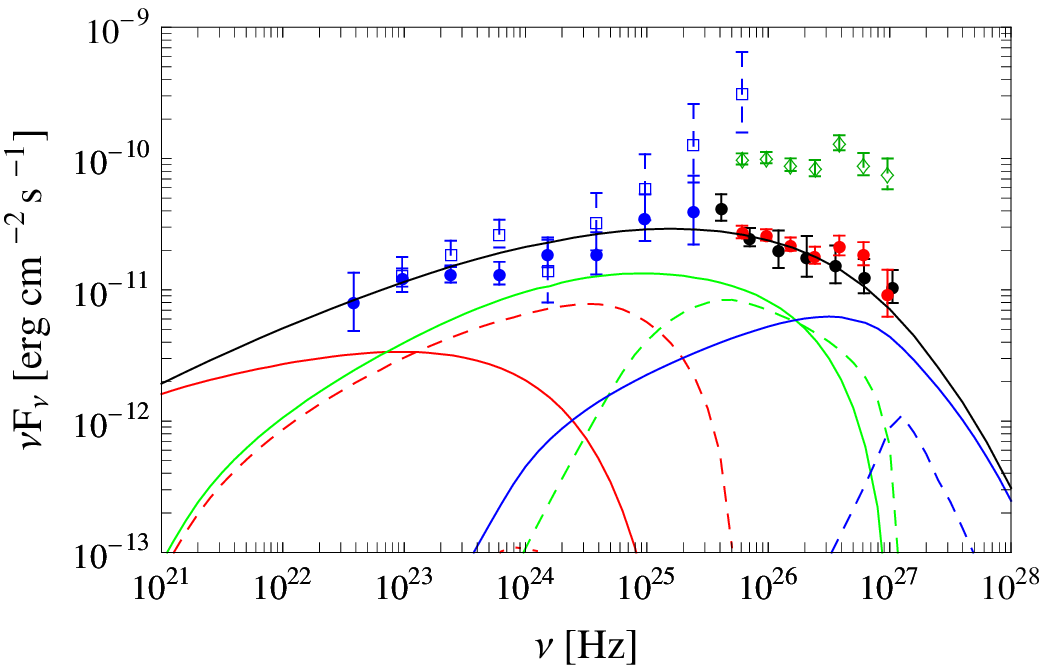}
  \caption{The decomposition of the SSC continuum for Mrk\,501. The data points are the same as in the bottom panel of Figure~\ref{fig:MWSEDGamma}. The SSC fit to the average spectrum is denoted by the solid black curve. {\bf Top:} Contributions of the different segments of the electron spectrum Comptonizing the whole synchrotron continuum (red curve: $\gamma_{min} < \gamma < \gamma_{br,\,1}$; green curve: $\gamma_{br,\,1} < \gamma < \gamma_{br,\,2}$; blue curve: $\gamma_{br,\,2} < \gamma$). {\bf Bottom:} Contributions of the different segments of the electron spectrum (as in the top panel) Comptonizing different segments of the synchrotron continuum (solid curves: $\nu < \nu_{br,\,1} \simeq 10^{15}$\,Hz; dashed curves: $\nu_{br,\,1} < \nu < \nu_{br,\,2} \simeq 6 \times 10^{17}$\,Hz; curves for $\nu > \nu_{br,\,2}$ do not appear in the plot because the corresponding flux levels are all less than $10^{-13}$\,erg\,cm$^{-1}$\,s$^{-1}$).}
  \label{fit2}
 \end{figure}

Figure\,\ref{fit2} (top panel) presents in more detail the GeV--TeV $\gamma$-ray spectrum of Mrk\,501, together with the decomposition of the SSC model continuum. Here the contributions of different segments of the electron energy distribution are indicated by different colors. As shown, the low-energy electrons, $\gamma_{min} \leq \gamma < \gamma_{br,\,1}$, which emit synchrotron photons up to $\simeq 10^{15}$\,Hz, dominate the production of $\gamma$-rays up to a few GeV (red line). The contribution of higher-energy electrons with Lorentz factors $\gamma_{br,\,1} \leq \gamma < \gamma_{br,\,2}$ is pronounced within the observed synchrotron range $10^{15}-10^{18}$\,Hz, and at $\gamma$-ray energies from a few GeV up to $\sim$\,TeV  (green line). Finally, the highest energy tail of the electron energy distribution, $\gamma \geq \gamma_{br,\,2}$, responsible for the observed hard-X-ray synchrotron continuum ($>2$\,keV) in the fast cooling regime, generates the bulk of $\gamma$-rays with observed energies $>$\,TeV (blue line). Interestingly, even though any sharp breaks in the underlying electron energy distribution are `smeared out' into a smoothly curved spectral continuum due to the nature of the synchrotron self-Compton emission, the average data set does support the presence of distinct low-energy and high-energy segments in the electron spectrum.

It therefore seems reasonable to argue that the spectral variability of Mrk\,501 observed by \FermiLAT may be explained by postulating that the low-energy segment of the electron energy distribution ($\gamma < \gamma_{br,\,1}$) is characterized by only small flux variations, while the high-energy electron tail ($\gamma > \gamma_{br,\,1}$) varies more substantially. In such a scenario, some correlation might be expected between the fluxes in the UV-to-soft-X-ray photon energies from the synchrotron bump and the GeV--TeV fluxes from the inverse-Compton bump. This expectation is not inconsistent with the fact that we do not see any obvious relation between the ASM/BAT fluxes and the LAT ($> 2$\,GeV) fluxes (see Figure\,\ref{fig:Lc30daysMW}), because the electrons producing X-ray synchrotron photons above $2-3$\,keV contribute to the SSC emission mostly at the highest photon energies in the TeV range (see Figure\,\ref{fit2}). This issue will be studied in detail (on timescales of 1 month down to 5 days) in a forthcoming publication using the data from this multifrequency campaign. 

The X-ray/TeV connection has been established in the past for many BL Lacs (and for Mrk\,501 in particular). However, the exact character of the correlation between the X-ray and TeV fluxes is known to vary from object to object, and from epoch to epoch in a given source, as widely discussed in, e.g., \cite{kra04,bla05,gli06,fos08}. Note that the data analyzed in those papers were obtained mostly during periods of high activity. Consequently, the conclusions presented were somewhat biased towards flaring activity, and hence they might not apply to the typical (average) behaviour of the source, which is the main focus of this paper. Moreover, the data set from our campaign includes UV fluxes, soft-X-ray fluxes (down to $0.3$\,keV; Swift), and $\gamma$-ray fluxes spanning a very wide photon energy range ($0.1$\,GeV$-10$\,TeV; \FermiLAT combined with MAGIC and VERITAS). This unique data set allows us to evaluate the multifrequency variability and correlations for Mrk\,501 over an unprecedented range of photon energies.

Considering only the data set presented in this paper, we note that by steepening the high-energy electron continuum above the intrinsic break energy $\gamma_{br,\,1}$ (and only slightly adjusting the other model parameters), one can effectively remove photons above $10$\,GeV in the SSC component, leaving a relatively steep spectrum below $10$\,GeV, similar to the one observed by \FermiLAT during the time interval MJD 54862--54892 (see \S\ref{FermiSpectrum}). Such a change should be accompanied by a decrease in the UV-to-soft-X-ray synchrotron fluxes by a factor of a few, but the data available during that time interval are not sufficient to detect this effect\footnote{The 4.5-month multifrequency campaign started 13 days after the end of the 30-day time interval MJD 54862--54892. Therefore, for this epoch, the only additional multifrequency data are from \RXTEc-ASM and \Swiftc-BAT, which have only moderate ability to detect Mrk\,501 on short timescales.}. This statement is further justified by the bottom panel in Figure\,\ref{fit2}, where the contributions of the different segments of electrons Comptonizing the different segments of the synchrotron bump to the average $\gamma$-ray emission of Mrk\,501 are shown. Note that the lowest-energy electron population ($\gamma < \gamma_{br,\,1}$) inverse-Compton upscattering only synchrotron photons emitted by the same population ($\nu < \nu_{br,\,1} \sim 10^{15}$\,Hz; solid red line) may account for the bulk of the observed steep-spectrum $\gamma$-ray emission. 

Another important conclusion from this figure is that Comptonization of the highest-energy synchrotron photons ($\nu > \nu_{br,\,2} \sim 6 \times 10^{17}$\,Hz) by electrons with arbitrary energies produces only a negligible contribution to the average $\gamma$-ray flux of Mrk\,501 due to Klein-Nishina suppression. Thus, the model presented here explains in a natural way the fact that the X-ray and TeV fluxes of TeV-emitting BL Lacs are rarely correlated according to the simple scaling $F_{\rm TeV} \propto F_{\rm keV}^2$ which would be expected from the class of SSC models in which the highest-energy electrons upscatter (in the Thomson regime) their own synchrotron photons to the TeV band \cite[see, e.g.,][]{gli06,fos08}. In addition, it opens a possibility for accommodating short-timescale variability ($t_{var} < 4$\,d) at the highest synchrotron and inverse-Compton frequencies (hard X-rays and TeV photon energies, respectively). The reason for this is that, in the model considered here, these high-energy tails of the two spectral components are produced by the highest-energy electrons which are deep in the strong cooling regime (i.e., for which $t'_{rad} \ll R/c$), and thus the corresponding flux changes may occur on timescales shorter than $R/c \,\delta$ \citep[see in this context, e.g.,][]{chi99,kat00}.

A more in-depth analysis of the multifrequency data set (including correlation studies of the variability in different frequency ranges) will be presented in a forthcoming paper. The epoch of enhanced $\gamma$-ray activity of Mrk\,501 (MJD 54952--54982; see \S\ref{FermiSpectrum}) may be more difficult to explain in the framework of the one-zone SSC model, because a relatively flat \FermiLAT spectrum above $10$\,GeV, together with an increased TeV flux as measured by the VERITAS and Whipple 10\,m telescopes around this time, may not be easy to reproduce with a set of model parameters similar to that discussed in previous sections. This is mostly due to Klein-Nishina effects, which tend to steepen the high-energy tail of the SSC component, thus precluding the formation of a flat power law extending beyond the observed TeV energies. Hence, detailed modeling and data analysis will be needed to determine whether the enhanced VHE $\gamma$-ray activity period can be accommodated within a one-zone SSC model, or whether it will require a multi-zone approach. 

\begin{deluxetable}{lll}
\label{TableModelresults}
%\rotate
\tabletypesize{\scriptsize}
\tablecolumns{3} 
\tablewidth{0pc}
\tablecaption{Parameters of the blazar emission zone in Mrk\,501.}
\tablehead{\colhead{Parameter}	&\colhead{Main SSC fit considered} &\colhead{Exemplary SSC fit}}  
\startdata 
Magnetic field			&	$B = 0.015$\,G	&	$B = 0.03$\,G\\
Emission region size		&	$R = 1.3 \times 10^{17}$\,cm	&	$R = 0.2 \times 10^{17}$\,cm\\
Jet Doppler and bulk Lorentz factors			& $\Gamma = \delta = 12$	& $\Gamma = \delta = 22$\\
Equipartition parameter			&	$\eta_e \equiv U'_e/U'_B = 56$	&	$\eta_e \equiv U'_e/U'_B = 130$\\
Minimum electron energy                 & $\gamma_{min} = 600$	& $\gamma_{min} = 300$ \\
Intrinsic electron break energy                 & $\gamma_{br,\,1} = 4 \times 10^4$	& $\gamma_{br,\,1} = 3 \times 10^4$ \\
Cooling electron break energy                 & $\gamma_{br,\,2} = 9 \times 10^5$	 & $\gamma_{br,\,2} = 5 \times 10^5$ \\
Maximum electron energy                 & $\gamma_{max} = 1.5 \times 10^7$	& $\gamma_{max} = 0.3 \times 10^7$ \\
Low-energy electron index			& $s_1 = 2.2$	& $s_1 = 2.2$\\
High-energy electron index			& $s_2 = 2.7$	& $s_2 = 2.7$\\
Electron index above the cooling break & $s_3 = 3.65$	& $s_3 = 3.5$\\
\\
\hline
\\
Mean electron energy                 & $\langle \gamma \rangle \simeq 2400$	& $\langle \gamma \rangle \simeq 1200$ \\
Main variability timescale			& $t_{var} \simeq 4$\,day	& $t_{var} \simeq 0.35$\,day\\
Comoving electron energy density	& $U'_e \simeq 0.5 \times 10^{-3}$\,erg\,cm$^{-3}$	& $U'_e \simeq 4.6 \times 10^{-3}$\,erg\,cm$^{-3}$\\
Comoving magnetic field energy density	& $U'_B \simeq 0.9 \times 10^{-5}$\,erg\,cm$^{-3}$	& $U'_B \simeq 3.6 \times 10^{-5}$\,erg\,cm$^{-3}$\\
Comoving energy density of synchrotron photons	& $U'_{syn} \simeq 0.9 \times 10^{-5}$\,erg\,cm$^{-3}$	& $U'_{syn} \simeq 3.1 \times 10^{-5}$\,erg\,cm$^{-3}$\\
Comoving electron number density	&	$N'_e \simeq 0.3$\,cm$^{-3}$	&	$N'_e \simeq 4.6$\,cm$^{-3}$\\
\\
\hline
\\
Luminosity of the host galaxy		&	$L_{star} \simeq 3 \times 10^{44}$\,erg\,s$^{-1}$ & $L_{star} \simeq 3 \times 10^{44}$\,erg\,s$^{-1}$\\
Jet power carried by electrons		&	$L_e \simeq 1.1 \times 10^{44}$\,erg\,s$^{-1}$	&	$L_e \simeq 0.85 \times 10^{44}$\,erg\,s$^{-1}$\\
Jet power carried by magnetic field		&	$L_B \simeq 2 \times 10^{42}$\,erg\,s$^{-1}$	&	$L_B \simeq 0.65 \times 10^{42}$\,erg\,s$^{-1}$\\
Jet power carried by protons$^{a}$		&	$L_p \simeq 3 \times 10^{43}$\,erg\,s$^{-1}$	&	$L_p \simeq 4.2 \times 10^{43}$\,erg\,s$^{-1}$\\
Total jet kinetic power		&	$L_j \simeq 1.4 \times 10^{44}$\,erg\,s$^{-1}$	&	$L_j \simeq 1.3 \times 10^{44}$\,erg\,s$^{-1}$\\
Total emitted power		&	$L_{em} \simeq 9.7 \times 10^{42}$\,erg\,s$^{-1}$	&	$L_{em} \simeq 2.7 \times 10^{42}$\,erg\,s$^{-1}$\\
Isotropic synchrotron luminosity		&	$L_{syn} \simeq 10^{45}$\,erg\,s$^{-1}$	&	$L_{syn} \simeq 10^{45}$\,erg\,s$^{-1}$\\
Isotropic SSC luminosity		&	$L_{ssc} \simeq 2 \times 10^{44}$\,erg\,s$^{-1}$	&	$L_{ssc} \simeq 2 \times 10^{44}$\,erg\,s$^{-1}$\\

\enddata
\tablecomments{$^{a}$ Assuming one electron-proton pair per electron-positron pair, and mean proton Lorentz factor $\langle \gamma_p \rangle \sim 1$.}
\end{deluxetable}

\section{Conclusions} 
\label{Conclusions}

We have presented a study of the $\gamma$-ray activity of Mrk\,501 as
measured by the LAT instrument on board the Fermi satellite during its
first 16 months of operation, from 2008 August 5 (MJD 54683) to 2009 November 27 (MJD 55162). Because of the large leap in capabilities of LAT with respect to its predecessor, EGRET, this is the most extensive study to date of the $\gamma$-ray activity of this object at GeV-TeV photon energies. The \FermiLAT spectrum (fitted with a single power-law function) was evaluated for 30-day time intervals. The average photon flux above $0.3$\,GeV was found to be $(2.15 \pm 0.11) \times 10^{-8}$\,ph\,cm$^{-2}$\,s$^{-1}$, and the average photon index  $1.78 \pm 0.03$. We observed only relatively mild (factor less than 2) $\gamma$-ray flux variations, but we detected remarkable spectral variability. In particular, during the four consecutive 30-day intervals of the ``enhanced $\gamma$-ray flux'' (MJD 54862--54982), the photon index changed from $2.51 \pm 0.20$ (for the first interval) down to $1.63 \pm 0.09$ (for the fourth one). During the whole period of 16 months, the hardest spectral index within the LAT energy range was $1.52 \pm 0.14$, and the softest one was $2.51 \pm 0.20$. Interestingly, this outstanding (and quite unexpected) variation in the slope of the GeV continuum did not correlate with the observed flux variations at energies above 0.3\,GeV.

We compared the $\gamma$-ray activity measured by LAT in two different energy ranges ($0.2-2$\,GeV and $> 2$\,GeV) with the X-ray activity recorded by the all-sky instruments \RXTEc-ASM ($2-10$\,keV) and \Swiftc-BAT ($15-50$\,keV). We found no significant difference in the amplitude of the variability between X-rays and $\gamma$-rays, and no clear relation between the X-ray and $\gamma$-ray flux changes. We note, however, that the limited sensitivity of the ASM and (particularly) the BAT instruments to detect Mrk\,501 in a 30-day time interval, together with the relatively stable X-ray emission of Mrk\,501 during the observations, precludes any detailed X-ray/$\gamma$-ray variability or correlation analysis.

In this paper we also presented the first results from a 4.5-month
multifrequency campaign on Mrk\,501, which lasted from 2009 March 15
(MJD 54905) to 2009 August 1  (MJD 55044). During this period, the source was systematically observed with different instruments covering an extremely broad segment of the electromagnetic spectrum, from radio frequencies up to TeV photon energies. In this manuscript, we have focused on the average SED emerging from the campaign. Further studies on the multifrequency variability and correlations will be covered in a forthcoming publication. 

We have modeled the average broadband spectrum of Mrk\,501 (from radio
to TeV) in the framework of the standard one-zone synchrotron
self-Compton model, obtaining a satisfactory fit to the experimental
data. We found that the dominant emission region in this source can be characterized by a size of $R \simeq 10^3 \, r_g$, where $r_g \sim 1.5 \times 10^{14}$\,cm is the gravitational radius of the black hole ($M_{BH} \simeq 10^9 M_{\odot}$) hosted by Mrk\,501. The intrinsic (i.e., not affected by cooling or absorption effects) energy distribution of the radiating electrons required to fit the data was found to be of a broken power-law form in the energy range $0.3$\,GeV$-10$\,TeV, with spectral indices 2.2 and 2.7 below and above the break energy of $E_{e,\,br} \sim 20$\,GeV, respectively. In addition, the model parameters imply that all the electrons cool predominantly via synchrotron emission, forming a cooling break at $0.5$\,TeV. We argue that the particular form of the electron energy distribution emerging from our modeling is consistent with the scenario in which the bulk of the energy dissipation within the dominant emission zone of Mrk\,501 is related to relativistic, proton-mediated shock waves. The low-energy segment of the electron energy distribution ($E_e < E_{e,\,br}$) formed thereby, which dominates the production of $\gamma$-rays observed below a few GeV, seems to be characterized by low and relatively slow variability. On the other hand, the high-energy electron tail ($E_e > E_{e,\,br}$), responsible for the bulk of the $\gamma$-rays detected above a few GeV, may be characterized by more significant variability.  

Finally, we found that ultrarelativistic electrons and
mildly-relativistic protons within the blazar zone of Mrk\,501, if
comparable in number, are in approximate energy equipartition, with
their energy dominating the energy in the jet magnetic field by about
two orders of magnitude. The model fit implies also that the total jet
power, $L_j \simeq 10^{44}$\,erg\,s$^{-1}$, constitutes only a small
fraction of the Eddington luminosity, $L_j/L_{Edd} \sim 10^{-3}$, but
is an order of magnitude larger than the bolometric, accretion-related
luminosity of the central engine, $L_j/L_{disk} \sim 10$. Finally, we
estimated the radiative efficiency of the Mrk\,501 jet to be at the
level of a few percent, $L_{em} / L_j \lesssim 0.1$, where $L_{em}$ is
the total emitted power of the blazar.  The results from this study could perhaps be extended to all HSP BL Lacs. 

\section{Acknowledgments} 

\acknowledgments

The \FermiLAT Collaboration acknowledges the generous support of a number of agencies and institutes that have supported the \FermiLAT Collaboration. These include the National Aeronautics and Space Administration and the Department of Energy in the United States, the Commissariat \`a l'Energie Atomique and the Centre National de la Recherche Scientifique / Institut National de Physique Nucl\'eaire et de Physique des Particules in France, the Agenzia Spaziale Italiana and the Istituto Nazionale di Fisica Nucleare in Italy, the Ministry of Education, Culture, Sports, Science and Technology (MEXT), High Energy Accelerator Research Organization (KEK) and Japan Aerospace Exploration Agency (JAXA) in Japan, and the K.\ A.\ Wallenberg Foundation, the Swedish Research Council and the Swedish National Space Board in Sweden. Additional support for science analysis during the operations phase is gratefully
acknowledged from the Istituto Nazionale di Astrofisica in Italy and the Centre National d'\'Etudes Spatiales in France.

The MAGIC collaboration would like to thank the Instituto de Astrof\'{\i}sica de Canarias for the excellent working conditions at the Observatorio del Roque de los Muchachos in La Palma.
The support of the German BMBF and MPG, the Italian INFN,  the Swiss National Fund SNF, and the Spanish MICINN is gratefully acknowledged. This work was also supported by the Marie Curie program, by the CPAN CSD2007-00042 and MultiDark CSD2009-00064 projects of the Spanish Consolider-Ingenio 2010 programme, by grant DO02-353 of the Bulgarian NSF, by grant 127740 of 
the Academy of Finland, by the YIP of the Helmholtz Gemeinschaft, by the DFG Cluster of Excellence ``Origin and Structure of the Universe'', and by the Polish MNiSzW Grant N N203 390834.

VERITAS is supported by grants from the US Department of Energy, the US National Science Foundation, and the Smithsonian Institution, by NSERC in Canada, by Science
Foundation Ireland, and by STFC in the UK. The VERITAS Collaboration also acknowledges the support of the Fermi/LAT team through the Guest Investigator
Program Grant  NNX09AT86G. 

We acknowledge the use of public data from the Swift and RXTE data
archive. The Mets\"ahovi team acknowledges the support from the
Academy of Finland to the observing projects (numbers 212656, 210338,
among others).  This research has made use of data obtained from the
National Radio Astronomy Observatory's Very Long Baseline Array
(VLBA), projects BK150, BP143 and MOJAVE. The National Radio Astronomy
Observatory is a facility of the National Science Foundation operated
under cooperative agreement by Associated Universities, Inc.
St.Petersburg University team acknowledges support from Russian RFBR
foundation via grant 09-02-00092. AZT-24 observations are made within
an agreement between  Pulkovo, Rome and Teramo observatories. This
research is partly based on observations with the 100-m telescope of
the MPIfR (Max-Planck-Institut f\"ur Radioastronomie) at Effelsberg,
as well as with the Medicina and Noto telescopes operated by INAF -
Istituto di Radioastronomia. The Submillimeter Array is a joint
project between the Smithsonian  Astrophysical Observatory and the
Academia Sinica Institute of Astronomy and  Astrophysics and is funded
by the Smithsonian Institution and the Academia Sinica. M. Villata
organized the optical-to-radio observations by GASP-WEBT as the
president of the collaboration. Abastumani Observatory team
acknowledges financial support by the Georgian National Science
Foundation through grant GNSF/ST07/4-180. The OVRO 40 m program was
funded in part by NASA (NNX08AW31G) and the NSF (AST-0808050).

\end{document}